\documentclass[journal,comsoc]{IEEEtran}

\usepackage{cite}
\usepackage{amsmath,amssymb,amsfonts}
\usepackage{algorithmic}
\usepackage{graphicx}
\usepackage{textcomp}
\def\BibTeX{{\rm B\kern-.05em{\sc i\kern-.025em b}\kern-.08em
    T\kern-.1667em\lower.7ex\hbox{E}\kern-.125emX}}
\usepackage{caption}
\usepackage{subcaption}
\usepackage{booktabs}
\usepackage{multirow}
\usepackage{threeparttable}
\usepackage{balance}

\usepackage{authblk}

\usepackage{afterpage}
\usepackage{multicol}
\newsavebox{\shortpagebox}
\usepackage[hyphens]{url}

\newtheorem{definition}{Definition}[section]

\usepackage{makecell}
\usepackage{pifont}
\newcommand{\cmark}{\ding{51}}%
\newcommand{\xmark}{\ding{55}}%

\usepackage{xcolor,colortbl}

\definecolor{Gray}{gray}{0.85}
\definecolor{LightCyan}{rgb}{0.88,1,1}
\definecolor{DarkOrange}{rgb}{0.97, 0.72, 0.47} 

\usepackage{color}
\definecolor{accessblue}{cmyk}{1,0.3,0,0.2}
\definecolor{greycolor}{cmyk}{0,0,0,1}

\usepackage[first=0,last=9]{lcg}

\newcommand*{\ik}[1]{\textcolor{black}{#1}}

\usepackage{soul}

\newcommand{\dor}{\cellcolor{DarkOrange}}

\usepackage{mathpazo}

\usepackage{float}

\usepackage[hidelinks]{hyperref}
\hypersetup{breaklinks=true}

\makeatletter
\def\ps@IEEEtitlepagestyle{%
  \def\@oddfoot{\mycopyrightnotice}%
  \def\@oddhead{\hbox{}\@IEEEheaderstyle\leftmark\hfil\thepage}\relax
  \def\@evenhead{\@IEEEheaderstyle\thepage\hfil\leftmark\hbox{}}\relax
  \def\@evenfoot{}%
}
\def\mycopyrightnotice{%
  \begin{minipage}{\textwidth}
  \centering \scriptsize
  Copyright~\copyright~2023 IEEE. Personal use of this material is permitted. Permission from IEEE must be obtained for all other uses, in any current or future media, including reprinting/republishing this material for advertising or promotional purposes, creating new collective works, for resale or redistribution to servers or lists, or reuse of any copyrighted component of this work in other works.”
  \end{minipage}
}
\makeatother

\title{\bf{Don't be a Victim During a Pandemic! Analysing Security and Privacy Threats in Twitter During COVID-19}}

\author[1]{Bibhas Sharma}
\author[1,*]{Ishan Karunanayake}
\author[1]{Rahat Masood}
\author[2]{Muhammad Ikram}
\affil[1]{University of New South Wales (UNSW) - Sydney, Australia}
\affil[2]{Macquarie University - Macquarie Park, Australia}
\affil[*]{Corresponding author - \textit{ishan.karunanayake@unsw.edu.au}}

\begin{document}

\maketitle

\begin{abstract}

There has been a huge spike in the usage of social media platforms during the COVID-19 lockdowns. These lockdown periods have resulted in a set of new cybercrimes, thereby allowing attackers to victimise social media users with a range of threats. This paper performs a large-scale study to investigate the impact of a pandemic and the lockdown periods on the security and privacy of social media users. We analyse 10.6 Million COVID-related tweets from 533 days of data crawling and investigate users' security and privacy behaviour in three different periods (i.e., before, during, and after the lockdown). Our study shows that users unintentionally share more personal identifiable information when writing about the pandemic situation (e.g., sharing nearby coronavirus testing locations) in their tweets. The privacy risk reaches 100\% if a user posts three or more sensitive tweets about the pandemic. We investigate the number of suspicious domains shared on social media during different phases of the pandemic. Our analysis reveals an increase in the number of suspicious domains during the lockdown compared to other lockdown phases. We observe that IT, Search Engines, and Businesses are the top three categories that contain suspicious domains. Our analysis reveals that adversaries' strategies to instigate malicious activities change with the country's pandemic situation.

\end{abstract}
 
\section{Introduction}
\label{sec:introduction}

\ik{Pandemics, such as COVID-19, always have a devastating impact on people's personal, social, and professional lives and lead to a series of cybersecurity threats for online users. During COVID-19, there has been a substantial increase in a range of different cyber attacks such as phishing, ransomware, spamming, and malicious messaging \cite{khan2020ten}. Employees were abruptly forced to work from home without proper training and arrangements when most companies did not have the necessary infrastructure and plans for such a drastic change. In addition, only a small percentage of companies had cybersecurity policies in place \cite{pranggono2021covid}. Moreover, school children and their parents became more frequent online users, and most were unaware of cybersecurity threats and their impact. These reasons considerably increased the playing field of cybercriminals by providing them with more attack vectors. For instance, German companies suffered around 53 billion Euros worth of damages due to Cyberattacks as a result of working from home \cite{look2021wfhattacks}.}

In addition, the problem became more devastating with the increasing use of social media platforms to share public and personal information related to the pandemic. According to one of the reports \cite{kemp2021smstats}, social media users increased by 13.2\% (+490 million) in 2020 and by 10.1\% (+424 million) in 2021. It has led to a massive increase in user generated-data that possess various privacy threats, thereby making social media platforms an appealing target for organisations to aggregate such information for legitimate or malicious intent \cite{beigi2018privacy}. According to a report by United State's Federal Trade Commission, \cite{emma2022smscam}, in 2020, there have been losses of up to 258 million dollars as a result of social media scams, while that number rose up to a massive 770 million in 2021. Another example of exploiting social media for financial fraud is the use of Twitter bots to trick users to make payments to illegitimate accounts using Paypal or Venmo \cite{jessica2021smScam}.

\ik{To this end, this research aims to investigate the impact of lockdown periods during a pandemic by taking COVID-19 as a case study on the security and privacy of social media users. To be more specific, in this paper, we mainly investigate the research question; \textit{How has the pandemic and the resulting lockdowns affected the privacy and security of Twitter users?} When considering privacy, we analyse the trends of sharing private information on Twitter and identify whether the pandemic has unintentionally caused people to share their private information. For example, we identified that some people had shared information about the location of their homes in relation to vaccination centres, while others have shared their medical conditions along with concerns about vaccinations. When it comes to security, in this work, we mainly consider the exposure to suspicious content. To do that, we identify the malicious URLs shared during several stages of the pandemic and try to ascertain different strategies used by adversaries to spread such URLs. For example, we noticed that during situations where people relied heavily on government announcements and news (e.g. border closures, lockdowns, social distancing restrictions), there were many suspicious domains that were categorised as related to Government and Law enforcement.} 

\ik{In addition to this main research question, we also intend to examine the influence of social media networks (in this case, Twitter) in managing the COVID-19 pandemic based on human behaviour and sentiments. In other words, as a secondary contribution, we try to answer the questions \textit{What are the most popular topics related to the pandemic during the lockdown periods, and what are the user sentiments regarding these topics?} There are many works that have done topic modelling and sentiment analysis on social media content \cite{WANG2018, NKOMO2020, ASLAN2020}. However, only a handful of works have focused on topic modelling and sentiment analysis during the pandemic \cite{boonitt2020, hussain2021, Huang2020, ChandraGuntuku2020, Visentin2021}. These works have a special focus, and their insights are related to a specific area. For example, \cite{hussain2021} investigates the sentiment on the COVID-19 vaccine, \cite{ChandraGuntuku2020} analyses the mental health concerns during the pandemic and \cite{Visentin2021} look into the sentiments regarding the spread of conspiracy theories. Meanwhile, our work focuses on identifying the most popular topics in general during the pandemic (and related to the pandemic) and analyses the public sentiment on all of them. In addition to these distinctions, our work includes two other considerations. For most of our analysis, the dataset is subdivided under two bases: the pandemic stage (i.e., before, during, and after the lockdown) and the country (Australia, India, UK, and the US). This classification allows us to conduct a systematic study to identify and analyse commonly discussed topics, public sentiments, privacy and security risks. We also investigate how these aspects vary across countries compared to global trends. Moreover, looking at the Infection Rates (IR) of the countries gives more insights into relating it to social posts and people's sentiments.}

In order to do all this work, we used 10 Million COVID-related tweets, which were posted on Twitter from 01 Jan 2020 to 21 June 2021 (533 days). In essence, this paper makes the following four main contributions:

{\bf Collection and characterisation of large-scale dataset.} We collect (cf. \S~\ref{sec:dccharacterisation}) a large dataset consisting of 10 million tweets from four different geolocations spanning over 533 days. We classify this dataset into three phases of the pandemic (i.e., before, during, and after the lockdown). We first perform \textit{Hashtag analysis} to identify the topics that people are mostly discussing on social media platforms about the pandemic. Our analysis indicates that \texttt{supporting businesses, politics,} and \texttt{latest news/updates} have been more frequent topics during all the stages of the pandemic. We also perform \textit{URL analysis} on the tweets and find that users share \texttt{social media} URLs from Twitter, Facebook, Instagram and \texttt{news and media} URLs to propagate information about the pandemic.
    
{\bf Perception analysis toward COVID-19.} We perform (cf. \S~\ref{sec:sentAnalysis})  \textit{Sentiment analysis} on the tweets to explore the people's perception (i.e., emotions and feelings of people) during the three phases of the pandemic. Our study shows that COVID-19 restriction rules such as \texttt{social distancing} received a high positive sentiment of approximately 70\% from the public. Similarly, \texttt{staying home} received a positive sentiment of approximately 45\% from the community. On the contrary, \texttt{political discussions} and \texttt{death tolls} have a highly negative sentiment of approximately 50\% for all three phases. Moreover, we observe that the user sentiments directly relate to the IR of a region. For example, people show negative sentiments on the \texttt{death toll} and positive sentiments on the \texttt{social distancing} topics when the IR of a country is higher.
    
{\bf Privacy risks exposure.} We investigate (cf. \S~\ref{sec:prvcAnalysis}) the trend of sharing private information on social media platforms during the pandemic. For example, we investigate whether or not people are more inclined to share their personal information, such as their names, addresses, or locations, during a lockdown. We use a probabilistic framework that quantifies the privacy of user tweets based on three privacy probabilities, i.e., {\it Uniqueness}, {\it Uniformity}, and {\it Linkability} (explained later in the section). Our results indicate that users' average privacy risk reaches 100\% after posting three sensitive tweets. Moreover, the average risk of predicting a user with just one sensitive tweet before the lockdown is 94\% (0.94). It is  95\% (0.95) during and after the lockdown.

{\bf Exposure to suspicious content.} Finally, we perform (cf. \S~\ref{sec:prvcAnalysis}) a security analysis on the social media tweets. We investigate the number of suspicious domains shared in social media during different phases of the pandemic. Our analysis reveals an increase in the number of suspicious domains during the lockdown compared to other lockdown phases. We also observe that \texttt{IT, Search Engines}, and \texttt{Businesses} are the top three categories that contain suspicious domains. Moreover, we notice that adversaries' strategies to instigate malicious activities change with the country's pandemic situation. For example, if a government has imposed a lockdown, people are more likely to watch and hear news from government agencies, allowing adversaries to design government look-alike malicious websites.

\section{Related Work}
\label{sec:rwork}
\begin{table*}
\centering
\caption{\ik{Comparison of related works.}}
\scalebox{1} {{\color{black}
\begin{tabular}{rlcccccc} 
\toprule
Publication & Primary Focus & \makecell{Privacy \\ Analysis} & \makecell{Security \\ Analysis} & \makecell{Sentiment\\ Analysis} & \makecell{Topic \\ Modelling} & \makecell{Pandemic \\ Impact}\\ 
\midrule
Hoeisini et al. \cite{hoseini2020demystifying} & Insights on Group Messaging Platforms & \cmark & \xmark & \xmark & \cmark & \xmark
\\
\midrule
Narayan et al. \cite{narayanan2008robust} & De-anonymise social media datasets & \cmark & \xmark & \xmark  & \xmark & \xmark
\\

\midrule
Masood et al. \cite{masood2018incognito} & Privacy Risks of user web data & \cmark & \xmark & \xmark  & \xmark & \xmark
\\

\midrule
Cengiz et al. \cite{CENGIZ2022} & Effect of user behaviour for security \& privacy & \cmark & \cmark & \xmark  & \xmark & \xmark
\\

\midrule
Wang et al. \cite{WANG2018} & Investigate Community Interests & \xmark & \xmark & \xmark  & \cmark & \xmark
\\

\midrule
Nkomo et al. \cite{NKOMO2020} & Student perspectives on Lecture Recordings & \xmark & \xmark & \cmark  & \xmark & \xmark
\\

\midrule
Aslan et al. \cite{ASLAN2020} & Behaviour of Website defacers & \xmark & \xmark & \cmark  & \cmark & \xmark
\\

\midrule
Alathur et al. \cite{ALATHUR2022} & Emotions of social media users & \xmark & \xmark & \cmark  & \cmark & \cmark
\\

\midrule
Boot-Itt et al. \cite{boonitt2020} & Public perception of Covid-19 & \xmark & \xmark & \cmark  & \cmark & \cmark
\\

\midrule
Hussain et al. \cite{hussain2021} & Public Sentiment on Covid-19 Vaccine & \xmark & \xmark & \cmark  & \xmark & \cmark
\\

\midrule
Huang et al. \cite{Huang2020} & Mobility Dynamics during pandemic & \xmark & \xmark & \xmark  & \xmark & \cmark
\\

\midrule
Guntuku et al. \cite{ChandraGuntuku2020} & Mental Health during the pandemic & \xmark & \xmark & \cmark  & \xmark & \cmark
\\

\midrule
Visentin et al. \cite{Visentin2021} & Spread of privacy and conspiracy theories & \xmark & \xmark & \cmark  & \xmark & \cmark
\\

\midrule
Xia et al. \cite{xia2021coviddomains} & Identifying malicious domain campaigns & \xmark & \cmark & \xmark  & \xmark & \cmark
\\

\midrule
Pattnaik et al. \cite{PATTNAIK2023} & Non-expert user perspectives on security \& privacy & \xmark & \xmark & \cmark & \cmark & \cmark
\\

\midrule
{\bf Our Paper} & {\bf Privacy \& security analysis during Pandemic} & {\bf \cmark} & {\bf \cmark} & {\bf \cmark} & {\bf \cmark} & {\bf \cmark}
\\

\bottomrule
\end{tabular}}}
\label{tab:relatedwork}
\end{table*}

\ik{Social media content analysis has been a popular research area for some time. The popularity of many social media platforms resulted in an increase in privacy and security concerns for their users. It motivated researchers to investigate how such platforms affected the security and privacy of their users. In this section, we discuss the existing literature and show how our work differs from theirs.}

Hoeisini et al. \cite{hoseini2020demystifying} analysed approximately 351K URLs on Twitter by modelling them based on different topics and performed a content-based analysis to determine differences between group messaging platforms such as WhatsApp, Telegram, and Discord. The study also analysed the level of PII exposure on the three platforms and collected over 34,000 phone numbers. Narayan et al. \cite{narayanan2008robust} discussed the possibilities of adversaries to de-anonymise social media datasets using different strategies. A more comprehensive study was performed by Masood et al. \cite{masood2018incognito} that used Hidden Markov Model (HMM) to predict the privacy risk based on the probabilities of uniqueness, uniformity and linkability of user's web data. Authors conducted experiments using \textit{AOL search queries} dataset and \textit{Android application reviews} dataset. The results show that with a minimum of only 10 sensitive web queries, a user's privacy risk reaches 100\%.

\ik{A survey-based study conducted by Cengiz et al. \cite{CENGIZ2022} to identify the effect of user behaviour on the security and privacy of social media users. Their work considered 700 online social network users in Turkey and Iraq, which enabled them to come to several conclusions based on the nationality of the social media users. They also try to investigate the relationship between different online threats, such as phishing, cyberbullying, fraud, etc., with user behaviour. The authors of \cite{CENGIZ2022} also look into how the frequency of Internet use, the use of pseudonyms, and the use of security tools relate to each other. Wang et al. \cite{WANG2018} investigated the degree of homophily in social media communities by using the ``Latent Dirichlet Allocation (LDA)" algorithm for modelling topics in different communities. They identified that the communities identified by two community detection algorithms, the clique-based clique augmentation algorithm and the infomap algorithm, show a strong community theme. Also, their work further proved the assumption that users form community groups based on shared interests.}

\ik{There have been several works that have focused on sentiment analysis for social media networks. For example, Nkomo et al. \cite{NKOMO2020} carried out experiments to identify the student perspectives on lecture recordings, while Aslan et al. \cite{ASLAN2020} used sentiment analysis to understand the behaviour of website defacers. In Nkomo et al.'s \cite{NKOMO2020} work, they used messages posted on the Student Union's Facebook page. They used Google's pre-trained NLP models, including the Google NLP Sentiment API, in their experiments. Their results show that the students value lecture recordings as supplementary resources for live lectures. Aslan et al. \cite{ASLAN2020} use Twitter data from a list of defacers they had identified. A defacer is a person who executes a website defacement attack in which a website's appearance or content is altered. Defacers usually seek publicity and leave traces that can identify them. Aslan et al. has collected the Tweet information of a selected set of defacers using the Twitter API for their research. They use LDA-based topic modelling and the sentiment analysis algorithm in TextBlob (a python library for processing textual data). Their results suggest that it may be possible to identify unknown hacker groups by social media analysis, and the defacers are interested in topics such as politics, cybersecurity, and relationships. In another study, Alathur et al. \cite{ALATHUR2022} analysed the awareness and capability of social media users with respect to their emotions. Their research identified that while positive emotions encourage users to share information on social media, negative emotions do the opposite. They also identified that discussions regarding infections are usually insinuating fear (a negative emotion) among users during infectious periods.}

Since the pandemic, more research works have focused on identifying user behaviour and perceptions using Twitter data (among other social media data). Boot-Itt et al. \cite{boonitt2020} identified three main topics of concern for Twitter users during the pandemic using topic modelling. The topics were COVID-19 emergency, COVID-19 control mechanisms and reports on COVID-19. Their sentiment analysis confirmed the common notion that people had a negative outlook toward COVID-19. Meanwhile, Hussain et al. \cite{hussain2021} used Twitter data to assess public opinion regarding the COVID-19 vaccine. Their research revealed more than 50\% positive sentiments from people in the UK and the US. Huang et al. \cite{Huang2020} proposed Twitter data analysis as an efficient, cost-effective, and privacy-preserving method to assess human mobility dynamics during the pandemic. Their results suggested that Twitter data is capable of quantifying mobility dynamics in various geographical scales. Guntuku et al. \cite{ChandraGuntuku2020} used Twitter data to analyse mental health and symptoms. Additionally, Visentin et al. \cite{Visentin2021} tried to identify the relationship of words, linguistics, styles, and emotions with privacy concerns and conspiracy theories on Twitter and how those elements contributed to the spread of such theories. To this end, they analysed tweets related to an Italian tracing app called ``Immuni''.

The security of social media users can be analysed with respect to different types of emerging security threats such as malware, phishing, spam email, ransomware, etc. Xia et al. \cite{xia2021coviddomains} conducted research to identify and characterise COVID-19 themed malicious domains. Authors aggregated a dataset containing 4,500 malicious COVID-19 themed domains from a number of different sources.  They differentiated the COVID-19 malicious campaigns based on the underlying network infrastructure such as subnet distribution, cloud IPs, geolocation, domain registration, WHOIS records and so on. They then constructed a network knowledge graph followed by clustering the nodes based on the relations in the graph (related IPs, name servers, etc.). The study concludes that adversaries are rapidly exploiting COVID-19 to facilitate cyber-attacks. \ik{Pattnaik et al. \cite{PATTNAIK2023} analysed Twitter discussions to identify the perspectives of non-expert users related to cybersecurity and privacy. They first develop two machine learning classifiers; one to detect tweets focusing on the above two topics and the other to detect non-expert accounts. Then they identify the main topics related to cybersecurity discussed in those tweets (e.g. VPNs, Wifi, smart home devices, financial security, etc.) and carry out a sentiment analysis for those topics. They also study the trends for those topics and their sentiments across three years (2019-2021).}

In this paper, we carry out large-scale study on the impact of COVID-19 pandemic on the security and privacy of social media users. This study is the first of its kind to comprehensively investigate the 10 Million tweets from various aspects that mainly include characterisation, sentiment analysis, security analysis, and privacy analysis, respectively. \ik{Table \ref{tab:relatedwork} highlights the uniqueness of our work compared to the related work. Our study complements all the prior research and delivers new contributions to the knowledge of privacy and security in social media.}

\section{Data Collection And Characterisation}
\label{sec:dccharacterisation}

\subsection{Data Collection Methodology}
\label{sec:datacollection}
We begin by presenting our methodology to collect and analyse COVID-related data from Twitter.  

\subsubsection{Data Collection:} 
We use a dataset provided by the Panacea Lab \cite{banda2021twitterdata} to collect COVID-related tweets. The Panacea Lab contains approximately 730 million COVID-related tweets, which can be used for scientific purposes. Since Panacea Lab only provides Tweet IDs, we need to hydrate (i.e., extracting the original content of the Tweet such as tweet text, geo-location, timestamp, likes, comments, etc.) those tweet IDs. We use the Twitter API \cite{twitterAPI}, and Twarc \cite{twarc} python library for that purpose. We run our data-collection framework on  High-Performance Cluster (HPC)--with over 4,000 CPU cores with multiple compute nodes, each having 1TB of memory--at our institute. Over the period of two months crawling, from June 2021 to August 2021, we collect tweets spanning over 533 days from January 1, 2020, to June 21, 2021. We further filter the collected tweets based on selected countries, \ik{the} English language, and lockdown periods. The filtering process is explained in \ik{detail} below.

\subsubsection{Data Filtering} 
{\it Next}, we filter tweets based on the \textit{Geolocation} and \textit{Language}. Specifically, we selected tweets from \texttt{Australia, India, United States of America (US)}, and \texttt{United Kingdom (UK)} from the dataset. We selected India, US, and UK because they had the highest IRs and death tolls during the pandemic. On the contrary, we selected Australia because its strategy to eliminate/contain COVID was different from other countries. Australia imposed international and domestic border closure and state lockdown for a prolonged period of time. We assume this would provide some interesting trends in Australia as compared to other countries. Next, we filtered out non-English tweets from the dataset of selected countries. That is because multilingual tweets can affect the accuracy of hashtag analysis (cf. \S~\ref{sub:hashtag analysis}) and sentiment analysis (cf. \S~\ref{sec:sentAnalysis}). For example, it has been previously shown by Boyd-Graber et al. \cite{boydgraber2009} that the topics learned by the Latent Dirichlet Allocation (LDA) algorithm are language-specific when used with bilingual datasets. They used LDA against a dataset of English and German Tweets, and their results indicated that English and German tweets were clustered independently. After filtering non-English tweets from the selected countries, our dataset contained 10.22 Million tweets for further analysis.

\subsubsection{Ethics Consideration:} 
We obtain a publicly available dataset from Panacea Lab. Prior to data collection, we obtained ethics approval from our organisation’s ethics board. Throughout data collection, we did not attempt to obtain the real identities of the participants via, for instance, a linkage study. We follow ethics guidelines \cite{rivers2014ethical} and do not use, track, or de-anonymise users from the collected dataset. The data collected was not released publicly. We did not store any identifying information other than the attributes such as user tweets, timestamps, and hashtags on our servers.

\subsection{Data Characterisation}
\label{sec:statAnalysis}
In this section, we discuss our findings after characterising COVID-19 tweets. We first discuss our findings on the hashtags analysis, followed by the discussion on the  URL analysis. For hashtag analysis, we identify the most discussed topics related to COVID-19 during the pandemic and discern their variation with respect to IRs in certain locations during different stages of the lockdown. We examine the most widely shared URLs in tweets for the URL analysis and identify which types of websites people are frequently visiting. We also try to identify any relationships among user behaviour with the IRs of countries during different phases of the lockdown.

\begin{table}[t]
  \small
  \centering
  \caption{Breakdown of number (\#) and percentage (\%) of tweets collected for each country in three different stages of COVID-19 pandemic.}
%   \vspace{-0.4cm}
  \tabcolsep=0.09cm
\scalebox{0.90} {
  \begin{tabular}{l*{8}{c}}
    \toprule
     & \multicolumn{2}{c}{\bf Australia} & \multicolumn{2}{c}{\bf India} 
      & \multicolumn{2}{c}{\bf UK} & \multicolumn{2}{c}{\bf US} \\
    \cmidrule{2-9}
    {\bf Period} & {\bf \#} & {\bf \%} & {\bf \#} & {\bf \%} & {\bf \#} & {\bf \%} & {\bf \#} & {\bf \%} \\
    \midrule
    Before & 95,416 & 17 & 350,638 & 20 & 1,030,106 & 21 & 1,260,757 & 41 \\
    \rowcolor{DarkOrange}
    During & 398,825 & 70 & 1,041,865 & 58 & 2,441,744 & 51 & 925,964 & 30 \\
    After & 75,404 & 13 & 400,090 & 22 & 1,324,941 & 28 & 870,943 & 29 \\
    \hline
    \textbf{Total} & \textbf{569,645} & 100& \textbf{1,792,593} & 100& \textbf{4,796,791} &100 &  \textbf{3,057,664} & 100 \\
    \bottomrule
  \end{tabular}
  }
  \label{table:statAnalysisSummary2}
\end{table}

\begin{figure*}[!htbp]
\centering
\subfloat[]{
\includegraphics[width=0.5\textwidth, keepaspectratio]{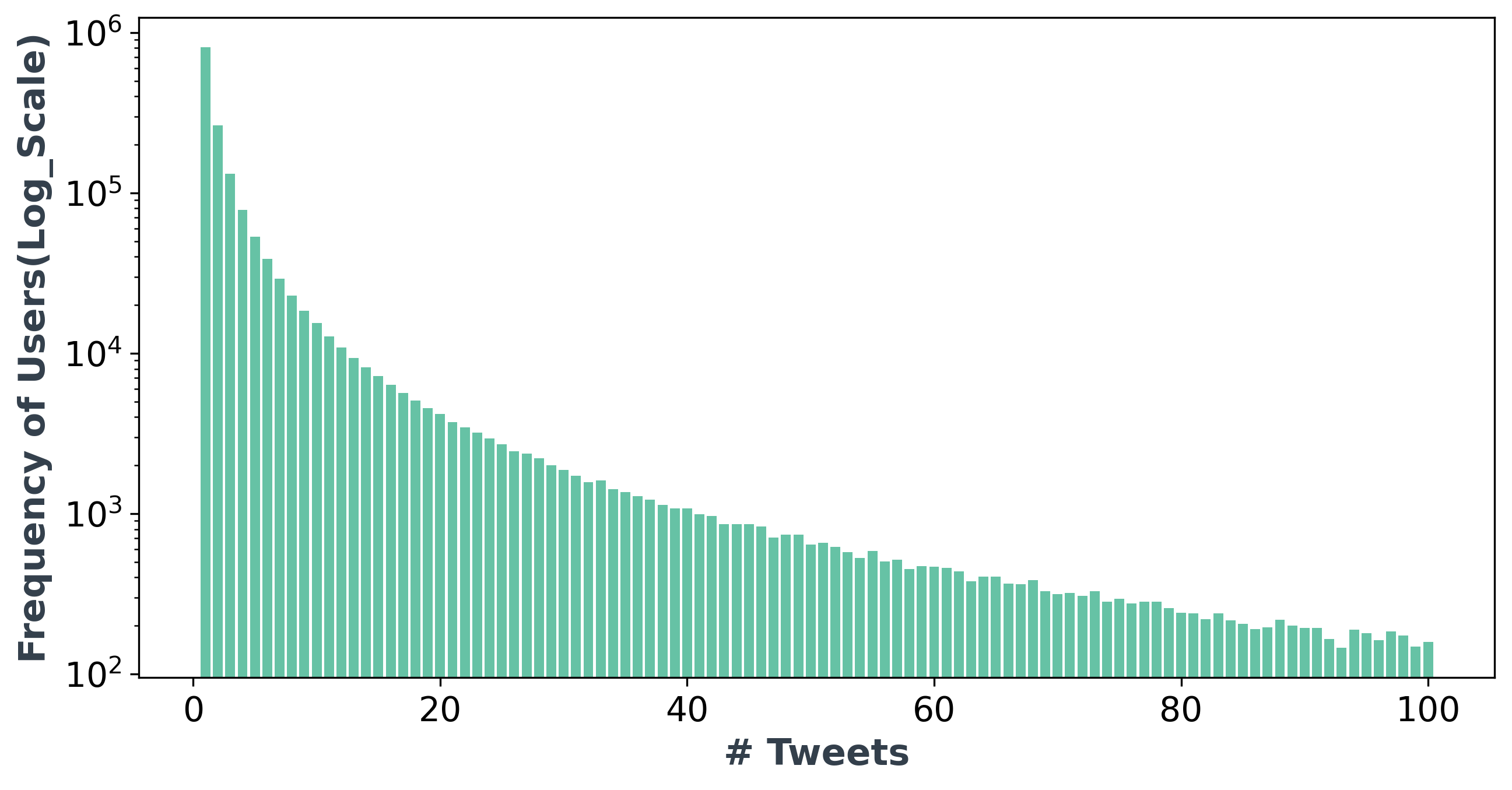} \label{fig:tweetDistribution}}
\subfloat[]{\includegraphics[width=0.5\textwidth, keepaspectratio]{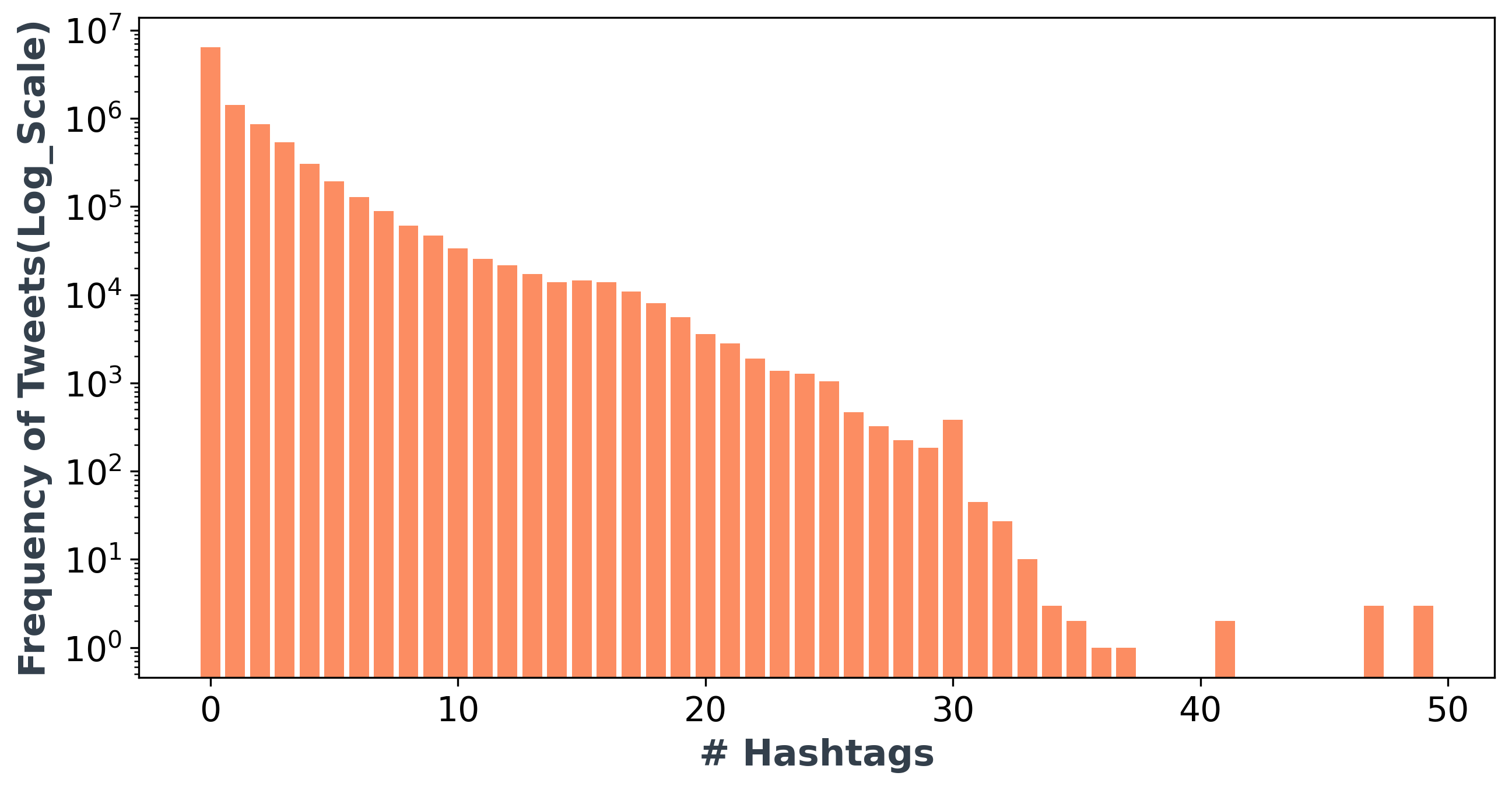} \label{fig:hashtagDistribution}}
%\vspace{-0.5cm}
\caption{\small Tweet distribution per User (a) and Hashtag distribution per Tweet (b).}
\label{fig:tweet and hashtag distribution}
\end{figure*}

\begin{table*}[!htbp]
  \small
  \centering
%   \vspace{-0.5cm}
  \caption{Lockdown Periods and Infection Rates (IR)--the ratio of total number of COVID-19 cases to the number of days \cite{owdata}. }%  as per Our World In 
%   \vspace{-0.4cm}
  \begin{threeparttable}
  \begin{tabular}{*{7}{r}}
    \toprule
     & \multicolumn{2}{c}{\bf Before Lockdown} & \multicolumn{2}{c}{\bf During Lockdown} 
      & \multicolumn{2}{c}{\bf After Lockdown} \\
    \cmidrule{2-7}
    {\bf Country} & {\bf Dates} & {\bf IR} & {\bf {\cellcolor{DarkOrange}{Dates}}} & {\bf {\cellcolor{DarkOrange}{IR}}} & {\bf Dates} & {\bf IR} \\
    \midrule
    
    \multirow{2}{*}{Australia} & 5 Mar - 20 Mar (2020) & 63.5 & \cellcolor{DarkOrange}{21 Mar - 15 May (2020)} & \cellcolor{DarkOrange}{229.4} & 16 May - 1 Jun (2020) & 10.4 \\
     & 21 Jun - 6 Jul (2020) & 80.1 & \cellcolor{DarkOrange}{7 Jul - 19 Oct (2020)} & \cellcolor{DarkOrange}{176.4} & 20 Oct - 5 Nov (2020) & 11.8 \\\cmidrule{2-7}
     
    India & 24 Feb - 23 Mar (2020) & 17.1 & \cellcolor{DarkOrange}{24 Mar - 31 May (2020}) & \cellcolor{DarkOrange}{2,754.7} & 1 Jun - 30 Jun (2020) & 12,903.7 \\\cmidrule{2-7}
    
    US & 29 Feb - 28 Mar (2020) & 4,392.8 & \cellcolor{DarkOrange}{29 Mar - 28 Apr (2020)} & \cellcolor{DarkOrange}{28,445.5} & 29 Apr - 27 May (2020) & 22,688.7 \\\cmidrule{2-7}
    
    \multirow{3}{*}{UK} & 10 Mar - 25 Mar (2020) & 1,059.6 & \cellcolor{DarkOrange}{26 Mar - 1 Jun (2020)} & \cellcolor{DarkOrange}{3,521.8} & 2 Jun - 17 Jun (2020) & 1,033.6 \\
     & 7 Oct - 22 Oct (2020) & 17,941.1 & \cellcolor{DarkOrange}{23 Oct - 7 Nov (2020)} & \cellcolor{DarkOrange}{21,297.5} & 8 Nov - 23 Nov (2020) & 22,281.1 \\
     & 20 Dec (2020) - 5 Jan (2021) & 43,326.4 & \cellcolor{DarkOrange}{6 Jan - 16 Mar (2021)} & \cellcolor{DarkOrange}{26,181.8} & 17 Mar - 1 Apr (2021) & 5,146.5 \\
    \bottomrule
  \end{tabular}
  \vspace{-0.4cm}
  \end{threeparttable}
  \label{table:lockdown_periods}
\end{table*}

\subsubsection{Data Statistics} Table~\ref{table:statAnalysisSummary2} shows the number of tweets collected from each country in different lockdown periods. 
We collect a majority of the tweets from the UK and the US. Of the 10.22 Million tweets, 4.8 Million and 3.06 Million tweets are from the UK and USA, respectively. We collect 1.79 Million tweets from India, while a meager 0.57 Million are from Australia.
For Australia, 70\% of the tweets are from \textit{during} the lockdown period, while this number is 58\%, 51\%, and 30\% for India, UK, and US, respectively. We have a lower number of US tweets because it has a shorter lockdown period, as given by the stringency level. We have a similar distribution in the number of tweets during the three stages for all the other countries. Furthermore, the hydrated tweets were stored as \texttt{jsonl} files at our HPC server. We use the following attributes from the {\tt jsonl} file for our analysis: \textit{anonymised user IDs, timestamp of the Tweet, Tweet ID, Tweet Text, URLs, geo-location of the Tweet} and \textit{hashtags}. Figure~\ref{fig:tweet and hashtag distribution} shows the distribution of tweets per user and the distribution of hashtags per tweet, respectively. Figure~\ref{fig:tweetDistribution} clearly shows that most users have posted less than 20 tweets; however, a sheer amount of users have also posted more than 40 tweets. Approximately 0.7\% of users have posted more than 100 Tweets, whereas approximately 50\% of users in the dataset have posted exactly one tweet. Similarly, we see a 3,812,764 (37.3\%) number of tweets with up to 20 hashtags in Figure~\ref{fig:hashtagDistribution}.

\subsubsection{Data Classification:} 
We classify the filtered dataset into three phases based on the country's lockdown dates, i.e., \textit{before lockdown, during \ik{the} lockdown,} and \textit{after lockdown}. This classification helps us analyse the variations in public sentiments and security and privacy risks across different pandemic stages.  Moreover, it also helps us identify various trends across the pandemic and come to better conclusions on potential drivers behind them. We refer to these phases as ``lockdown Periods'' throughout the paper. To determine the lockdown phases (dates) for each country, we use the stringency level given by the COVID-19 stringency level dashboard \cite{stringency2021}. The {\it stringency index measures} how strict the government restrictions have been in response to COVID-19. We select 65 (100 is the maximum) as our stringency index for determining the lockdown dates. We select this number by first checking the lockdown dates of the selected countries on various news articles. We then put these dates into the stringency website to get the stringency index. For most of the lockdown dates, the index was >=65, hence giving us a clear indication of the lockdown stringency index. Table~\ref{table:lockdown_periods} shows the lockdown periods of the selected four countries. 

We also calculate the average number of infections per day for each country and each period. %In our paper, 
We define this value as the \textit{Infection Rate (IR)}, which is calculated by dividing the total number of COVID-19 cases for a specific period by the total number of days for that period. This value provides us with a high-level idea about the COVID-19 situation \ik{in} a certain country before, during, and after a lockdown. The number of cases for each period was retrieved from the \textit{Our World in Data} website\cite{owdata}. We can observe a few interesting details in Table \ref{table:lockdown_periods}. For example, let's consider the first lockdown period between 21st March to 15th May in Australia. We can see that the IR is higher (229.38)  during the lockdown than before the lockdown (63.5) and eventually reduces after the lockdown (10.41). It shows the impact of strict COVID containment strategies followed by Australia. The states imposed \ik{lockdowns} when the cases were rising and eased restrictions when the transmission was under control. In India, we can see the IR is still high even after the lockdown (2,754.68 during the lockdown and 12,903.7 after the lockdown), which indicates that they haven't been able to get ahead of the virus and prevent community transmission. In the US, although we can observe a slight decline \ik{in} the IR after the lockdown, the number itself (22,688.66) is very high and suggests that community transmission must have been happening. However, the restrictions seem to have slowed down the rate of transmission. In the UK, we can see some contradicting relationships between lockdown periods and IR. The first lockdown between 26th March and 1st June seems to have controlled the spread of the disease considerably, while the second lockdown seems to have only managed to slow the rate of transmission. After two lockdowns, it seems that the UK was a bit late to impose a third lockdown, resulting in an extremely high IR of 43,326.41 before the lockdown. 

% {\it Insight.} 
The restrictions seem to have helped control the transmission as the number has fallen to 26,181.82 during the lockdown. It has kept falling, which suggests the possibility of herd immunity--{\it takes place when a substantial population of a community becomes immune to a disease}--in the UK.

\subsubsection{Hashtag Analysis} \label{sub:hashtag analysis}

A hashtag is a metadata tag prefaced by a hash sign (\#) and is used on microblogging and other photo-sharing websites to identify digital content on a specific topic. \ik{On} Twitter, the hashtag indicates the topic associated with the tweet. The hashtags are also used to index keywords and help users follow a specific topic they are interested in. We analyse hashtags in our dataset to identify the widely discussed COVID-related topics during the lockdown periods and their relation with IRs in specific countries. 

{\it Methodology:} We use K-Means clustering to group the tweets that contain similar hashtags. This algorithm identifies $k$ number of centroids for a given dataset and assigns every data point to the nearest cluster. \ik{We selected K-Means clustering for our work over other algorithms (e.g., Hierarchical Clustering and Gaussian Mixture Models (GMMs)) as it is fast, efficient, and can be easily scaled to handle large datasets by parallel/distributed computing.} 
For our analysis, we first selected tweets that contained hashtags. Next, we duplicated each tweet by the number of hashtags it contained. For example, if a tweet had 3 hashtags, we created two additional duplicates of that tweet. This approach gave us a dataset of 10.6 million tweets. We then processed this new dataset by removing URLs, mentions, punctuations, and stopwords\footnote{Since all the tweets in the dataset are covid related, we removed words such as COVID, COVID-19, coronavirus to avoid redundancy.} using {\tt NLTK} library \cite{nltk}. Generally, a single hashtag is a concatenated text of multiple words with different semantics, e.g. \texttt{stayHomeSayNoToDrugs}. Therefore, we cannot extract \ik{the} lexical features of hashtags as they have significant noise. We argue that since a hashtag represents the main content of a tweet, we can use the main content to represent the hashtag. Therefore, we remove the hashtags from the selected tweets for this analysis and only consider the main content.

\begin{figure*}[t]
\centering
\subfloat[]{
\includegraphics[width=0.4\textwidth, keepaspectratio]{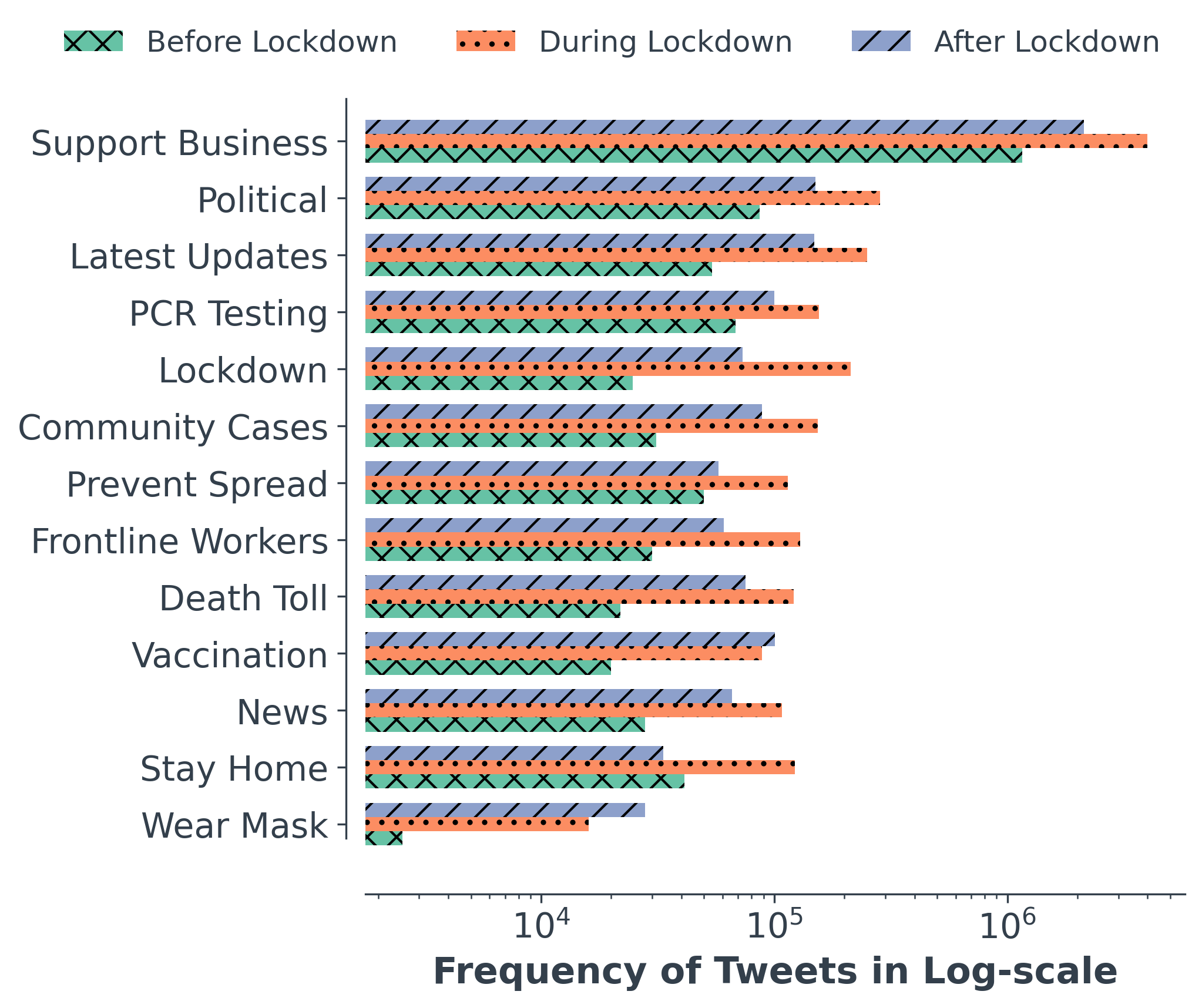} \label{fig:frequentHashtags}}
\subfloat[]{
\includegraphics[width=0.55\textwidth, keepaspectratio]{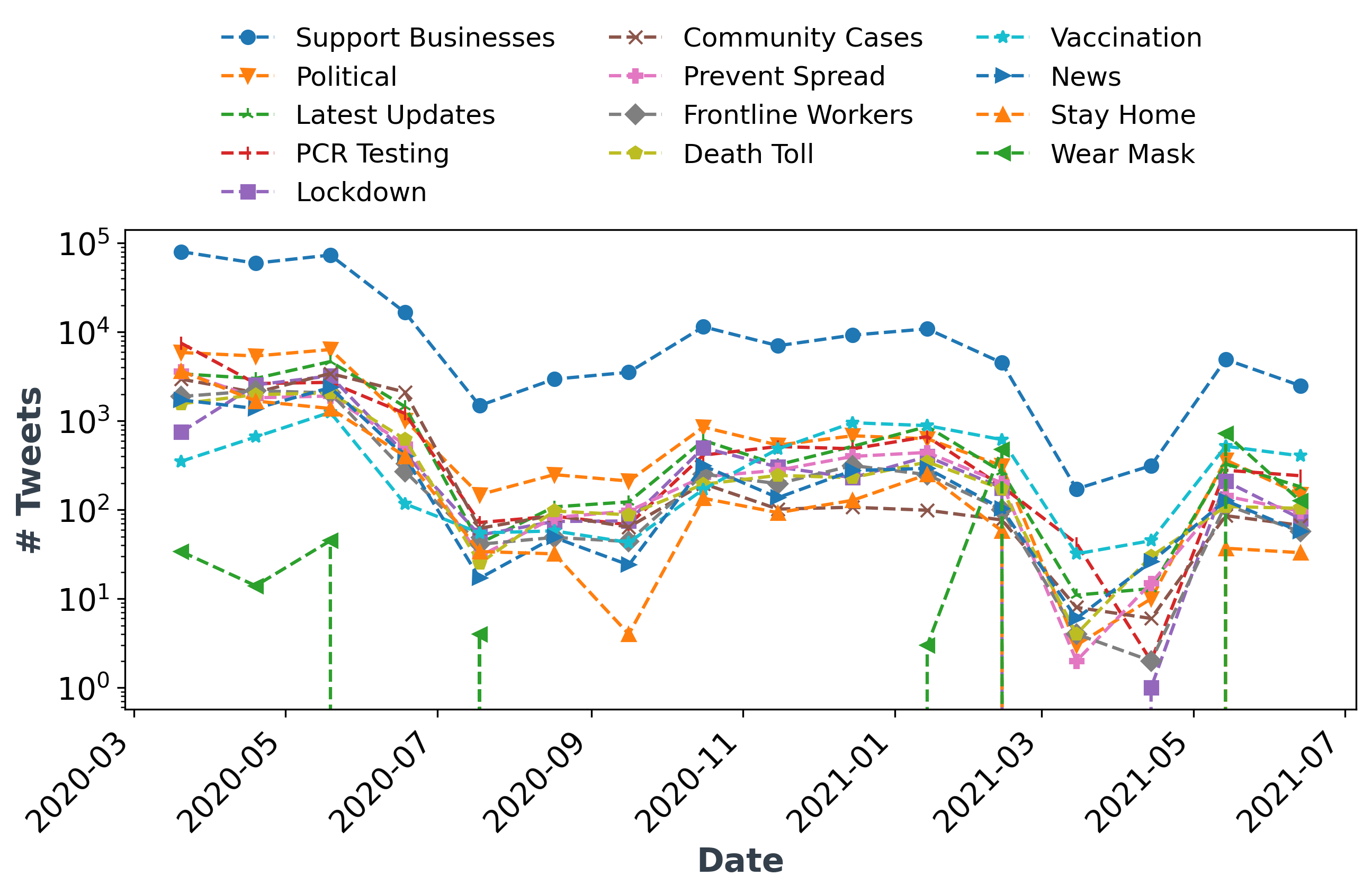} \label{fig: generalTrendHashtag}}
% \vspace{-0.5cm}
\caption{\small General Trend of Hashtags over time and the top 10 frequent Hashtags.}
%\vspace{-0.5cm}
\label{fig:generalHashtagsDomains}
\end{figure*}

After the initial processing, we tokenised and lemmatised the tweets to extract numerical features from the text. Tokenisation is used to protect private data, while lemmatisation removes redundancy and converts the words into their lemma (the root word in vocabulary). We used Sklearn {\tt TF-IDF vectorizer} \cite{sklearn} for this task. \ik{We used TF-IDF as it typically results in a sparse representation of text data reducing its dimensionality and improving computational efficiency. Also, TF-IDF does not require training and can be computed directly from the text data (unlike alternative options such as word embeddings), significantly reducing our training time.} Finally, we used the Sklearn library \cite{sklearn} to fit the features into a K-Means model with 15 clusters. We experimented with the number of clusters ranging from 5 to 35 and found 15 to produce the best accuracy. The cluster names were assigned manually after inspecting the top 10 words in each cluster. Some of the clusters having words with similar semantics were merged to eliminate redundancy, which left us with 13 hashtag clusters in total.

% {\it Insight 1.}  
Our hashtag analysis reveals that people mostly talk about \texttt{supporting businesses} during the three stages of the pandemic, as shown in Figure~\ref{fig:frequentHashtags}. We find approximately 7.2 Million hashtags related to \texttt{supporting businesses}. For instance, people are frequently using hashtags such as \textit{\#fundraising, \#charities, \#our\_work\_is\_our\_identity} in their tweets. It indicates that the economic disaster was also immense during the pandemic, apart from death and sickness. Tourism, which was one of the most profitable industries before 2020, was almost brought to its knees. People, from small roadside sellers at tourist attractions to commercial airline pilots, lost their jobs and main sources of income. Almost every business that thrived on close human interactions or large numbers of people, including salons, massage parlours, pubs, nightclubs, gyms, restaurants, and cafes, had to be shut down during lockdowns. This situation affected a lot of livelihoods and directly impacted basic human needs, making \texttt{supporting businesses} the most frequent topic of discussion during the pandemic.

% {\it Insight 2.} 
We also observe that topics such as \texttt{politics}, \texttt{latest updates}, \texttt{PCR testing}, and \texttt{lockdown} have been quite frequently discussed on Twitter. We found approximately 285K tweets related to politics with hashtags such as \textit{\#BorisJohnson, \#PM, \#Trump} during the lockdown. Similarly, we observe that the topic \texttt{latest updates} (with hashtags such as \textit{\#LIVE, \#WATCH, \#currentaffairs}) was tweeted approximately 54K, 250K, and 148K times before, during and after the lockdown, respectively. Another noteworthy insight is that the topic ``Mask Wearing'' has been discussed less before the lockdown. However, the proportion increases by 84.1\%, i.e. from 2,547 to 15,986 during the lockdown for face covering. 

% {\it Insight 3.}  
Another interesting insight from hashtag analysis is that the frequency for most of the topics increased during the lockdown and slightly decreased after the lockdown, as shown in Figure~\ref{fig:frequentHashtags}. This pattern is only different for the \texttt{vaccination} and the \texttt{face mask} topics, which had kept rising even after lockdowns. The reasons can be that the governments kept pushing people to get vaccinated and mandated face masks most of the time, even after lockdowns. On the contrary, \texttt{staying home} related tweets had declined considerably after the lockdown. That is quite reasonable, as staying home is not a relevant topic after a lockdown.

% {\it Insight 4.} 
From Figure \ref{fig:generalHashtagsDomains}, we can observe that the general trend of Hashtags compared to the frequency based on lockdown periods is quite consistent. An interesting insight is that hashtags relating to wearing masks have been discussed more during early 2021 than in 2020. This could be because people realised that wearing a mask is a very effective preventive measure to contain the spread of COVID-19. We can see s similar trend for staying home, where people have been posting these hashtags more during lockdown compared to other periods. The volumetric trend seems to be consistent for the rest of the hashtags. 

\begin{figure*}[h]
\centering
\subfloat[]{\includegraphics[width=0.4\textwidth, keepaspectratio]{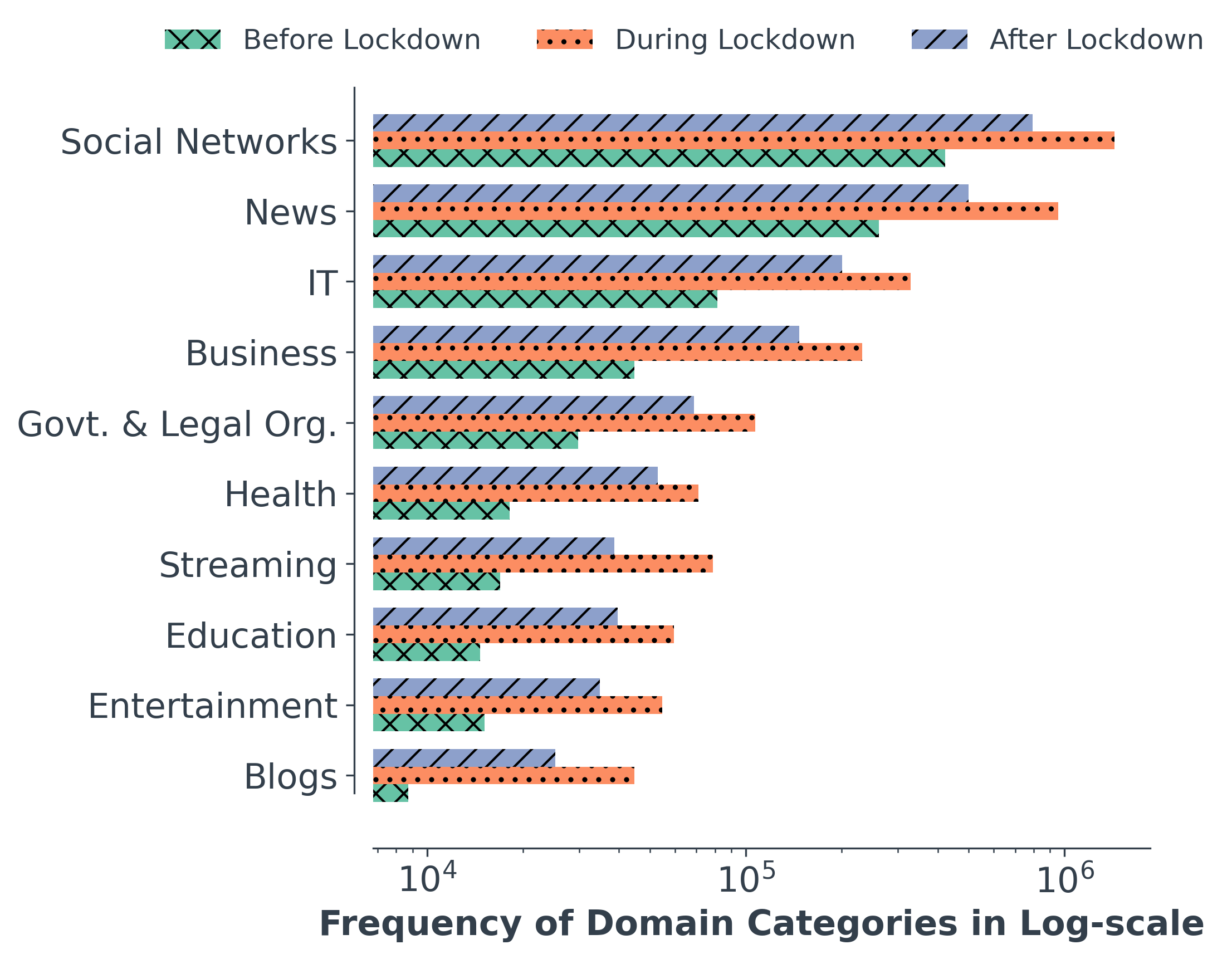}\label{fig:top domain categories}}
\subfloat[]{\includegraphics[width=0.55\textwidth, keepaspectratio]{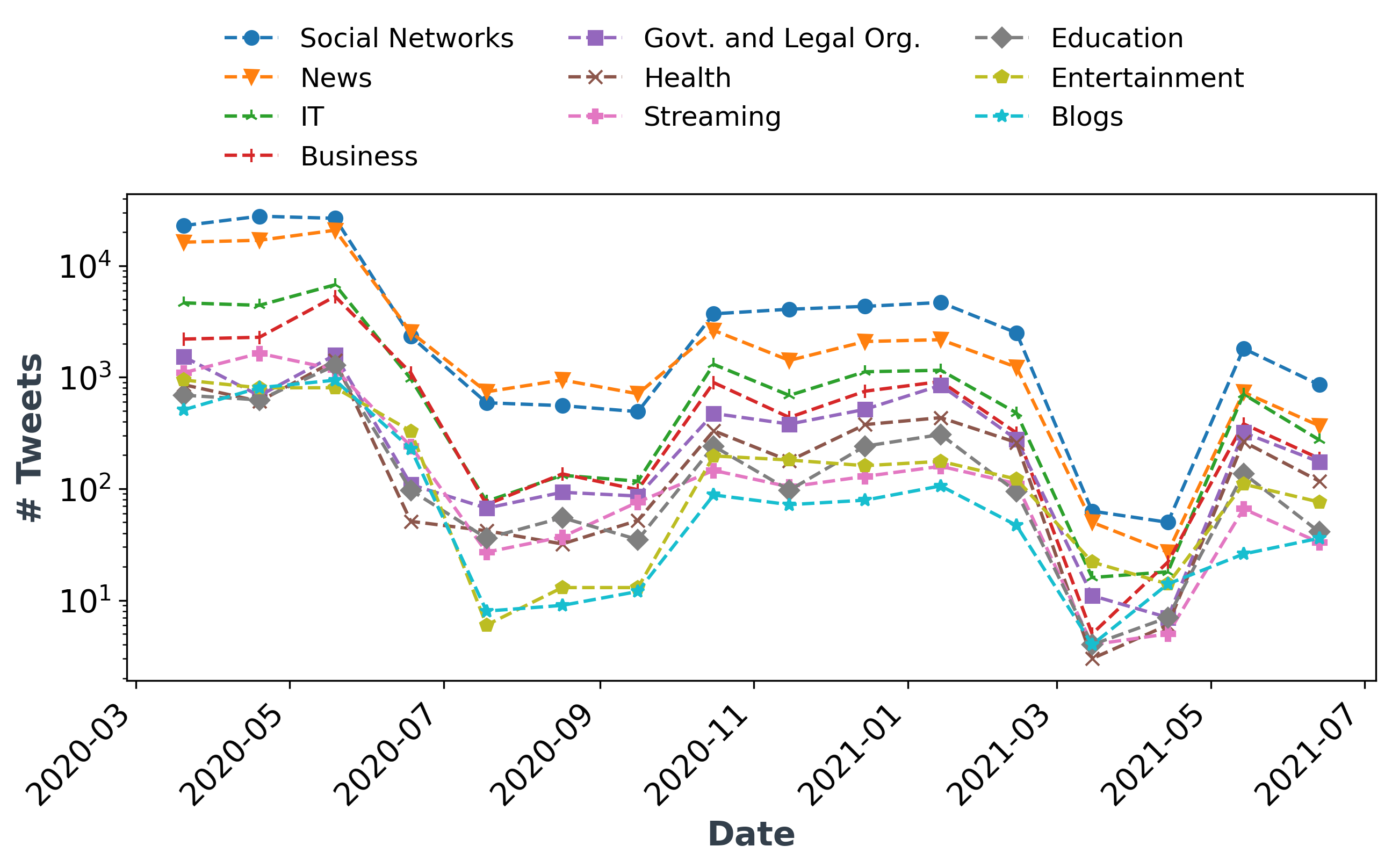}\label{fig:domainCatsGeneralTrend}}
% \vspace{-0.5cm}%
\caption{\small General Trend of Domain Categories over time and the top 10 domain categories.}
%\vspace{-0.5cm}
\label{fig:frequentHashtagsandURLs}
\end{figure*}

% {\it Insight 5.} 
When observing these trends for each of the countries considered in our paper (refer Appendix \ref{appendix A}), we notice that \texttt{supporting businesses} are the main topic in each of the countries during all the lockdown periods. Before the lockdown, common discussion topics are more or less the same for Australia, the UK, and the US. However, India seemed to have more discussions related to COVID-19 prevention (e.g., topics such as \texttt{PCR testing, preventing spread, staying home}). Considering India's low IR (17.1) before the lockdown, we can assume that the people were extremely concerned about the virus, which may be due to the devastating news they were receiving from other countries. During the lockdown period, we notice that topics such as \texttt{front-line workers} and \texttt{death toll} are highly discussed in countries such as USA and UK. The extremely high IRs in these countries (Table~\ref{table:lockdown_periods}), which subsequently caused an increase in death rates, must be the reason for this surge in topics. The trends we see in Australia are more consistent than other countries, resulting from its low infection rates throughout the pandemic. For further analysis, we refer readers to Table~\ref{table:statAnalysisSummary}.

\subsubsection{URL Analysis}\label{sub:url analysis}

People use Twitter as a medium to share articles and resources from other websites. The restricted character length in tweets encourages someone to write a short text with a piece of news or an opinion and share some supporting material. As a result, we can find a large number of URLs in tweets. An analysis of these URLs, their domains, and categories can give important insights \ik{into} widely discussed topics at a particular period in the Twitter community. With the objective of further investigating global trends, we did an URL analysis on the Twitter dataset.

{\it Methodology:} To perform URL analysis, we first extract all the URLs from the tweets. The Twitter API provides an attribute \textit{URL}, which can be used to extract URLs from the tweets while hydrating. Twitter usually shortens URLs using its URL shortening tool, which causes all URLs to have a Twitter domain. Nevertheless, Twitter API also provides the \textit{expanded\_url} attribute to extract the original URLs. Using \textit{expanded\_url}, we collected a corpus of 6.95 Million fully resolved URLs for our study.

We then used the python {\tt tld} library \cite{tld} to extract the domains of these URLs. For each domain, we use {\tt Fortiguard} \cite{fortiguard} to classify them into a specific category. Figure \ref{fig:top domain categories} depicts the top 10 domain categories we identified using the above technique.

% {\it Insight 1.} 
The URL analysis provides a slightly different perspective on global trends. According to Figure~\ref{fig:top domain categories}, \texttt{social networking} related URLs from domains such as \url{twitter.com}, \url{instagram.com} and \url{facebook.com} have been mostly shared on Twitter. Users shared approximately 1.43 Million social URLs during the lockdown and 793K after the lockdown. 
This suggests that overall social media usage across multiple platforms increased during the pandemic. However, as we only consider the domain category of the URL for our analysis, we do not examine the underlying content in those articles or posts. The topics of the shared articles can be anything, although we can assume that they are more or less similar to the results of our hashtag analysis. The same limitation applies to the \texttt{news and media} related URLs. There are 953K and 500K news-related URLs during and after the lockdown, respectively. For example, some of the most widely shared news-related URL domains are \url{https://subscribe.theepochtimes.com/} \textit{(during lockdown: 1307, after lockdown: 1309)}, \url{https://theconversation.com/} \textit{(during lockdown: 1445, after lockdown: 965)}, and \url{https://ncbi.nlm.nih.gov/} \textit{(during lockdown: 1441, after lockdown:963)}.

% {\it Insight 2.} 
We also notice that URLs related to \texttt{information technology (IT)} were also frequently shared on Twitter, especially during the lockdown period (approximately 330K number of times). That suggests that COVID-19 %is because COVID 
has significantly transformed business operations by forcing organisations to switch to remote working, increasing the load on IT equipment and network traffic. This transformation
hence forced people to share \texttt{IT} related URLs frequently. Some of the IT-related URLs include \url{https://apps.apple.com/} \textit{(during lockdown: 3110, after lockdown:  95)}, \url{https://play.google.com/store/apps/} \textit{(during lockdown: 3469, after lockdown: 99)}, and \url{https://dailym.ai/ios} \textit{(during lockdown: 1436, after lockdown: 586)}. We observe that approximately 329K IT-related URLs were shared during the lockdown, followed by 200K after the lockdown. Other most frequent URL categories include: {\t business, government, \& legal organisations, streaming media, health \& wellness, entertainment}, and {\it education}.

% {\it Insight 3.} 
The number of Tweets containing News related URLs seems to have been the highest from August 2020 to October 2020, which falls during the lockdown phase (see Table \ref{table:lockdown_periods}). Other times Social Network related URLs \ik{have} been shared the most, which can be observed in Figure \ref{fig:domainCatsGeneralTrend}. That can be because people relied more on the News to get updated information about COVID-19 during lockdown phases. The other clusters have similar trends and are consistent with Figure \ref{fig:frequentHashtags}.

\begin{figure*}[t]
\centering
\subfloat[Before]{
\includegraphics[width=0.480\textwidth, keepaspectratio]{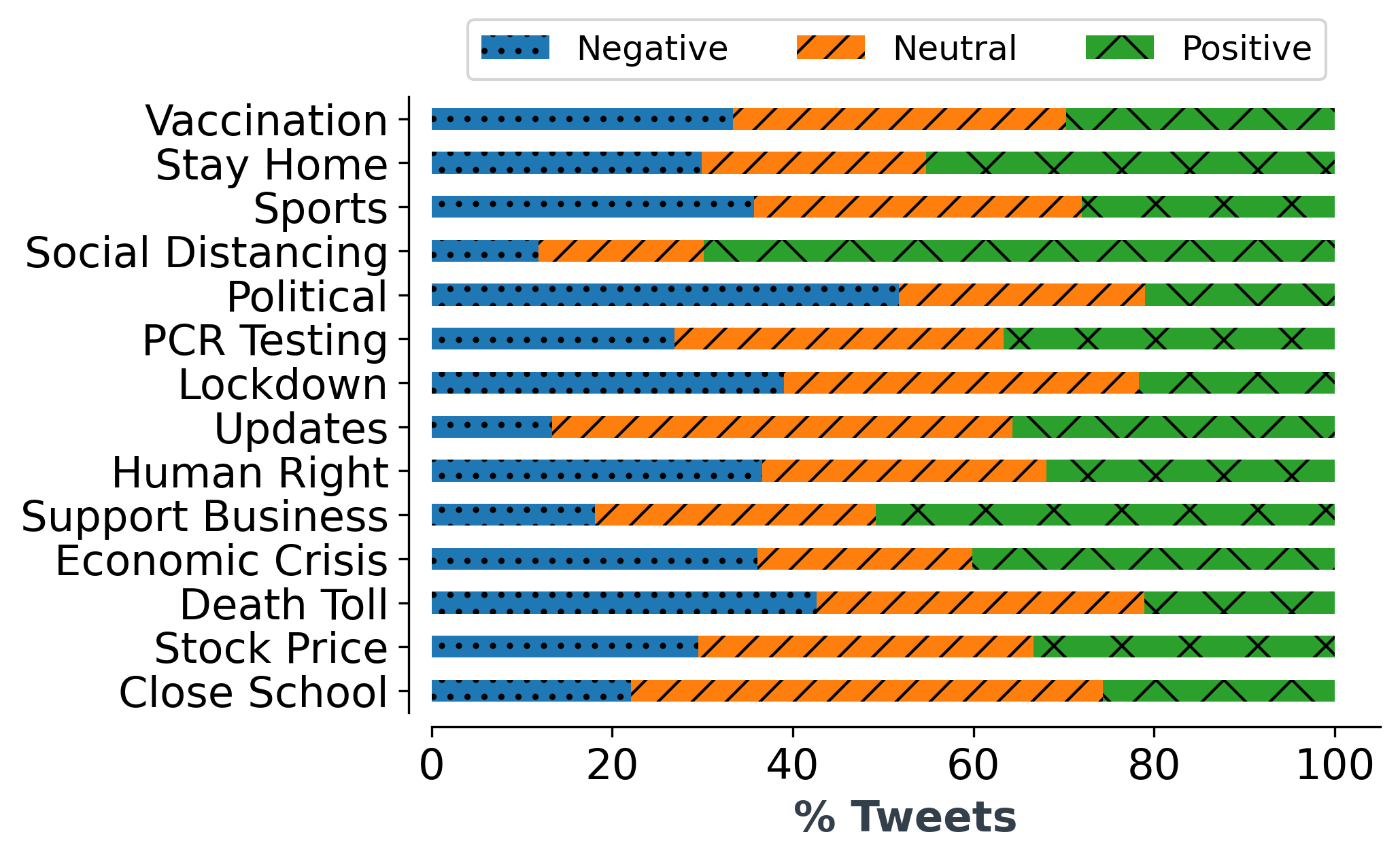}
}
\subfloat[During]{\includegraphics[width=0.480\textwidth, keepaspectratio]{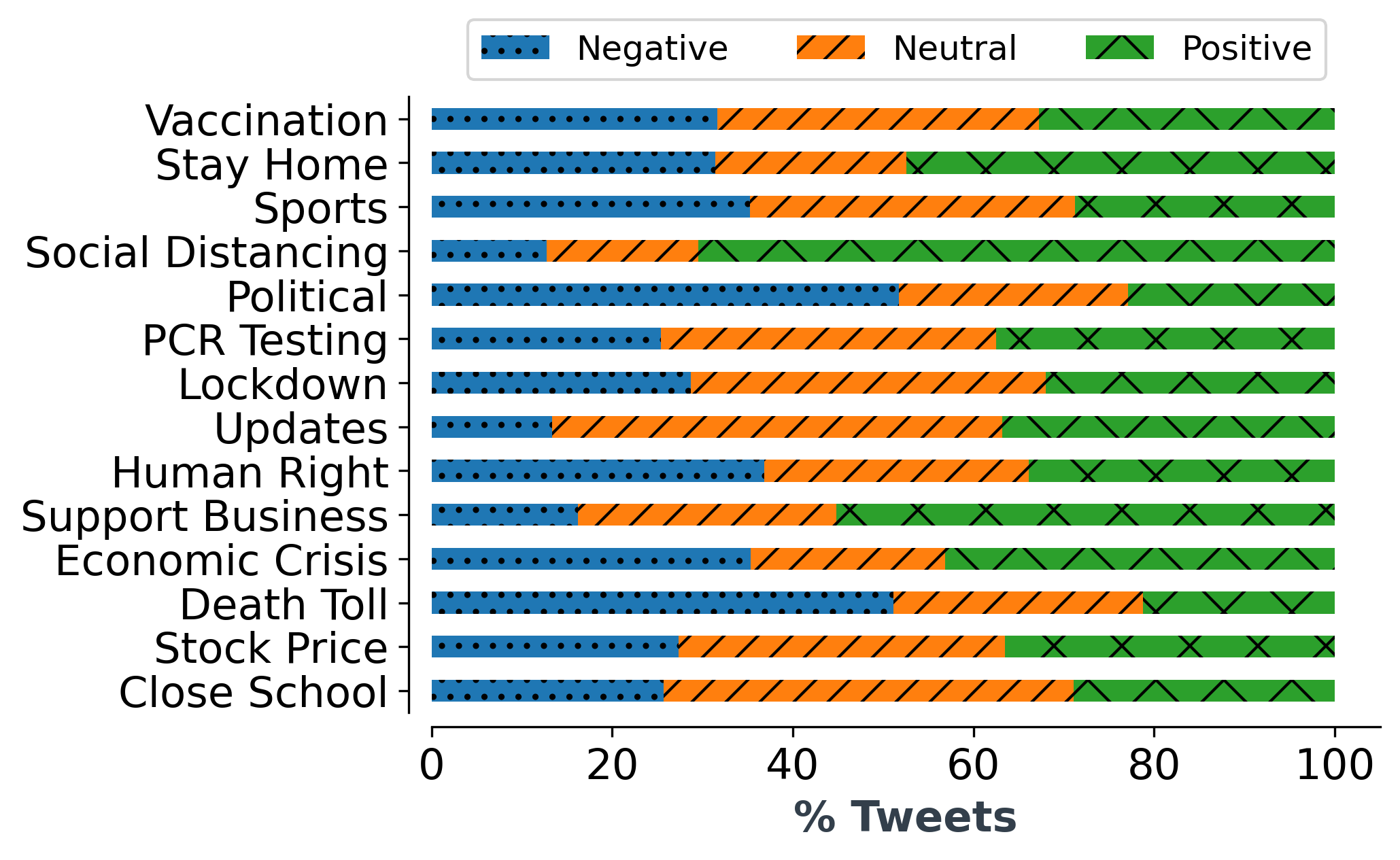}}

\subfloat[After]{\includegraphics[width=0.48\linewidth, keepaspectratio]{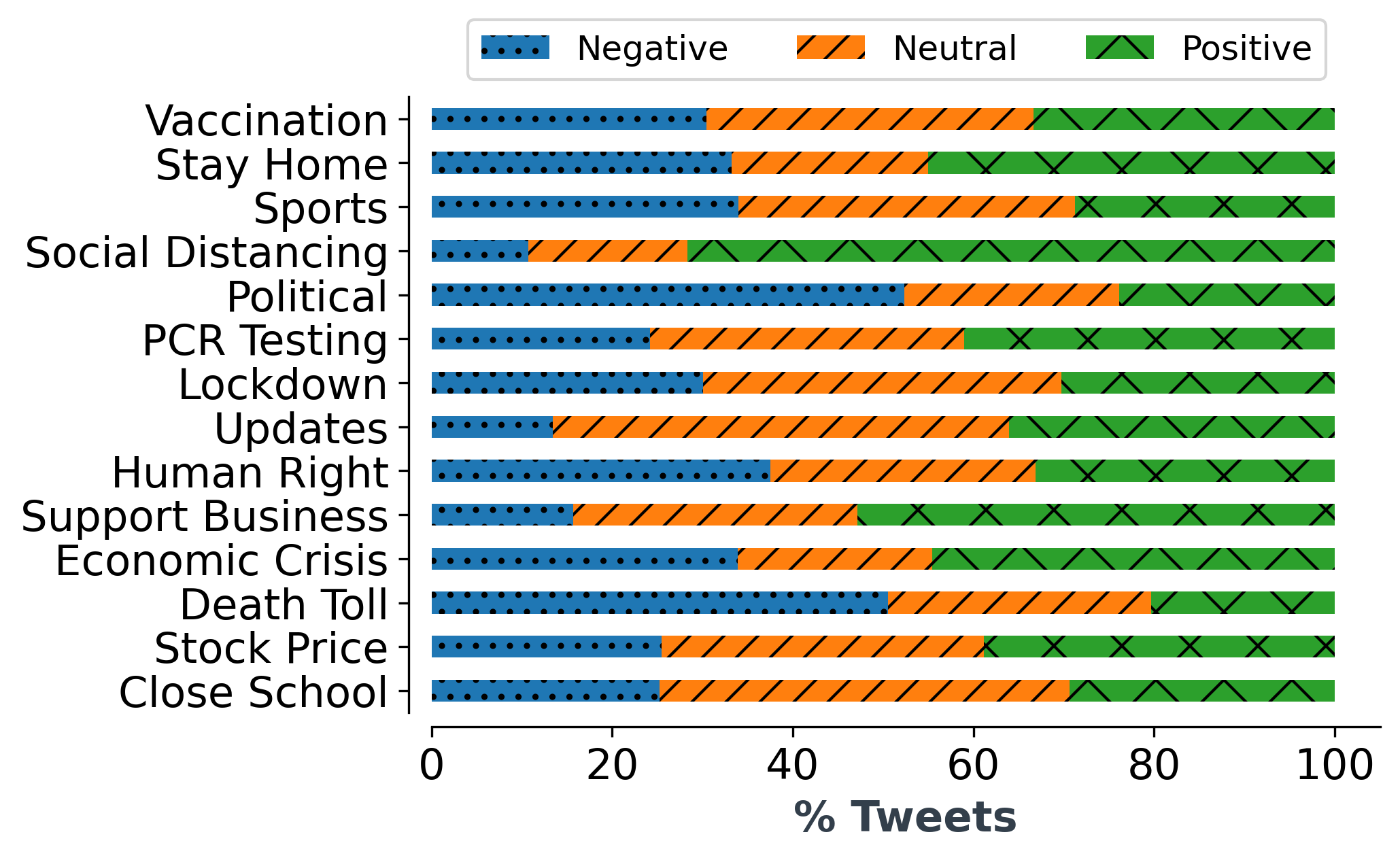}}
% \vspace{-0.5cm}
\caption{\small User Sentiment in each Topic from gelocations in three different period: {\bf Before}, {\bf During}, and {\bf After} lockdown.}
% \vspace{-0.5cm}
\label{fig:sentimentAnalysis}
\end{figure*}

% {\it Insight 4.} 
The URL domain categorisation for individual countries (see Table~\ref{table:statAnalysisSummary} in Appendix \ref{appendix A}) is most consistent with Figure \ref{fig:top domain categories}. For example, in the UK, the top 5 domain categories before, during, and after the lockdown are the same as the order of categories in Figure \ref{fig:top domain categories}. In India and USA, \texttt{streaming} related URLs \ik{seem} to be more popular than \texttt{government and law} related URLs. Meanwhile, in Australia, we can see that \texttt{News and Media} and \texttt{government} related URLs have been shared more times than \texttt{social media} and  \texttt{business} URLs. That is because Australia adopted a ``zero COVID-19 strategy'' and strictly implemented lockdowns, forcing people to share government announcements and news to keep up with the changes in restrictions. Another important observation is that Australians share \texttt{Health and Wellness} related URLs after the lockdown as compared to other countries.

\section{Perception  Analysis  Toward  COVID-19}
\label{sec:sentAnalysis}
{\it Next}, by topic modelling and sentiment analysis of \ik{people's tweets}, we illuminate \ik{people's} perceptions (i.e., feeling and emotion) \ik{during} different stages of the COVID-19 pandemic. 

\subsection{Topic Modelling}\label{sub: topic_modelling}

Topic modelling is the first step towards sentiment analysis. It is a clustering approach that helps in discovering some abstract topics in the dataset. For hashtag analysis, we only considered tweets with hashtags and duplicated them to represent multiple hashtags. However, for the sentiment analysis, we consider all the tweets from our dataset. We applied Latent Dirichlet Allocation (LDA) to generate 15 prominent topics. After obtaining the top 15 clusters, we manually inspected the top 15 words in each cluster and labelled them with a suitable name. We merged two similar topics related to \texttt{politics} hence ending up with 14 topics for our sentiment analysis.

As shown in Figure~\ref{fig:sentimentAnalysis}, most of these topics are consistent with the clusters we obtained in our hashtag analysis (e.g. \textit{vaccination, stay home, political, lockdown,} and \textit{PCR testing}). Some interesting topics identified additionally are \textit{sports, human rights, economic crisis, stock prices}, and \textit{closing schools}. \textit{Sports} and \textit{stock prices} are commonly discussed topics \ik{on Twitter,} regardless of the pandemic. 

% {\it Insight.} 
At the start of the pandemic in 2020, almost all sporting events were cancelled. However, gradually they resumed in controlled environments (e.g. bio-secure bubbles). The pandemic also caused major changes in the business world, collapsing many businesses while skyrocketing the valuation of others. For instance, Video conferencing tool {\tt Zoom}, pharmaceutical company Pfizer Inc., which developed an effective vaccine, and the e-commerce giant Amazon are some of the companies that had considerable increases in their stock prices as a result of the pandemic. Two other important topics we can observe in our topic modelling results are \textit{Human rights} and \textit{Economic crisis}. Due to some border restrictions, families were separated for prolonged periods in some countries. At the same time, many world leaders directly or indirectly mandated that people take vaccination to enjoy their freedom out of lockdown periods. Unvaccinated people even had to resign from their jobs in certain situations. Some people believe these actions violate human rights. Moreover, the closing of businesses due to lockdown periods, \ik{the} lack of seasonal and migrant workers due to border restrictions, and the huge decline in tourism have forced many countries into an economic crisis.

\subsection{Public Sentiment during the Pandemic}

For each topic, we performed sentiment analysis of the tweets using the {\tt VADER} sentiment library \cite{vader} in Python. The main goal of sentiment analysis is to evaluate a body of text and comprehend its viewpoint. Usually, we measure this feeling by assigning the text a positive or negative number known as polarity. The sign of the polarity score is then used to determine whether the prevailing emotion is positive, neutral, or negative. Finally, we normalised the count of each sentiment in each topic to produce results shown in Figure \ref{fig:sentimentAnalysis}.

% {\it Insight 1.} 
We observe that the \ik{topics} with the highest positive sentiments are \texttt{social distancing, support businesses}, and \texttt{stay home}. \texttt{Social distancing} has a positive sentiment of 69.85\%, 70.5\%, and 71.65\% before, during, and after the lockdown periods, respectively. Similarly, \texttt{support businesses} has a positive polarity score of more than 50\% for all periods, while this number is greater than 45\% for \texttt{stay home}. It shows that people were happy with preventative methods and restrictions even though those measures limited their freedoms to some extent. The most negative comments \ik{seem} to be towards the \texttt{death toll} (>50\% for during and after lockdown) and \texttt{politics} (>50\% for all periods). The increase in the number of cases and the resulting deaths were very upsetting to everyone worldwide. 
Moreover, the pandemic is a challenge for politicians as they have to implement strategies that are not welcoming to the public. For example, closing the borders affected families and businesses, making people angry with the government. However, if open borders increase the number of cases and deaths in a country, citizens become angry with the government for allowing COVID-19 and its variants into the country. This suggests that these kinds of situations lead to negative sentiment for political tweets. Meanwhile, \textit{latest updates} seems to have a highly neutral sentiment (\char`\~ 50\% for all periods) along with \textit{closing schools} (>45\% for all periods). Topics such as \textit{vaccination, sports, lockdown, human rights, economic crisis}, and \textit{stock prices} seem to have more or less balanced between the positive and the negative sentiments.

% {\it Insight 2.} 
When inspecting the sentiments of individual countries (please see Figure~\ref{fig:sentiment_countries} in Appendix \ref{appendix A}), we observe similar trends with some quite noticeable results: The negative sentiment for the death toll in India is significantly lower (31.06\%), while it is comparatively higher (57.31\%) in the UK. One reason is that the IR is considerably lower in India than in the UK for the lockdown dates we considered in our analysis. Moreover, it can be the same reason for the UK's highly positive sentiment (74.91\%) for social distancing compared to other countries.

\section{Privacy Risks Exposure Analysis}
\label{sec:prvcAnalysis}
\ik{In this section, we discuss our findings on privacy risks associated with COVID tweets that could lead to privacy leakages, such as sensitive information disclosure and user identification and tracking. We use the methodology given in~\cite{masood2018incognito} that quantifies privacy risks of web data based on three key aspects: \textit{uniqueness, uniformity,} and \textit{linkability} of the web data. Considering what we intend to investigate in this study and the generic nature of the model in \cite{masood2018incognito}, we apply this framework to the Twitter dataset for our privacy risk analysis.}

\subsection{Privacy Threat Model}

Our privacy risk quantification and estimation \textcolor{blue}{is} based on a defined threat model. The model considers an anonymised dataset of tweets that do not contain any user identification, i.e., all the user identity attributes have been removed from the dataset. We assume an adversary as a third party who has been given access to the dataset for non-malicious purposes (e.g., checking aggregated statistics). However, the adversary can analyse the tweets and identify the user based on their tweets. We assume that an adversary has sufficient resources to execute the privacy attack on the dataset. \\

\ik{\begin{definition}[\textbf{Privacy Risk in Anonymised Data}]
We define privacy risk in (anonymised) tweet data as a risk of identifying users and thereby learning their sensitive/private information through; {\it (1)} \textbf{uniqueness} in the sequences of a user's tweets from other users' tweets, {\it (2)} \textbf{uniformity} of the user in his tweets, and {\it (3)} \textbf{linkability} of the user using his personal identifiable information (PII)\footnote{Users often share or search for PII in the tweets including names, contact details, address/location details of people, and ego-surfing).} available in tweet data. \\
\end{definition} }

The user identification of an anonymised dataset is possible \ik{from} three different scenarios. 1. \textit{Uniqueness in Tweets} refers to a unique sequence of tweets posted by a user \ik{(e.g., posting about home quarantine after testing positive for COVID-19)}, 2. \textit{Uniformity in Tweets} refers to a set of similar tweets posted by a user \ik{(e.g., continuously posting about air travel from one country to another during a pandemic)}, and 3. \textit{Linkability in Tweets} refers to Personal Identifiable Information (PII) \ik{given} in the tweets (e.g., giving the location of a COVID-19 vaccination clinic). Our proposed threat model assumes that the continuous flow of information in the form of the above three scenarios could lead to user tracking and identification, even if the data is anonymised.

\begin{figure}[b!]
\centering
\includegraphics[width=1.0\columnwidth, keepaspectratio]{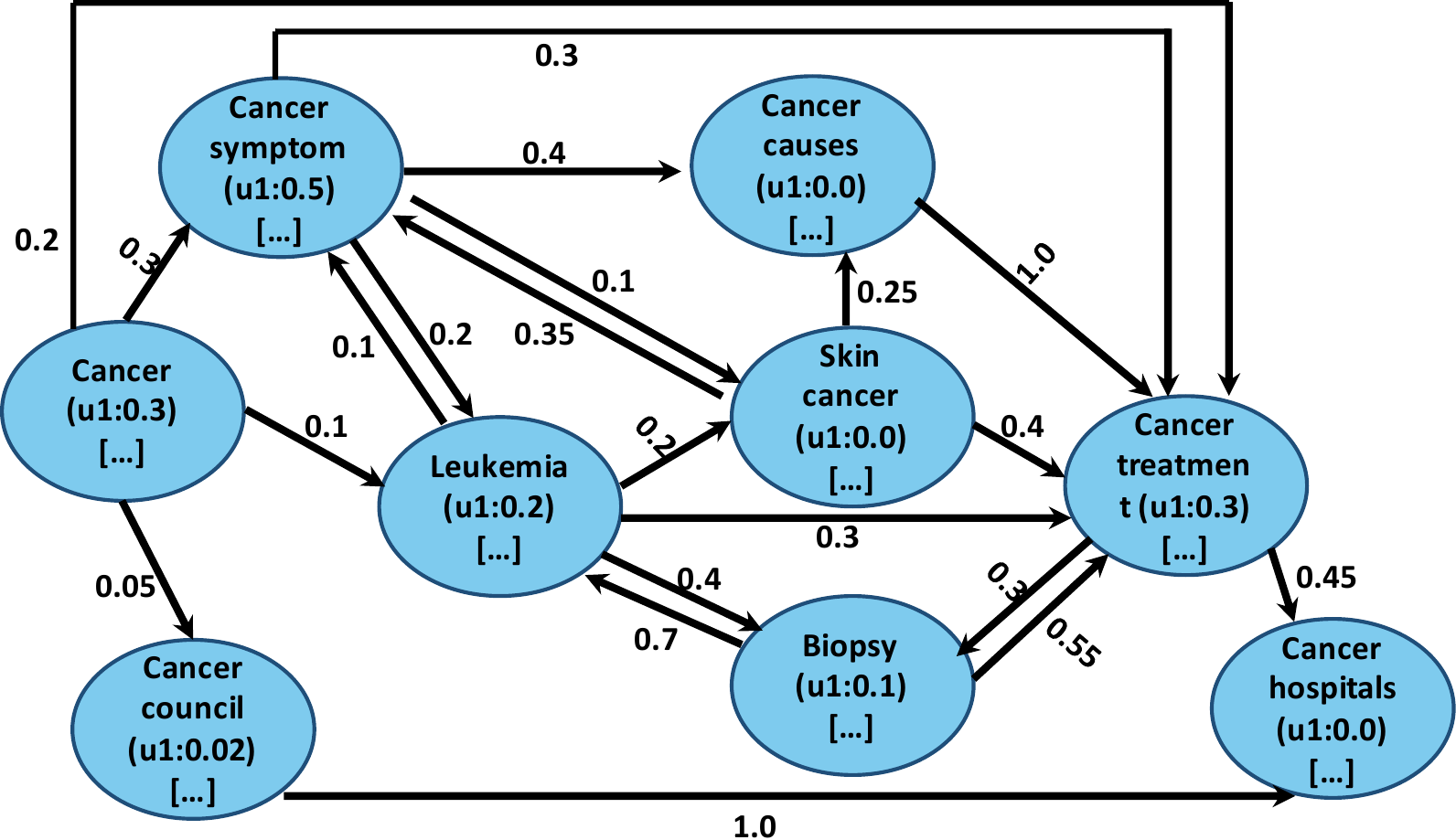}
\caption[An Example of an HMM model for Cancer Topic in Tweets Data]{\ik{An Example of HMM model for Cancer Topic in Tweets Data. Nodes are cancer-related posts/topics, and edges between nodes represent the transition (conditional) probabilities. Each node contains observation probabilities for different users (in this example, these probabilities are shown only for user $u1$)}.
}
\label{fig:HMM_priv}
\end{figure}

\subsection{Privacy Risk Quantification Method}
For our work, we define \textit{privacy risk} as \textbf{the probability of identifying social media users by learning their private or sensitive information through their tweets}. The three key probabilities that are involved in risk quantification are \textit{(1) Probability of Uniqueness:} measured as the non-likelihood of a user's tweets sequence being similar to tweets of other users such that the sequence is unique or distinguished to reveal the user's identity. \textit{(2) Probability of Uniformness:} measured as the likelihood of a user entering a specific tweet (and thereby \ik{being} interested in a specific topic) based on the user's previous tweet history. The more the user has entered a certain type of tweet, the more confidence in the inference that the user is interested in that topic. \textit{(3) Probability of Linkability:} measured by how much PII is available from a user's tweet data. PII could reveal the identity of \ik{a} user and allow linking the corresponding data to the user.

The overall privacy risk is calculated as the joint probability of identifiability (uniqueness and uniformity) and linkability. \ik{The probability of inference from a sequence of user tweets is often conditional on previous tweets. Therefore the risk of inference becomes higher along with a user's sequence of tweets (i.e., the probability of privacy preservation becomes lower with the sequence of a user's tweets/data). The reason behind this intuition is that a user reveals more with the sequence of his posts and the data become more refined or specified to a certain topic enabling the tweets sequence to be highly linkable (less anonymised) to an individual. Therefore, the inference probability becomes higher, and the following tweet data by the user might be at an even higher risk of disclosure.}

\begin{figure}[t!]
\centering
\includegraphics[width=1.0\columnwidth, keepaspectratio]{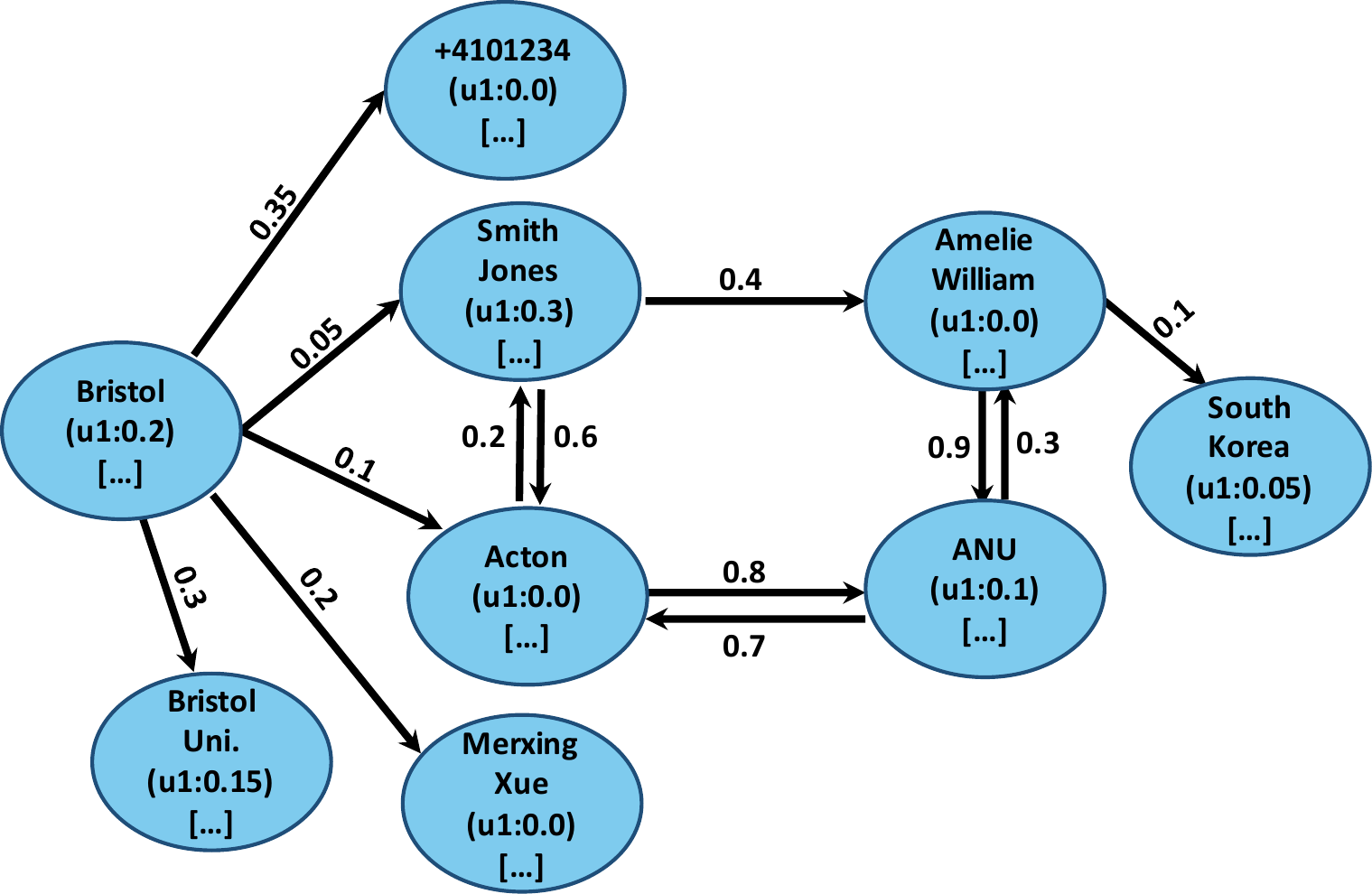}
\caption[An Example of HMM model for PII Topic in Tweets Data]{\ik{An Example of an HMM model for PII topic in Tweets Data. Nodes are tweets/posts containing PII, and edges between nodes represent the transition (conditional) probabilities. Each node contains observation probabilities for different users (in this example, these probabilities are shown only for user $u1$).}
}
\label{fig:HMM_PII}
\end{figure}

\begin{figure*}[t]
\centering
\subfloat[Before]{
\includegraphics[width=0.480\textwidth, keepaspectratio]{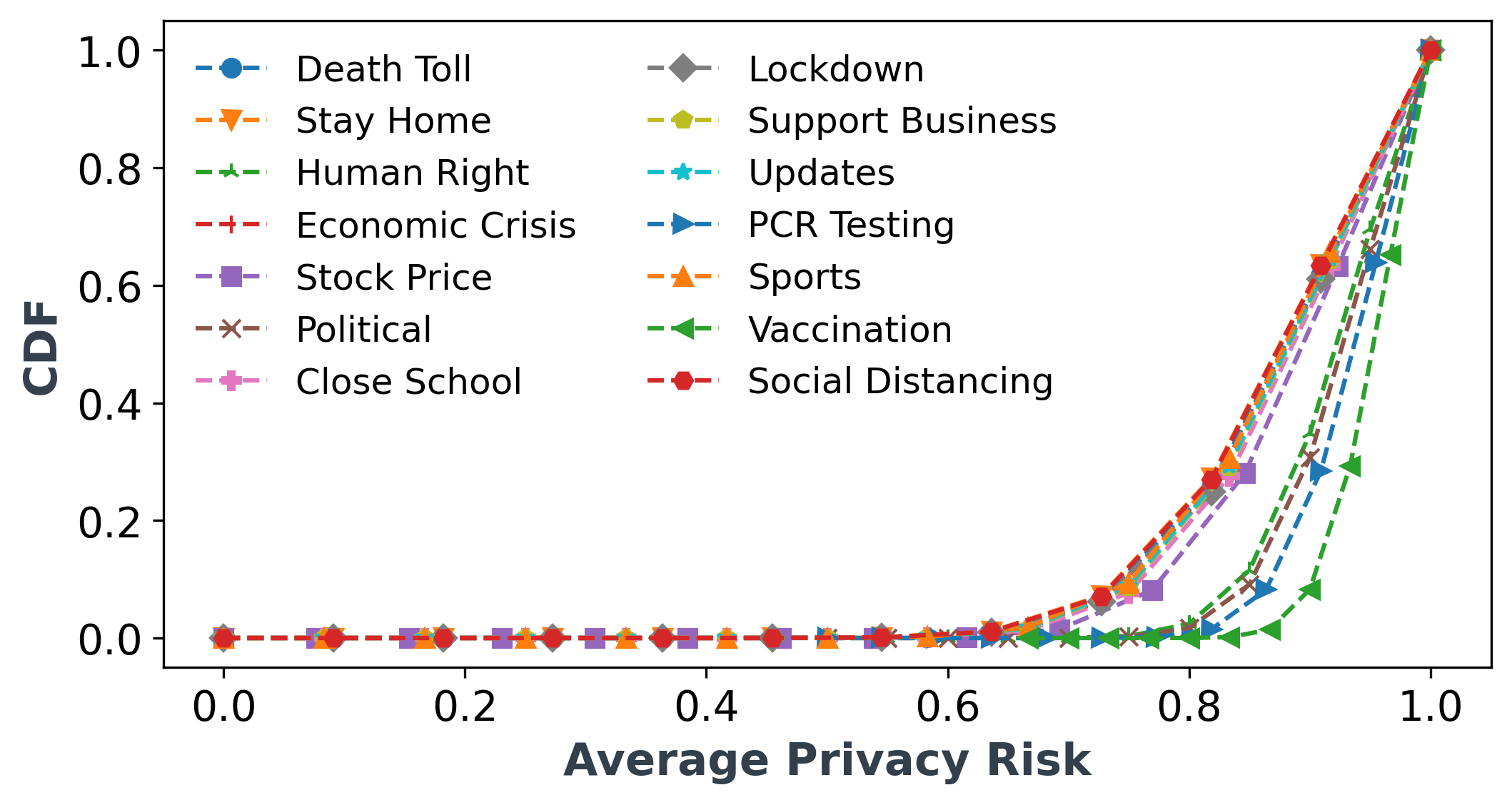}
}
\subfloat[During]{\includegraphics[width=0.480\textwidth, keepaspectratio]{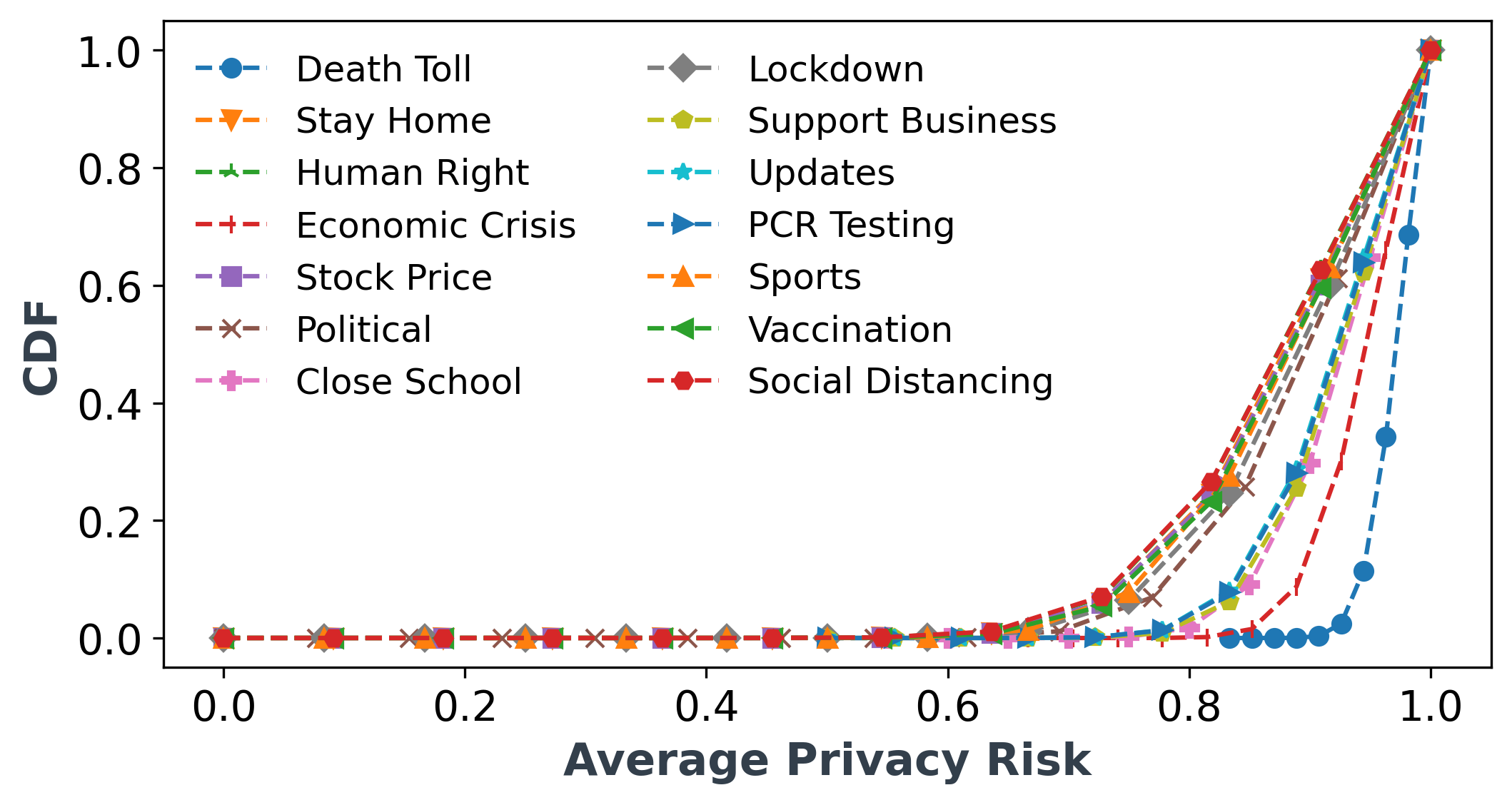}}

\subfloat[After]{\includegraphics[width=0.48\linewidth, keepaspectratio]{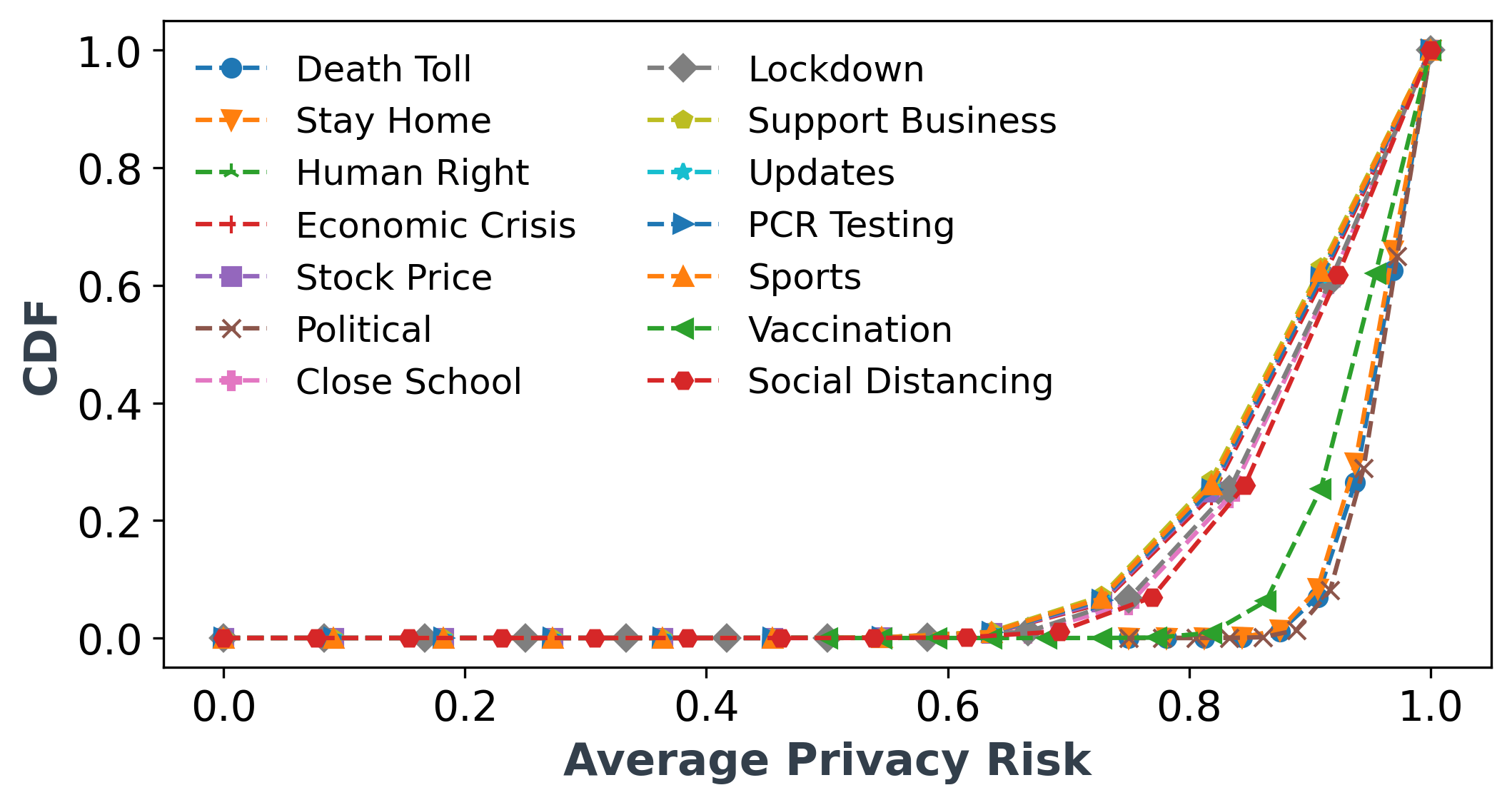}}
% \vspace{-0.5cm}
\caption{\small Average privacy risk per user in three different periods: {\bf Before}, {\bf During}, and {\bf After} lockdown.}
\label{fig:avgPrivacyRiskPerUserLinkable}
% \vspace{-0.4cm}
\end{figure*}

\subsubsection{\ik{Risk Prediction}}

In order to measure the \ik{uniqueness, uniformity, and linkability} probabilities, we use the Hidden Markov Model (HMM). We train the HMM model using previous tweets of a user in order to predict the privacy risk of that user's current tweet. \ik{HMM is a probabilistic model for representing probability distributions over sequences of observations. This model is used in speech recognition systems, computational molecular biology applications, computer vision applications, and other applications of artificial intelligence and pattern recognition~\cite{Bun01}.} Let us consider \ik{a user is represented by $u_i$} and $X_t$ represents a tweet at time $t$. Also, assume a sequence of events (i.e., tweets) by a user at time $t$ is $X_1, X_2, ..., X_t$, respectively. \ik{These events satisfy the (first-order) Markov property, i.e., the current event $X_t$ is independent of all the events prior to $X_{t-1}$. Each of these events $X_t$ outputs observations $Y_t$, which also satisfy the Markov property, i.e., $X_t$ and $Y_t$ are independent of the events and observations at all other time indices. These Markov properties state that the joint distribution of a sequence of events and their observations can be factored as:}

\ik{
\begin{equation}
p(X_{1:T},Y_{1:T}) = p(X_1)P(Y_1|X_1) \prod_{t=2}^{T} p(X_t|X_{t-1})p(Y_t|X_t).
\label{eq:markov}
\end{equation}
}
 
\ik{Figure~\ref{fig:HMM_priv} and Figure~\ref{fig:HMM_PII} show examples of an HMM trained for tweets related to PII on the sensitive topic \textit{cancer}. The tweets entered by a user become a node, and the probabilities of uniqueness, uniformity, and linkability are modelled in the HMM. The three probabilities modelled are:} 

\textbf{Uniqueness} is modelled as transition probabilities in the HMM. \textit{Transition probability} is a conditional probability of a tweet by all users given previous tweet sequences from all users. \ik{This is required to calculate the indistinguishability or non-uniqueness of a user's data from other users' data. The risk of a piece of data being distinguishable depends on the previous data. The reason is that the information gain from a piece of data becomes higher if the previous data on the same topic are considered.} In HMM, edges contain the transition probabilities between nodes $(p(X_t|X_{t-1}))$. These transition probabilities are weighted ($w_T$) by their confidence in terms of how many transitions have occurred, i.e., $w_T = 1/count(X_t|X_{t-1})$. Hence, the weighted transition probabilities are considered as $w_T \times p(X_t|X_{t-1})$.

\textbf{Uniformity} is modelled as observation probabilities in the HMM. \textit{Observation probability} is a probability of a tweet found in the previous tweet history of different users ($u_i$ ), including the user whose risk is to be predicted (if available). \ik{In HMM, each node contains a set of observations with observation probabilities.} We model the observation probabilities as different users’ probabilities of the given tweet $X_t$, found in previous tweet entries $(p(u_i |X_t ))$. \ik{It is required to incorporate the non-uniformity aspect of a user as the frequency of the data entered by that user. The more a user has entered specific data, the more confidence (and therefore higher risk) in the inference that the user is interested in this data.} Again, these probabilities are weighted by $w_O = 1/count(u_i |X_t )$ and then inversed (as the more uniform a user is higher the privacy risk is and therefore lower privacy probability), i.e., $(1 - w_O \times p(u_i |X_t ))$.

\textbf{Linkability} is measured from the prior probabilities of a user based on previous tweets that include PII (names, locations, and organisations). The privacy risks of user tweets that include PII are modelled in a separate HMM. For a given user $u_i$, the prior risk probability is calculated by getting the minimum privacy probability (maximum privacy risk) from all the paths in the PII HMM, which include nodes $X_t$ that contain an observation probability for the user, i.e., $p(u_i |X_t ) > 0$. \ik{For users who do not have revealed any PII in previous tweets, the prior privacy probability becomes $1.0$.}

\begin{figure*}[h]
\centering
\subfloat[Before]{
\includegraphics[width=0.480\textwidth, keepaspectratio]{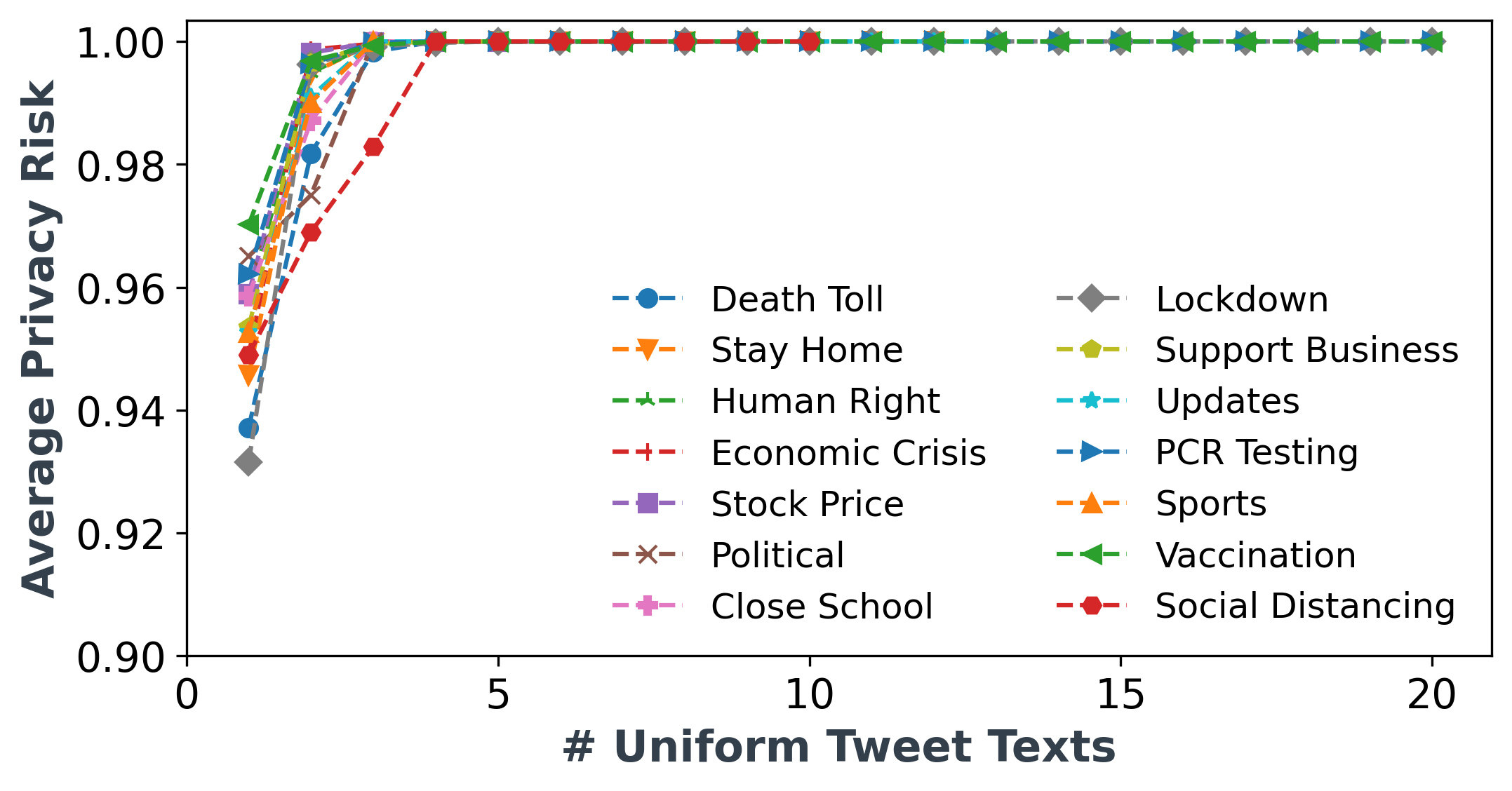}
}
\subfloat[During]{\includegraphics[width=0.480\textwidth, keepaspectratio]{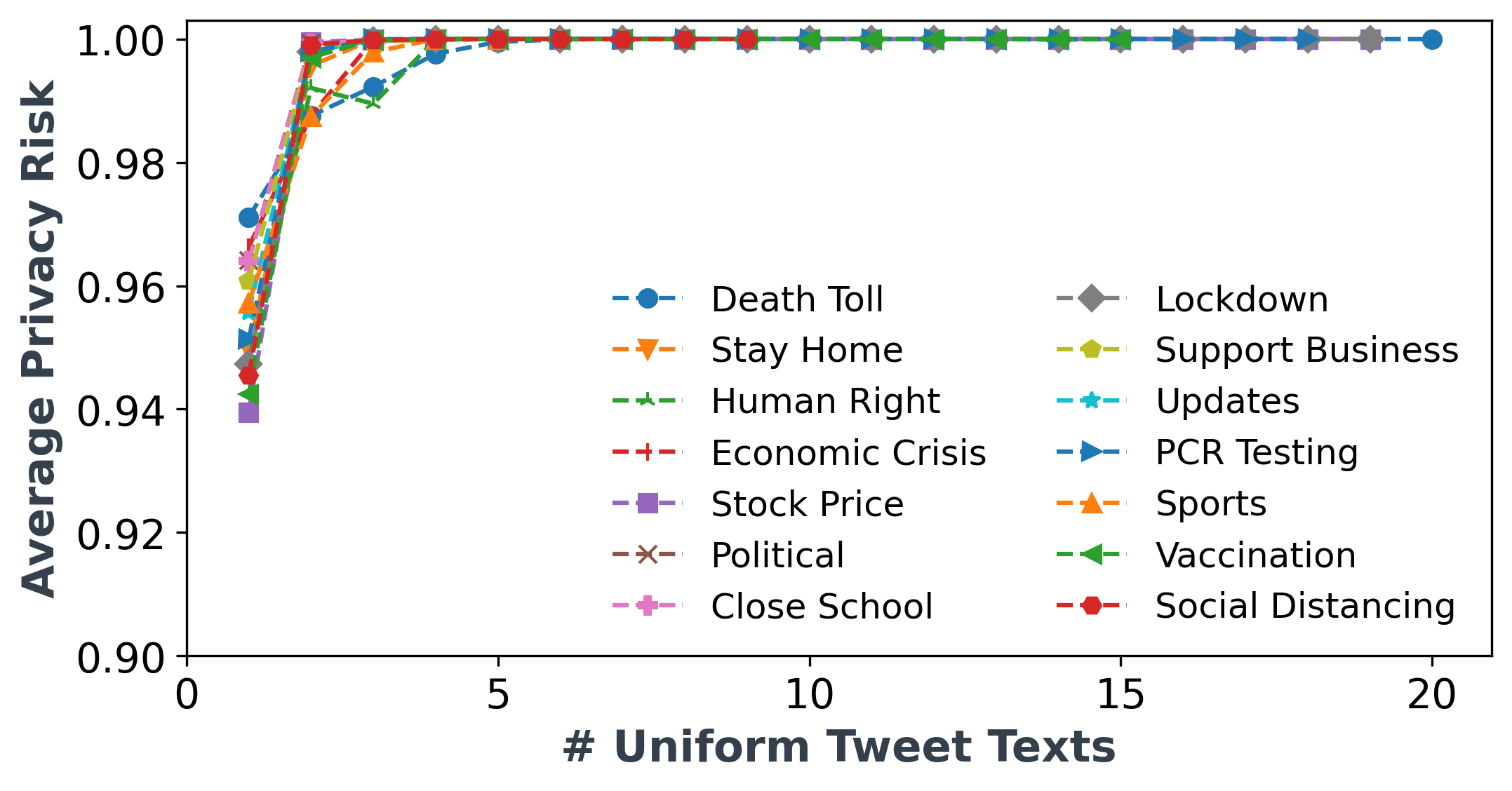}}

\subfloat[After]{\includegraphics[width=0.48\linewidth, keepaspectratio]{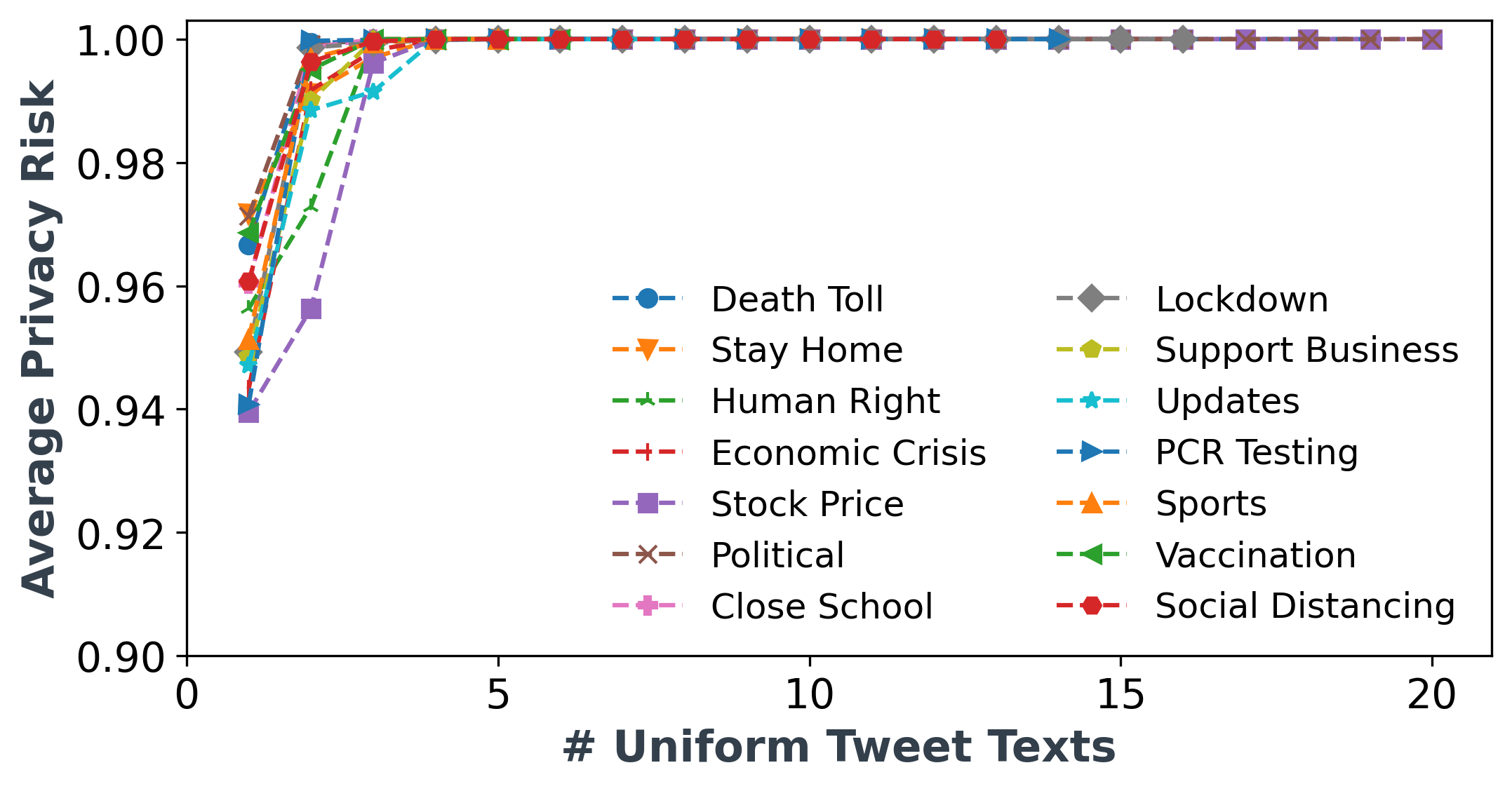}}
% \vspace{-0.5cm}
\caption{\small Risk prediction results of uniform Tweet texts in three different periods: {\bf Before}, {\bf During}, and {\bf After} lockdown.}
% \vspace{-0.4cm}
\label{fig:avgPrivacyRiskUniformity}
\end{figure*}

The overall privacy probability of a user $u_i$ for a sequence of tweets $X_1 \rightarrow X_2 \rightarrow ...\rightarrow X_t$ is calculated as:

\begin{equation}
\begin{split}
p(X_1, \cdots, X_t|u_i) = min(HMM_{PII}|u_i) \times w_T \times p(X_1)  \\
\times (1 - w_O \times p(u_i|X_1)) \times \prod_{x=2}^{t} w_T \times p(X_x | X_{x-1}) \\ \times (1- w_O \times p(u_i|X_x)),
\end{split}
\label{eq:hmm_overall}
\end{equation}

where $HMM_{PII}|u_i$ returns a list of privacy probabilities calculated from the PII HMM for all paths that include nodes where the user has an observation probability $>0.0$.

\subsection{Privacy Analysis}
We apply the privacy risk quantification methodology to the Twitter dataset and analyse the results from the three aspects of uniqueness, uniformity, and linkability. We also present the overall risk prediction results by combining all three of them.

\begin{figure*}[h]
\centering
\subfloat[Before]{
\includegraphics[width=0.480\textwidth, keepaspectratio]{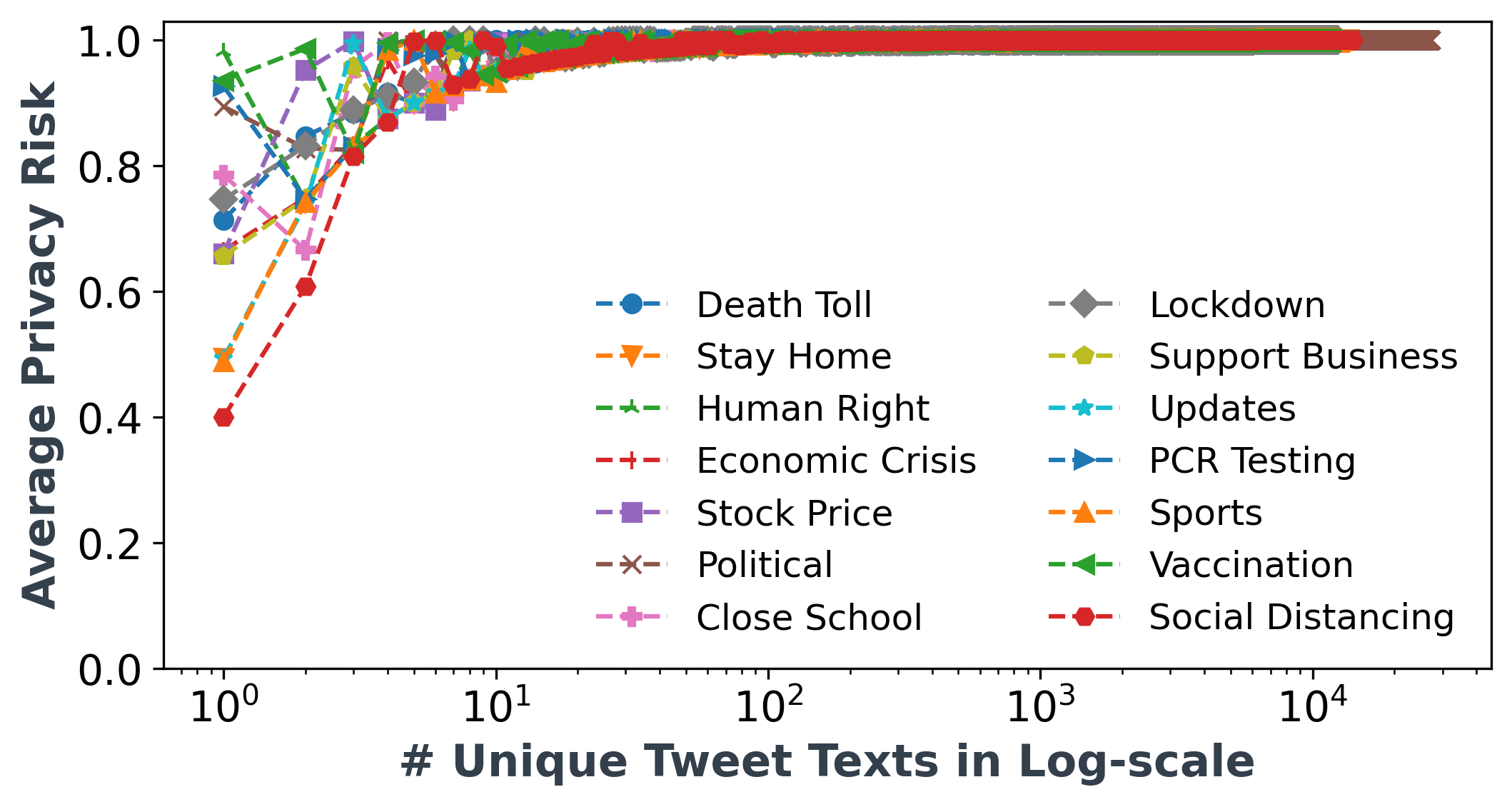}
}
\subfloat[During]{\includegraphics[width=0.480\textwidth, keepaspectratio]{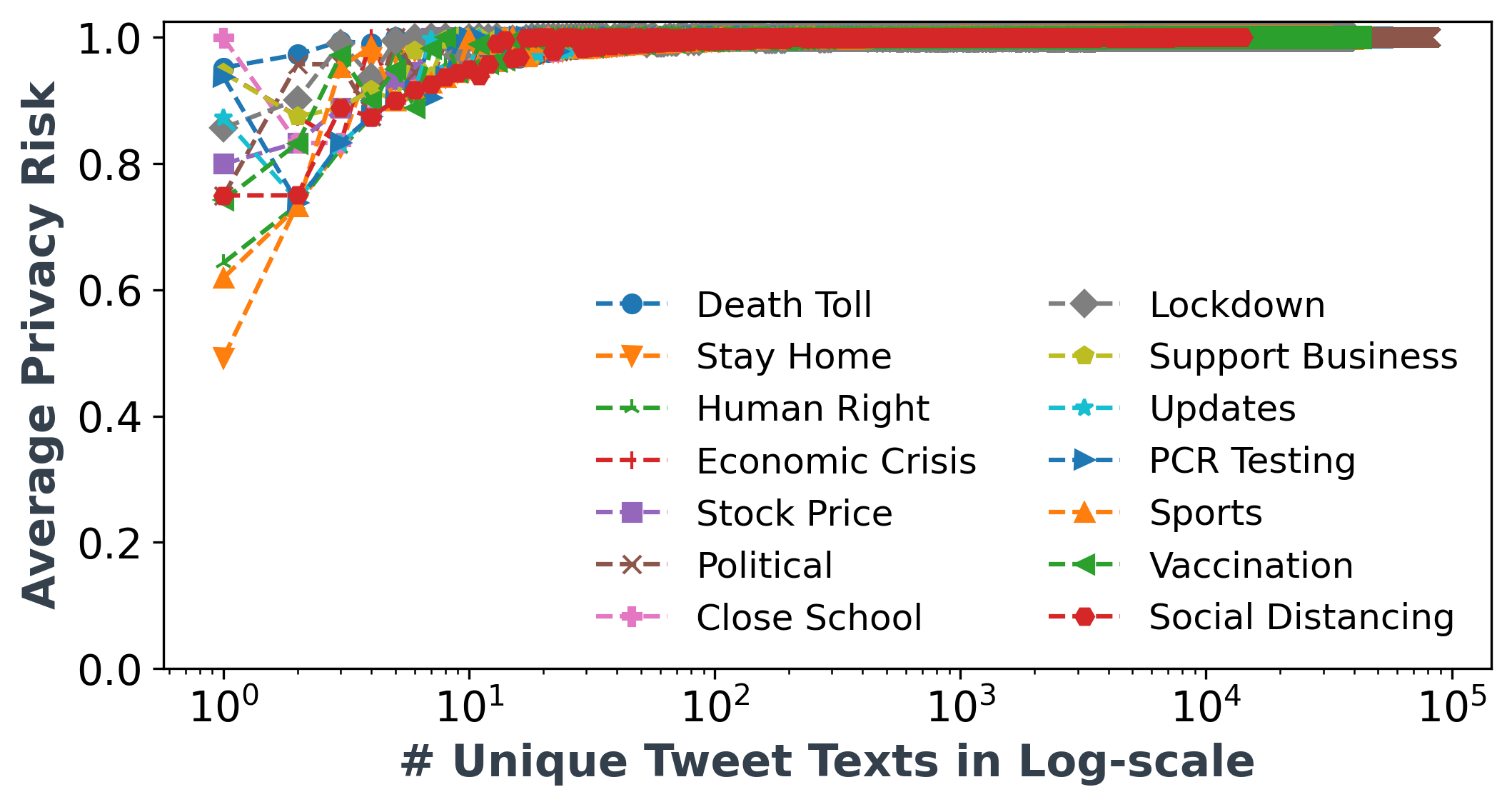}}

\subfloat[After]{\includegraphics[width=0.48\linewidth, keepaspectratio]{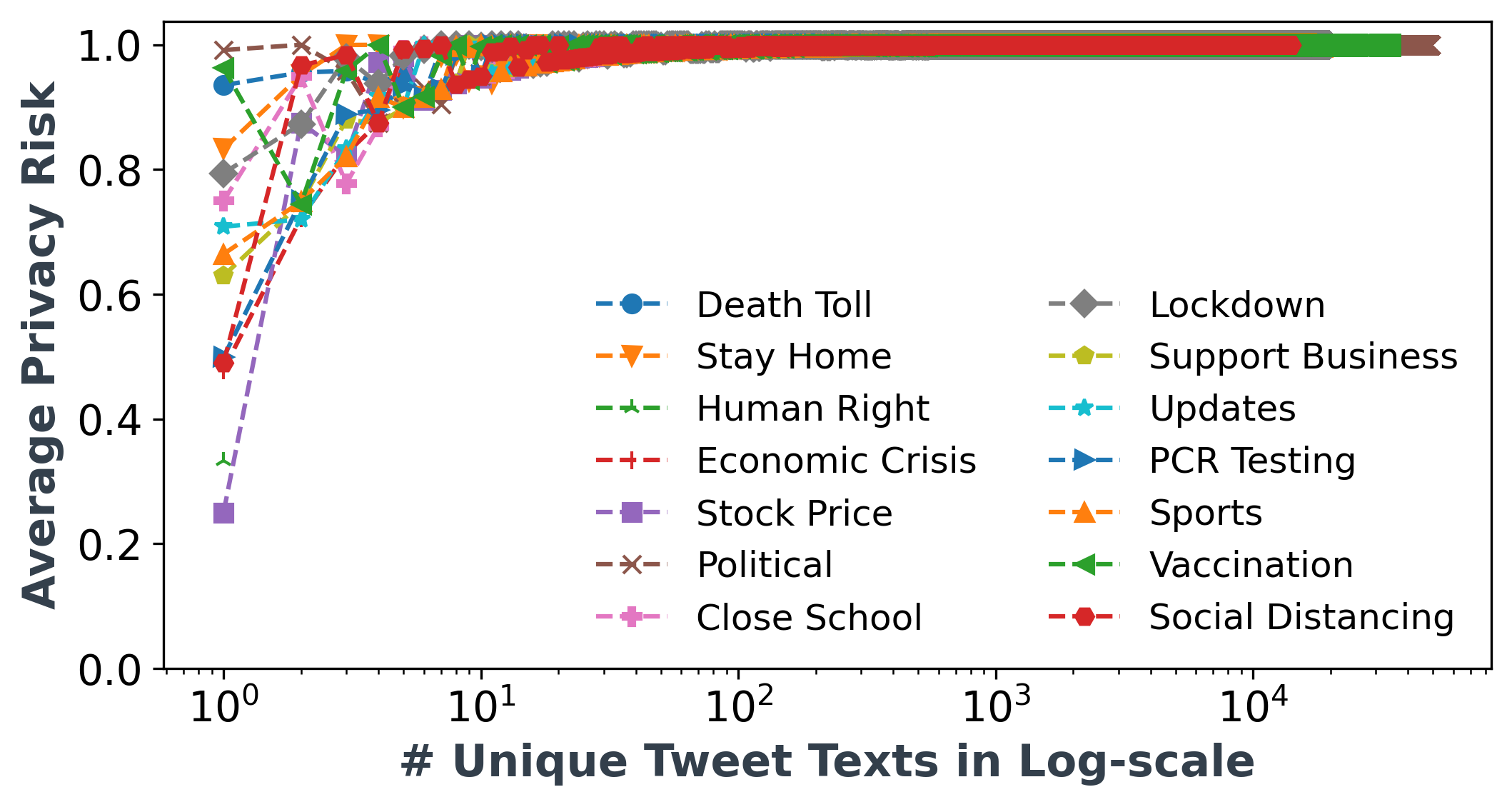}}
% \vspace{-0.5cm}
\caption{\small Risk prediction results of unique Tweet texts in three different periods: {\bf Before}, {\bf During}, and {\bf After} lockdown.}
% \vspace{-0.4cm}
\label{fig:avgPrivacyRisk_Unique}
\end{figure*}

Before applying the quantification method, we first split the data into 20:80, where 20\% of the dataset was used for testing, while 80\% was used to train the HMM model. Furthermore, to reduce training time, we applied k-means clustering that partitions the training data into k clusters and then used a multi-processing technique to run each training cluster simultaneously. As mentioned earlier, the k-means algorithm helps group similar tweets based on the nearest mean (centroid). For our datasets, we selected 14 clusters based on our topic modelling (see Section~\ref{sub: topic_modelling}). Results from each multi-processed cluster are then combined to create one training model. We use cosine similarity to find similar tweets.

% {\it Insight 1.} 
Our results indicate that an average privacy risk reaches 100\% (1.0 privacy risk) when a user enters 3 tweets for all three lockdown periods. Surprisingly, the above result holds true for most COVID-related topics. For instance, \texttt{vaccination} and \texttt{lockdown} topics reach 100\% identification rate after posting just 3 tweets. Figure~\ref{fig:overallPrivacyRisk} in Appendix~\ref{appendix A} shows the average privacy risk when users post 40 tweets on 14 different topics. We also illustrate some specific examples of tweets where the risk becomes 100\% after entering 3 queries in Table~\ref{tab2:case1} of Appendix~\ref{appendix A}. Moreover, the average risk of predicting a user with just 1 sensitive tweet is 94\% (0.94) before the lockdown and 95\% (0.95) during and after the lockdown. Comparing our results with \cite{masood2018incognito}, we observe that COVID-related tweets have 70\% higher privacy risk than normal web data. The higher privacy risk is perhaps because the quantification framework calculates risks based on three aspects, i.e., uniformity, uniqueness, and linkability. Even if a user does not have uniformity in his tweets, he might be identified through the unique pattern of tweets and vice versa. For instance, we can predict after 4 tweets of the user in Table~\ref{tab2:case1} of Appendix~\ref{appendix A} with user ID '168973' that a person has a 6-year-old daughter and is currently stuck in the UK without her parents. Similarly, we observe that another user (with user ID `666231') has a blood clotting condition and cannot have vaccination because of a pre-medical condition. We find similar cases for all the topics and observe that users can be identified through their unique tweet patterns. For instance, we discover that the user with ID '905643' is a male whose wife is 33 weeks pregnant and is concerned about giving coronavirus to an unborn baby. Likewise, the user with ID '369225' informs on Twitter that her daughter, \textit{Miha} is coming to Dubai-UAE after getting a travel exemption.

% {\it Insight 2.}
Figure~\ref{fig:avgPrivacyRiskPerUserLinkable} shows the CDF of users with their predicted privacy risks in three lockdown periods. Before the lockdown, topics such as \texttt{vaccination, lockdown}, and \texttt{PCR Testing}, have a risk higher than 0.85 for more than 50\% of users, while \texttt{Stock Price, Human Right}, and \texttt{Economic Crisis} has a prediction rate of 0.8 for more than 50\% of users. During the lockdown, we observe that \texttt{Death Toll} and \texttt{Economic Crisis} have an average privacy risk of 0.95 for more than 50\% of users, followed by \texttt{Support Business} and \texttt{School Close} topics with a 0.85 prediction rate for 50\% of users. This data clearly indicates that people share more information regarding their personal situation during the lockdown. After the lockdown, we see that topics such as \texttt{Politics}, \texttt{Stay Home}, and \texttt{Death Toll} have the highest privacy risk with a 0.95 prediction rate for 50\% of users.

% {\it Insight 3.} 
{\it on Uniformity:} We now discuss our results on the uniformity of users' tweets during different lockdown periods. Before the lockdown, people are consistently discussing \texttt{vaccination} and \texttt{PCR Testing}, which results in an average privacy risk of 0.97 with just 1 tweet. For instance, we observe that a user enters the tweet \textit{`my son should be returning to \#school today but @stocktoncouncil has withdrawn his transport with no plans to restart whilst there are \#COVID19 whats the plan from'} twice, which makes her 97\% identifiable. Similarly, during the lockdown, topics such as \texttt{death toll} and \texttt{economic crises} have been discussed consistently by the users making them 97\% identifiable with just 1 tweet. After the lockdown, \texttt{politics} and \texttt{vaccination} have an identification rate of 97\% with just 1 tweet. Figure~\ref{fig:avgPrivacyRiskUniformity} shows the average risk for uniform queries. Overall, our results indicate that users are 100\% identifiable after posting 5 uniform tweets for all the topics and all the lockdown periods.

% {\it Insight 4.} 
{\it on Uniqueness:} Figure~\ref{fig:avgPrivacyRisk_Unique} shows the results of posting unique tweets each time. Our analysis shows that around 95\% of tweet, sequences are unique and can lead to 100\% privacy risk for all the topics before the lockdown periods. For example, we observe that out of 82,069 unique sequences during the lockdown, 81,859 unique Tweets are 100\% identifiable. Finally, 48,927 unique sequences are 100\% identifiable out of 49,011 after the lockdown state. 

\begin{figure*}[t]
\centering
\subfloat[Before]{
\includegraphics[width=0.48\textwidth, keepaspectratio]{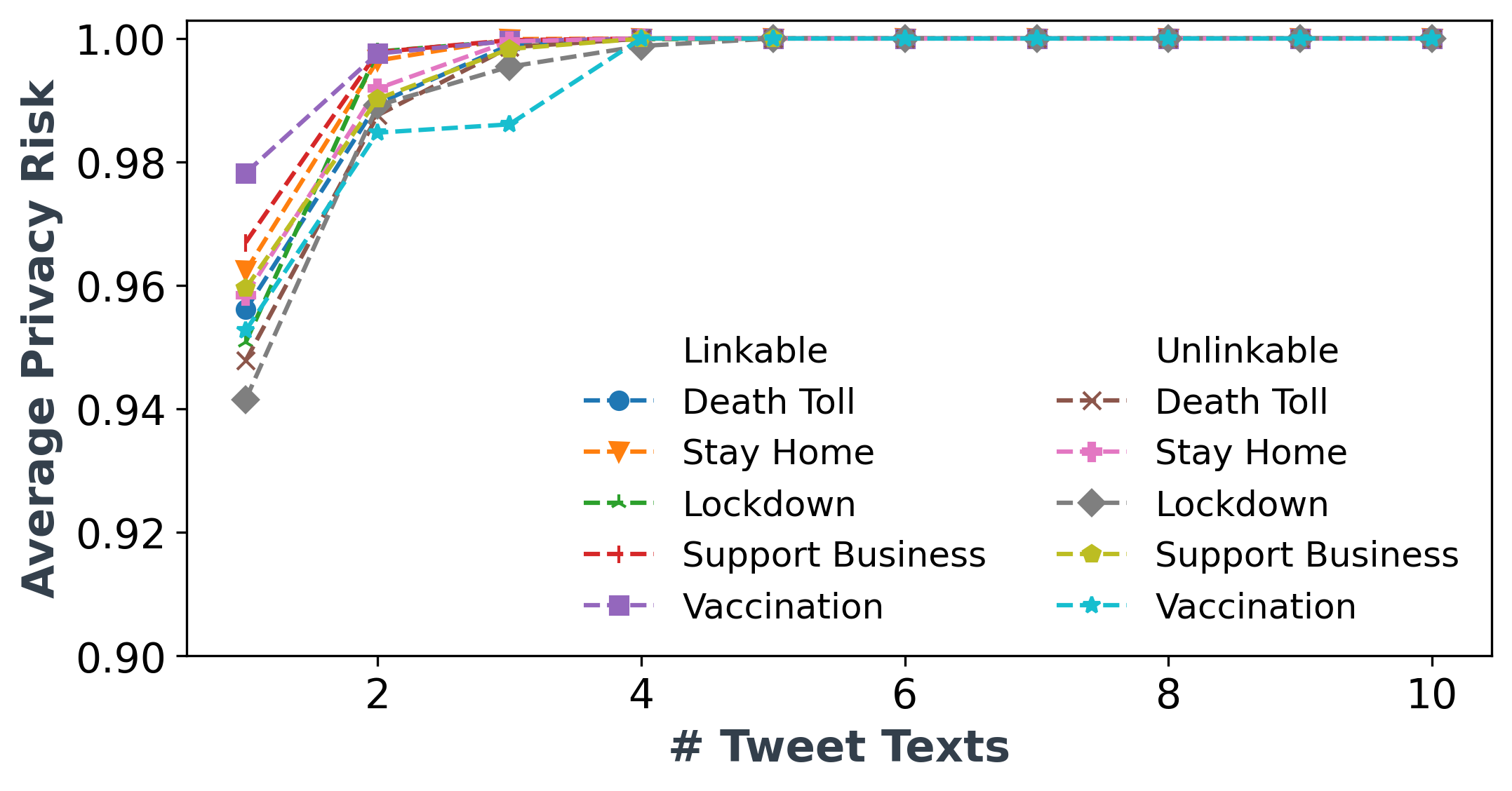}
}
\subfloat[During]{\includegraphics[width=0.48\textwidth, keepaspectratio]{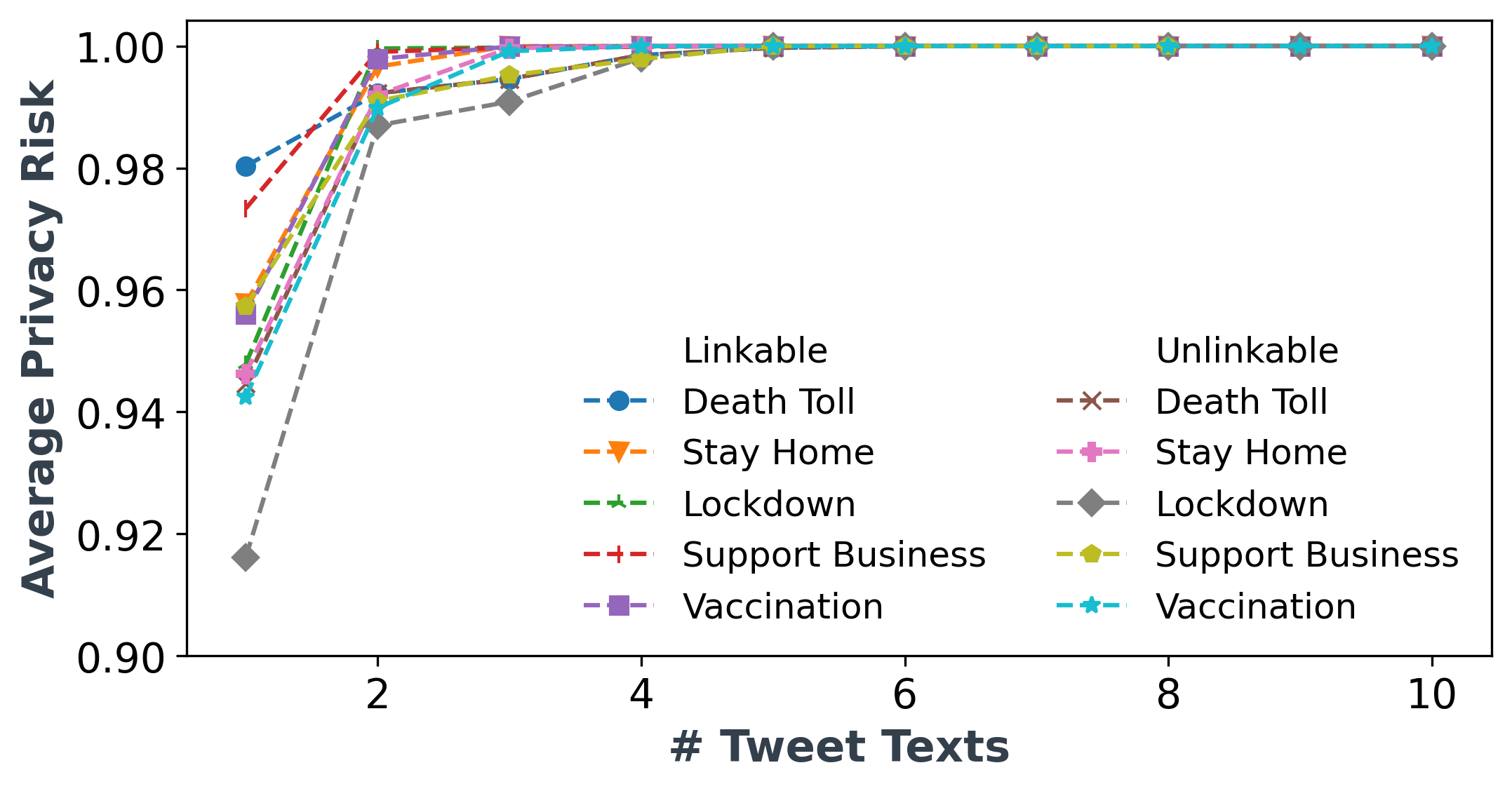}}

\subfloat[After]{\includegraphics[width=0.48\linewidth, keepaspectratio]{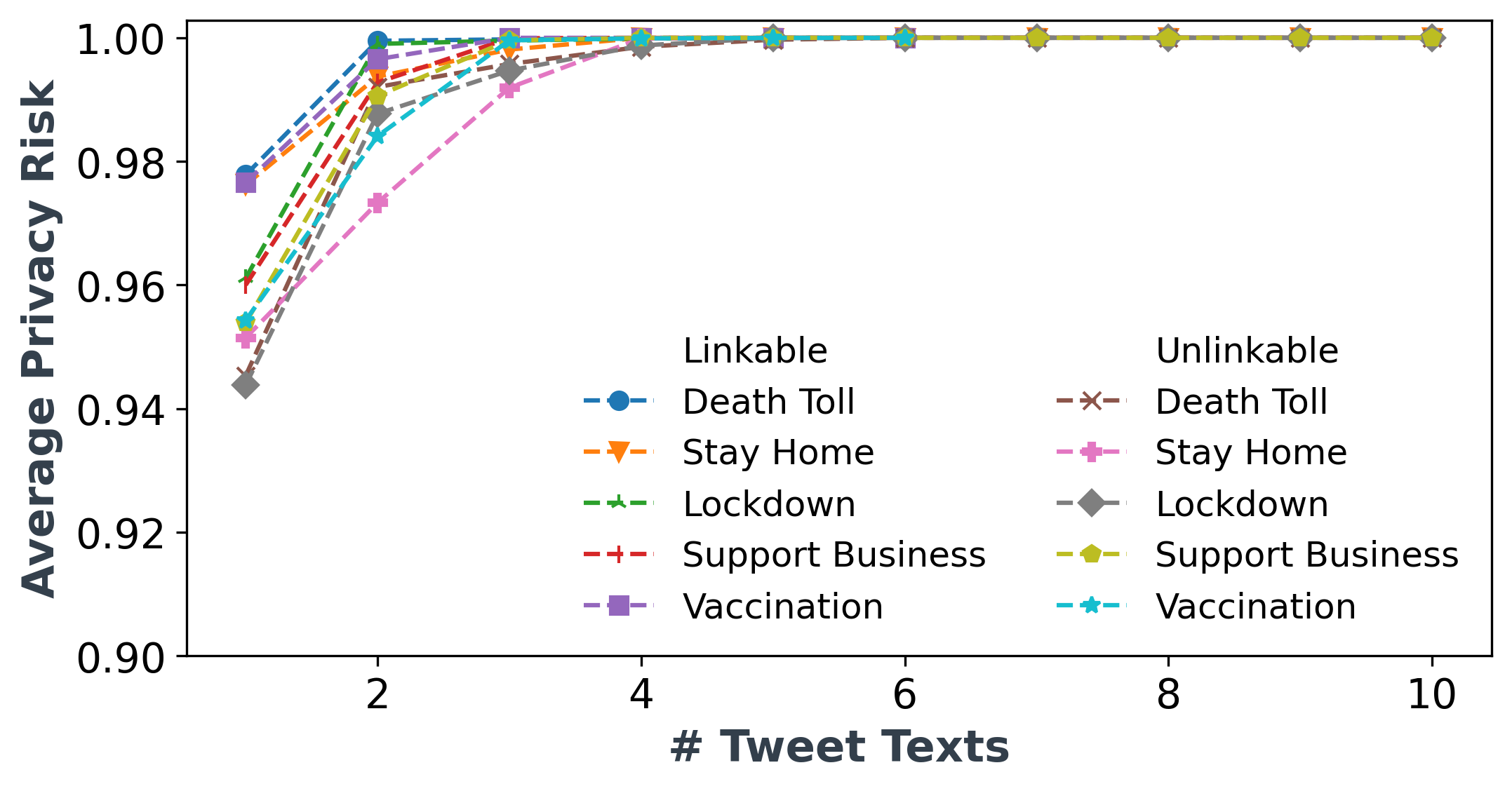}}
% \vspace{-0.5cm}
\caption{\small Linkable and unlinkable average privacy risk against Tweet texts having PII in three different periods.} %: {\bf Before}, {\bf During}, and {\bf After} lockdown.}
%\vspace{-0.5cm}
\label{fig:avgPrivacyRiskLinkvsUnLink}
\end{figure*}

% {\it Insight 5.} 
{\it on Linkability:} We now investigate the linkability of users' tweets using their PII. We found \ik{a} few users who have PII information available in their tweets. For instance, a user in a \texttt{Lockdown} topic shared a tweet \textit{'As fate would have it, I was scheduled to fly to Cairo this evening (tickets were canceled weeks ago). I havent m been in one place for this long in over six years. Today, of all days, I wish the skies were fully open and I could go and see my family.'}. Another user in \texttt{Social Distancing} topic entered PII query \textit{'Great family day out to the Chester Zoo today - great outdoors walk with socially-distanced measures in place all the way around the park. Congrats @chesterzoo for making it work so well within these COVID-19 times'}. Figure~\ref{fig:avgPrivacyRiskLinkvsUnLink} shows the average privacy risk for the queries having PII available for the three lockdown periods. We also present results without linkability information, i.e., we remove PII and evaluate the privacy risk for the same set of entries. Our results indicate that linking tweets with PII has a higher privacy risk compared to tweets with no PII. For instance, before the lockdown, \texttt{Vaccination} topic has the minimum average risk of 98\% for linkability, which reduces to 94\% if we remove PII. Similarly, during the lockdown, \texttt{Death Toll} has 98\% minimum privacy risk with PII and 95\% without PII. After the lockdown, the \texttt{Support Business} topic, for example, has a 96\% of minimum average risk with PII but reduces to 95\% without PII. However, we found that tweets with or without PII can eventually reach up to 100\% identifiability (uniqueness and uniformity) for all the topics and all the stages of lockdown, respectively.

\section{Exposure to Suspicious Content}
\label{sec:secAnalysis}
In this section, we aim to investigate people's exposure to suspicious content. We analyse the suspicious domains and the associated security risks for four individual countries (Australia, India, the US, and the UK) and three different periods of the COVID-19 pandemic (before, during, and after lockdown periods). We use the URLs shared in tweets and utilise VirusTotal \cite{virustotal} to determine whether or not the second-level domains of those URLs are involved in any malicious activities. We also use the URL categories to analyse the most suspicious categories of second-level domains.

\begin{table*}[t]
\small
\centering
\caption{\small Number of suspicious URLs and domains in our dataset (VT Score >= 3).}
\scalebox{0.95} {
\begin{tabular}{lcccc}
\toprule
{\bf Time} & {\bf Total} & {\bf Unique} & {\bf Suspicious} & {\bf Unique Suspicious} \\
{\bf Period} & {\bf URLs} & {\bf URLs} & {\bf Domains} & {\bf  Domains} \\ 
\midrule
    Before    & 1,061,853 & 586,787 & 15,411 & 140\\
    \rowcolor{DarkOrange}
    During  & 3,771,312	& 2,219,095 & 66,441 & 259\\
    After   & 2,116,832 & 1,254,057 & 35,139 & 196\\
    \bottomrule
 \end{tabular}
 }
 \label{tab:suspicious_urls}
 \vspace{-0.5cm}
\end{table*}

\textit{Methodology:} After extracting the expanded URLs as explained in Section \ref{sub:url analysis}, we removed duplicate URLs. It left us with 4.06 million URLs out of the 6.95 million total URLs. Next, we queried VirusTotal to get reports on each domain in our dataset. VirusTotal is an information aggregator, which presents a combined output of different antivirus products, file and website characterisation tools, website scanning engines, etc. For a URL or a domain, we can obtain a report from VirusTotal. This report provides a number (of positives), which indicates the number of tools that find the URL or the domain suspicious. Using this information, we calculate a parameter called VirusTotal Score (VTScore) for our analysis.  

For every unique domain, we query VirusTotal to get all the reports between Jan 1st, 2020, and Nov 6th, 2021. Then, for each domain with positives >= 1, we take the sum of positives in all the reports and divide it by the total number of reports. We use the VTScore as a metric to identify how suspicious a particular domain is. The higher the VTScore, the domain is deemed more suspicious. Table \ref{tab:suspicious_urls} shows the number of suspicious domains we obtained for a VTScore greater than or equal to three. We draw the following insights from our analysis.

% {\it Insight 1.} 
Firstly, we observe that the number of URLs shared during the lockdown periods is higher than before or after lockdown periods. It has caused a proportionate increase in the number of malicious domains. We found 345 unique suspicious domains overall (some domains are found in more than one period). For example, {\tt docsquiffy.com}, and {\tt peoples.it} are two suspicious domains flagged during the lockdown periods, while {\tt ccp.it} and {\tt comapnycsr.com} are flagged before and after lockdown, respectively. Meanwhile, {\tt buzzsawpoilitics.com} is flagged before and in lockdown periods. Table \ref{table:VirusTotalMaliciousnessScore} is complementary to Table \ref{tab:suspicious_urls}, where we show the number of suspicious domains and the number of unique domains out of them based on different VTScores. We can observe that as the VTScore increases, the number of suspicious domains decreases considerably. For a VTScore greater than and equal to 55, which means the domains are extremely suspicious, we obtained 9 unique domains. {\tt cjsa.org} and {\tt ccp.it} are flagged before the lockdown, and {\tt dudmc.com}, {\tt geitpl.com}, {\tt vietnam.travel}, and {\tt itcslimited.com} are flagged during the lockdown. Moreover, {\tt india.org} is flagged during and after lockdown periods, while {\tt begadistrictnews.com.au} and {\tt grantuk.com} are flagged during all three periods.

\begin{figure*}[h]
\centering
\subfloat[Before]{
\includegraphics[width=0.48\textwidth, keepaspectratio]{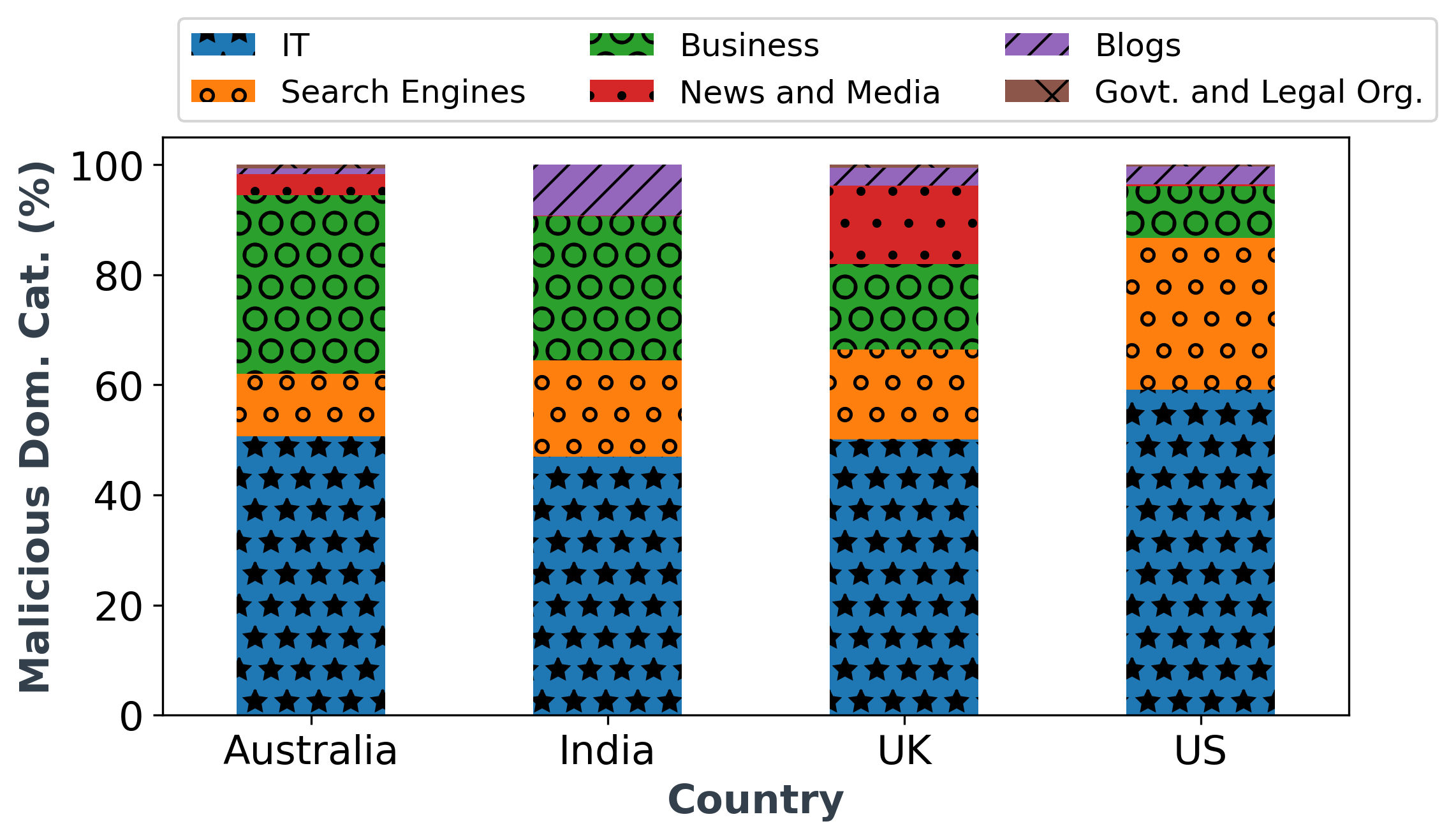}
}
\subfloat[During]{\includegraphics[width=0.48\textwidth, keepaspectratio]{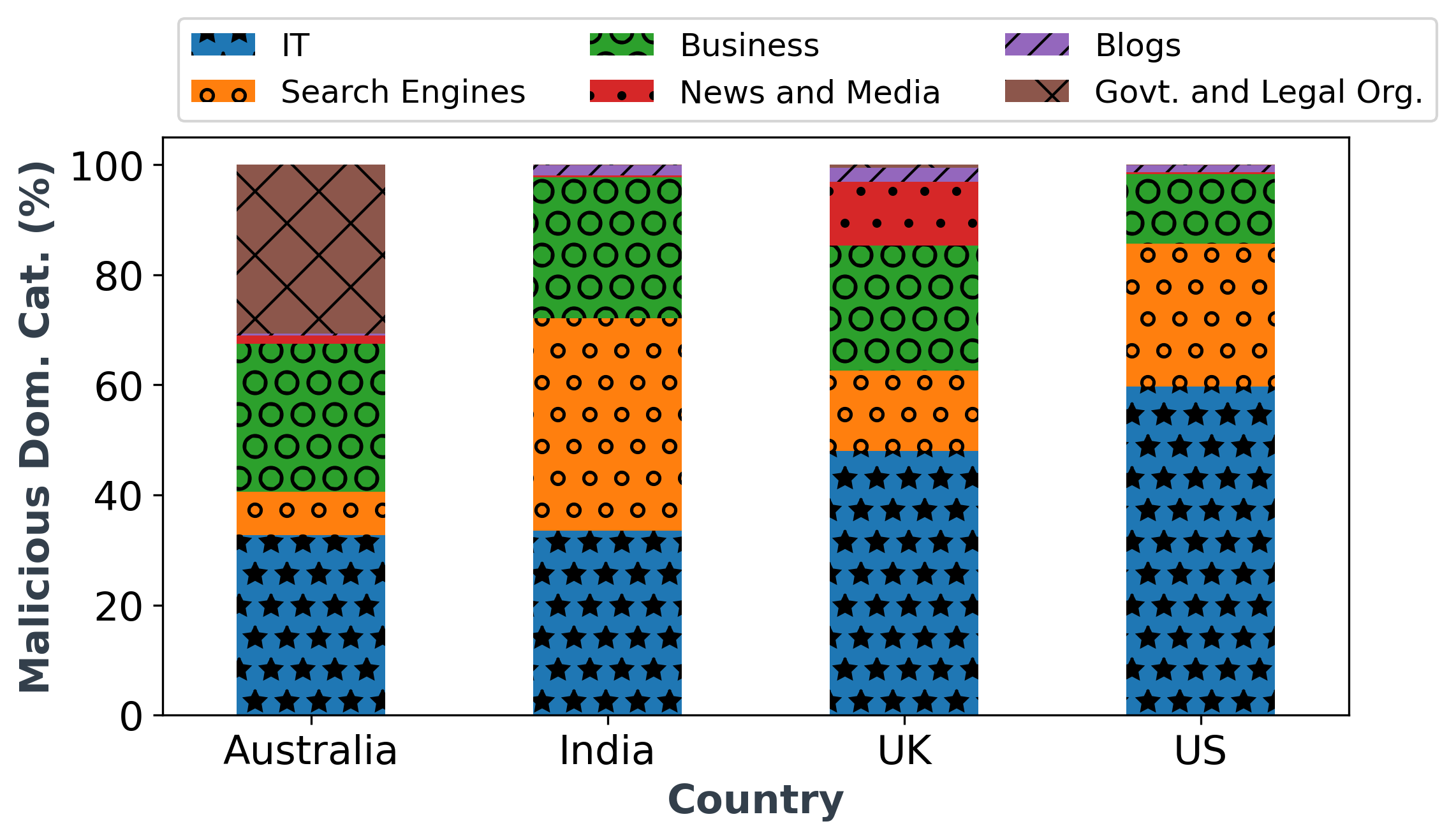}}

\subfloat[After]{\includegraphics[width=0.48\linewidth, keepaspectratio]{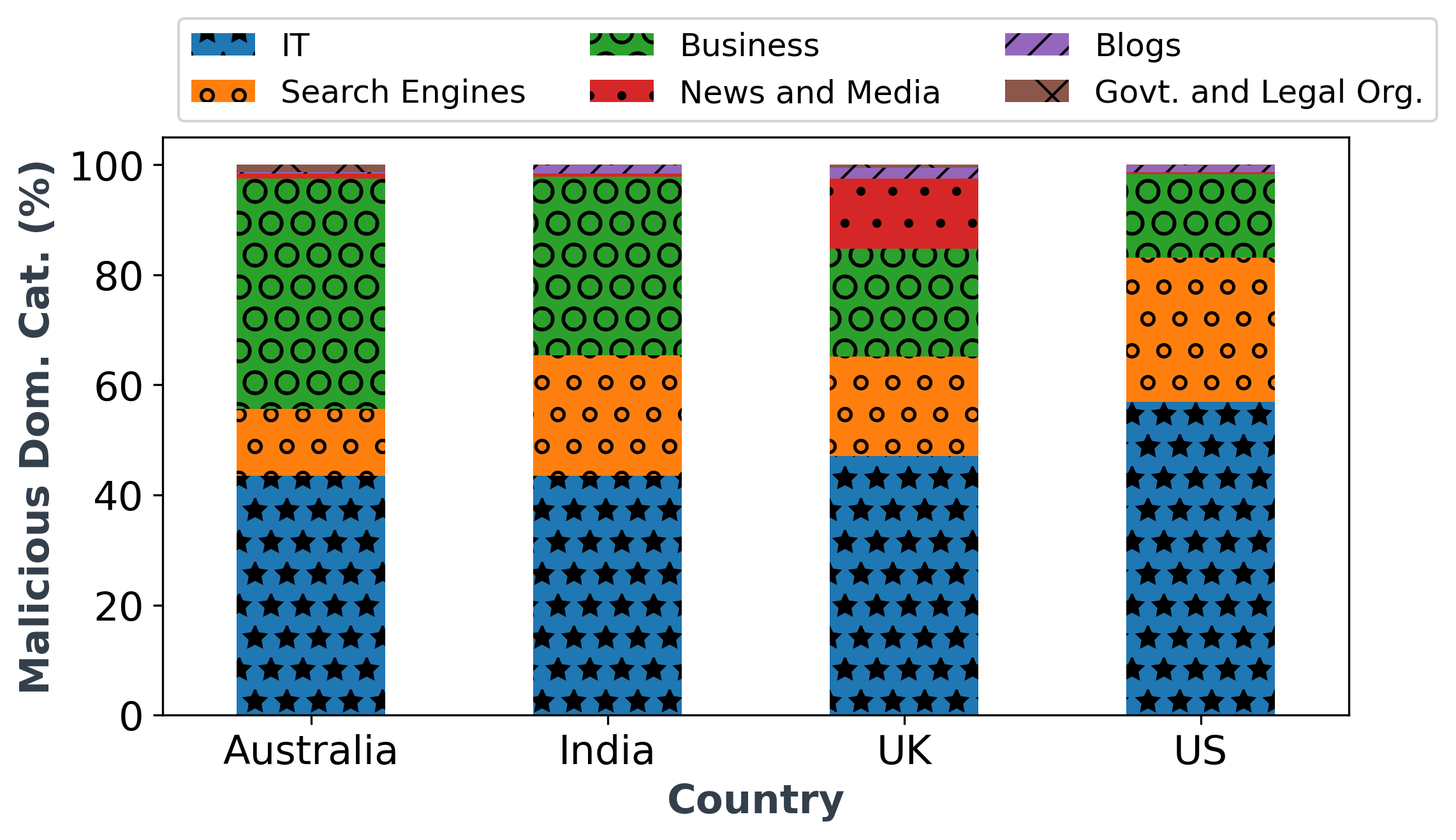}}
% \vspace{-0.4cm}
\caption{\small Distribution of suspicious domains' (with VTScore $\geq$ 3) categories distilled across different geolocations and lockdown periods.}
\label{fig:countrywise malicious domains}
% \vspace{-0.4cm}
\end{figure*}

We also analysed the domain categories of suspicious URLs and their distribution among Australia, India, the UK, and the US. We present our results in Figure \ref{fig:countrywise malicious domains}, and we can notice some interesting traits there. The most significant point is that we can see a domain category called Search Engines representing a considerable number of suspicious domains. We used the same methods described in \S~\ref{sub:url analysis} for the domain categorisation here. However, we do not see the Search Engines category as one of the widely shared domain categories in Figure \ref{fig:top domain categories}. 

% {\it Insight 2.} 
It indicates that URLs with a search engine-related domain have a higher chance of being malicious compared to other domain categories. Moreover, we can see that Social Network related domains are not included in the top 6 most suspicious domain categories, even though it was the most widely shared domain \ik{category} according to our URL analysis. We can assume that the main reason for this is that people mostly share URLs from major social media networks such as {\tt facebook.com}, {\tt instagram.com}, and twitter.com, which are legitimate domains. IT and Business related domains contribute to a majority of suspicious domains. Since both these categories can be work-related most of the time, people tend to click URLs with these domains without much hesitancy. This behaviour can encourage malicious actors to use such domains to distribute malicious URLs.

% {\it Insight 3.} 
When observing the distribution of suspicious URLs in individual countries, we can observe several interesting facts. One of the most significant observations is the high number of suspicious domains related to Government and Legal Organisations during the lockdown periods in Australia. As discussed earlier, due to how the Australian government handled the pandemic, individuals in Australia had to rely continuously on announcements from the government and legal authorities. This situation must have motivated malicious entities to act upon the government-related domains. If we consider India, we can notice two unique features in the distribution. First, we can deduce that the domain category of Blogs contains a noticeable portion of suspicious domains before the lockdown. We can only assume that \ik{suspicious blog-related} domains are widely shared in India during regular times, while COVID-19 has shifted people's focus to other topics of interest. Second, we can see a significant increase in search Engine related domains during \ik{the lockdowns in} India. During lockdown periods, people are mostly confined to their homes, which can increase Internet usage, which could have caused this peculiarity. However, it is difficult to understand why it has not happened in other countries as well. For the UK, we can observe that News and Media have taken prominent places in the charts, while it is not the case \ik{in} other countries. Being a country with high infection rates and many lockdown periods, we can assume that the people in the UK mostly relied on news and media-related domains to get updated about the situation in the country. This situation may have resulted in attackers sharing more suspicious URLs on Twitter, which belong under that category. For the US, we cannot observe any significant traits in Figure \ref{fig:countrywise malicious domains}.

\begin{table*}[t]
  \small
  \centering
  \caption{Number of total and unique domains with different VTScores across three pandmic periods.} %Malicious Domain stats for different VTScores in different stages of the pandemic.}
%   \vspace{-0.4cm}
  \scalebox{0.95} {
  \begin{threeparttable}
  \begin{tabular}{*{11}{l}}
    \toprule
    \textbf{Lockdown} & \multicolumn{2}{c}{\textbf{VTScore $\geq$ 3}} & \multicolumn{2}{c}{\textbf{VTScore $\geq$ 10}} 
      & \multicolumn{2}{c}{\textbf{VTScore $\geq$ 20}} & \multicolumn{2}{c}{\textbf{VTScore $\geq$ 40}} & \multicolumn{2}{c}{\textbf{VTScore $\geq$ 55}}\\
    \cmidrule{2-11}
    % & \# Suspicious & \% Suspicious & \# Suspicious & \% Suspicious & \# Suspicious & \% Suspicious & \# Suspicious & \% Suspicious & \# Suspicious & \% Suspicious & \# Suspicious & \% Suspicious \\
    \textbf{Period} & Total & Unique (\%) & Total & Unique (\%)  & Total & Unique (\%)  & Total & Unique (\%)  & Total & \# Unique (\%)  \\
    \midrule
    \multirow{1}{*}{Before}
     & 15,411 & 140(0.4\%) &	1,560 & 59(0.2\%) & 778 & 39(0.1\%)	& 125 & 21(0.1\%) &  4 & 4(0.01\%) \\\cmidrule{2-11}
    \multirow{1}{*}{During} & \cellcolor{DarkOrange}66,441 & \cellcolor{DarkOrange}259(0.3\%) & \cellcolor{DarkOrange}9,181 & \cellcolor{DarkOrange}108(0.2\%) & \cellcolor{DarkOrange}2,051 & \cellcolor{DarkOrange}72(0.1\%) & \cellcolor{DarkOrange}443 & \cellcolor{DarkOrange}34(0.04\%) & \cellcolor{DarkOrange}18 & \cellcolor{DarkOrange}7(0.01\%)\\
    \cmidrule{2-11}
    \multirow{1}{*}{After}
    & 35,139 & 196(0.3\%) & 5,391 & 88(0.2\%) & 985 & 53(0.1\%) &	277 & 26(0.1\%) & 7 & 3(0.01\%)\\
    \bottomrule
  \end{tabular}
  \end{threeparttable}
  \label{table:VirusTotalMaliciousnessScore}
  }
\end{table*}

\section{Conclusion and Future Research}
\label{sec:conclusion}
COVID-19 brought many changes to the lives of each and every person on the planet. As most of these changes were unprecedented, the impact of most such changes is still unknown. However, this situation encouraged many researchers to go beyond their established bounds and engage in impactful research. Under these circumstances, we decided to conduct a comprehensive study on user behaviour on social media with the major objective of understanding privacy and security risks. We try to identify the main topics of discussion during the pandemic related to COVID-19 and the generic user sentiment towards them. In addition, we try to extend our analysis to examine the impact of different phases during the pandemic and different countries, their infection rates and COVID-19-related policies on our results. Hence our study consists of statistical, sentiment, privacy, and security analyses. All analyses are based on the three lockdown periods (before, during, and after) and consider Australia, India, the UK, and the US.

Our statistical analysis revealed that \texttt{supporting businesses} and \texttt{politics} are the most widely discussed topics on Twitter. At the same time, URLs related to \texttt{social networks} and \texttt{news and media} domains have been widely shared. At the same time, the sentiment analysis shows that people have a highly positive sentiment for COVID-19 preventative methods, while they display highly negative sentiments towards discussions on \texttt{politics} and \texttt{death tolls}. These sentiments seem to be impacted by the infection rates in certain countries as well. Meanwhile, the privacy analysis revealed \ik{how} people share more information about their personal circumstances on social media networks. Users who posted just 3 sensitive tweets become 100\% identifiable. Finally, the security analysis showed that a major portion of suspicious URL domains belonged to \texttt{IT, business}, or \texttt{search engines}.

As for future work, there are a few aspects that we would like to extend this work. For example, we can extend the VirusTotal analysis to individual URLs instead of domains. It will provide an in-depth URL analysis and a better characterisation of suspicious URLs. In addition, we can utilise Gaussian distribution, maximum entropy Markov model, etc., to quantify the privacy risk instead of using the basic HMM model. We also aim to extend our work on identifying, characterising, and analysing the impact of spreading rumours and misinformation on social networks about the pandemic. 

\Urlmuskip=0mu plus 1mu\relax
\bibliographystyle{IEEEtran}

\bibliography{main.bib}

\appendix
\section{Appendix A}
\label{appendix A}
\begin{table*}[htb]
  \centering
  \caption{Top categories of discussion in each country in different stages of the pandemic}
  \tabcolsep=0.05cm
    \scalebox{0.82} {
  \begin{tabular}{*{13}{l}}
    \toprule
     & \multicolumn{4}{c}{\bf Before} & \multicolumn{4}{c}{\dor \bf During} 
      & \multicolumn{4}{c}{\bf After} \\
    \cmidrule{2-13}
    {\bf Country} & {\bf Hashtag} & {\bf \#} & {\bf URL} & {\bf \#} &  \dor \bf Hashtag & \dor\bf \# & \dor \bf URL & \dor \bf \# & {\bf Hashtag} & {\bf \#} & {\bf URL} & {\bf \#}\\
    \midrule
    
    \multirow{5}{*}{Australia} & Support Business & 70,066 & News \& Media & 17,940 & \dor Support Business & \dor 420,189 & \dor News \& Media & \dor 119,728 & Support Business & 91,748 & News \& Media & 22,941\\
     & Political & 4,246 & Social Networks & 15,156 & \dor Political & \dor 28,396 & \dor Social Networks & \dor 99,852 & Political & 5,108 & Social Networks & 19,046\\
     & PCR Testing & 3,230 & IT & 3,276 & \dor Latest Updates & \dor 19,506 & \dor IT & \dor 19,551 & Latest Updates & 3,903 & IT & 4,031\\
     & Latest Updates & 2,982 & Gov \& Legal Org & 2,257 & \dor PCR Testing & \dor 11,194 & \dor Gov \& Legal Org & \dor 14,899 & Lockdown & 2,089 & Gov \& Legal Org & 4,012 \\
     & Community Cases & 2,003 & Business & 1,965 & \dor Community Cases & \dor 10,256 & \dor Business & \dor 12,642 & Frontline Workers & 1,987 & Health \& Wellness & 2,904\\
     
     \cmidrule{2-13}
     
     \multirow{5}{*}{India} & Support Business & 208,252 & Social Networks & 27,489 & \dor Support Business & \dor 1,145,364 & \dor Social Networks & \dor 230,476 & Support Business & 453,118 & Social Networks & 74,828\\
     & PCR Testing & 17,026 & News \& Media & 22,388 & \dor Community Cases & \dor 118,381 & \dor News \& Media & \dor 186,274 & Community Cases & 64,939 & News \& Media & 74,708\\
     & Prevent Spread & 15,726 & IT  & 4,874 & \dor Lockdown & \dor 91,813 & \dor IT & \dor 67,187 & Latest Updates & 33,296 & IT & 29,720\\
     & Stay Home & 15,454 & Business & 3,356 & \dor Latest Updates & \dor 86,267 & \dor Business & \dor 54,620 & PCR Testing & 32,409 & Business & 26,965 \\
     & Community Cases & 12,383 & Gov \& Legal Org & 1,905 & \dor Political & \dor 72,249 & \dor Streaming & \dor 20,285 & Political & 29,191 & Entertainment & 6,950\\
    
    \cmidrule{2-13}
    
    \multirow{5}{*}{UK} & Support Business & 522,843 & Social Networks & 201,185 & \dor Support Business & \dor 1,848,765 & \dor Social Networks & \dor 762,958 & Support Business & 1,011,303 & Social Networks & 404,736\\
    & Political & 36,245 & News \& Media & 107,457 & \dor Political & \dor 125,763 & \dor News \& Media & \dor 418,109 & Vaccination & 86,625 & News \& Media & 186,675\\
    & Latest Updates & 29,165 & IT & 45,378 & \dor Latest Updates & \dor 110,149 & \dor IT & \dor 180,600 & Latest Updates & 75,306 & IT & 107,818\\
    & PCR Testing & 23,542 & Business & 24,647 & \dor Lockdown & \dor 104,482 & \dor Business & \dor 103,520 & Political & 65,568 & Business & 65,349\\
    & Prevent Spread & 18,026 & Gov \& Legal Org & 18,438 & \dor Frontline Workers & \dor 71,975 & \dor Gov \& Legal Org &  \dor 64,717 & PCR Testing & 46,558 & Gov \& Legal Org & 44,459\\
    
    \cmidrule{2-13}
    
    \multirow{5}{*}{US} & Support Business & 359,658 & Social Networks & 177,730 & \dor Support Business & \dor 586,671 & \dor Social Networks & \dor 340,671 & Support Business & 570,772 & Social Networks & 294,266\\
    & Political & 35,219 & News \& Media & 104,524 & \dor Political & \dor 58,866 & \dor News \& Media & \dor 191,389 & Political & 50,181 & News \& Media & 195,409\\
    & PCR Testing & 24,294 & IT & 25,862 & \dor Latest Updates & \dor 34,886 & \dor IT & \dor 54,082 & Latest Updates & 36,058 & IT & 57,521\\
    & Prevent Spread & 14,478 & Business & 7,972 & \dor Frontline Workers & \dor 22,577 & \dor Business & \dor 20,762 & Death Toll & 24,549 & Business & 27,533\\
    & Latest Updates & 13,885 & Streaming & 6,830 & \dor Death Toll & \dor 19,262 & \dor Streaming & \dor 18,296 & News \& Media & 22,136 & Streaming & 14,793\\
    \bottomrule
  \end{tabular}
  }
  \label{table:statAnalysisSummary}
\end{table*}

\subsection{Data Characterisation}

In Section~\ref{sec:statAnalysis}, we perform data characterisation by discussing the trends in the URLs and hashtags of tweets. We identified the top most discussed COVID-related topics in hashtags and URLs during all three lockdown periods. In Table~\ref{table:statAnalysisSummary}, we provide a more detailed view \ik{of} data characterisation by breaking \ik{it} down into country level. We can clearly observe that \texttt{Support Business} is the most discussed hashtagged topic among all the four countries, followed by \texttt{Politics} and \texttt{Latest Updates} topics. These topics are also common across the lockdown periods. Similarly,  URLs related to \texttt{News \& Media} and \texttt{Social Networks} are mostly shared among all the countries and all the periods. 

\begin{figure*}[h]
\centering
\subfloat[Before]{
\includegraphics[width=0.45\textwidth, keepaspectratio]{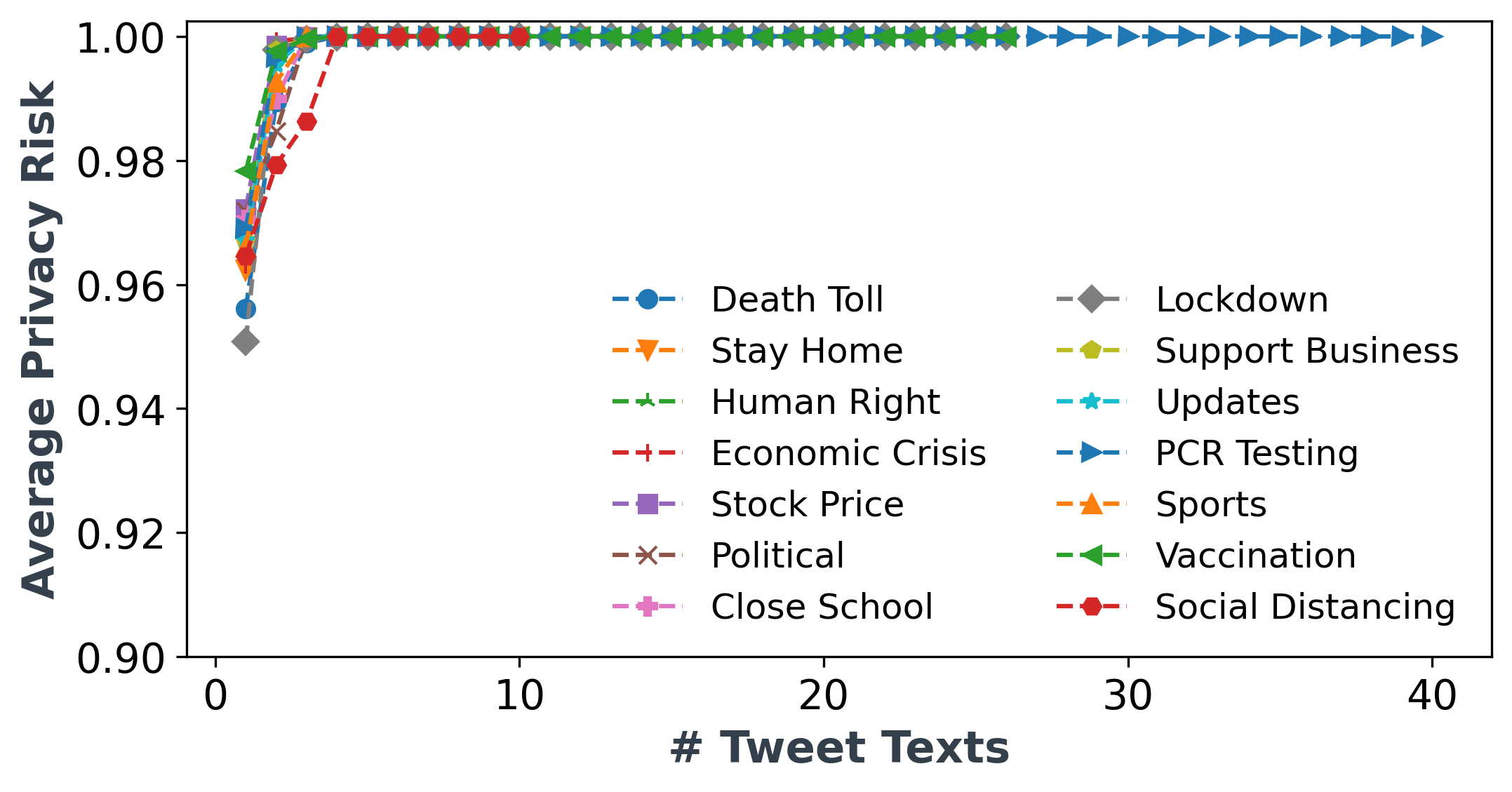}
}
\subfloat[During]{\includegraphics[width=0.45\textwidth, keepaspectratio]{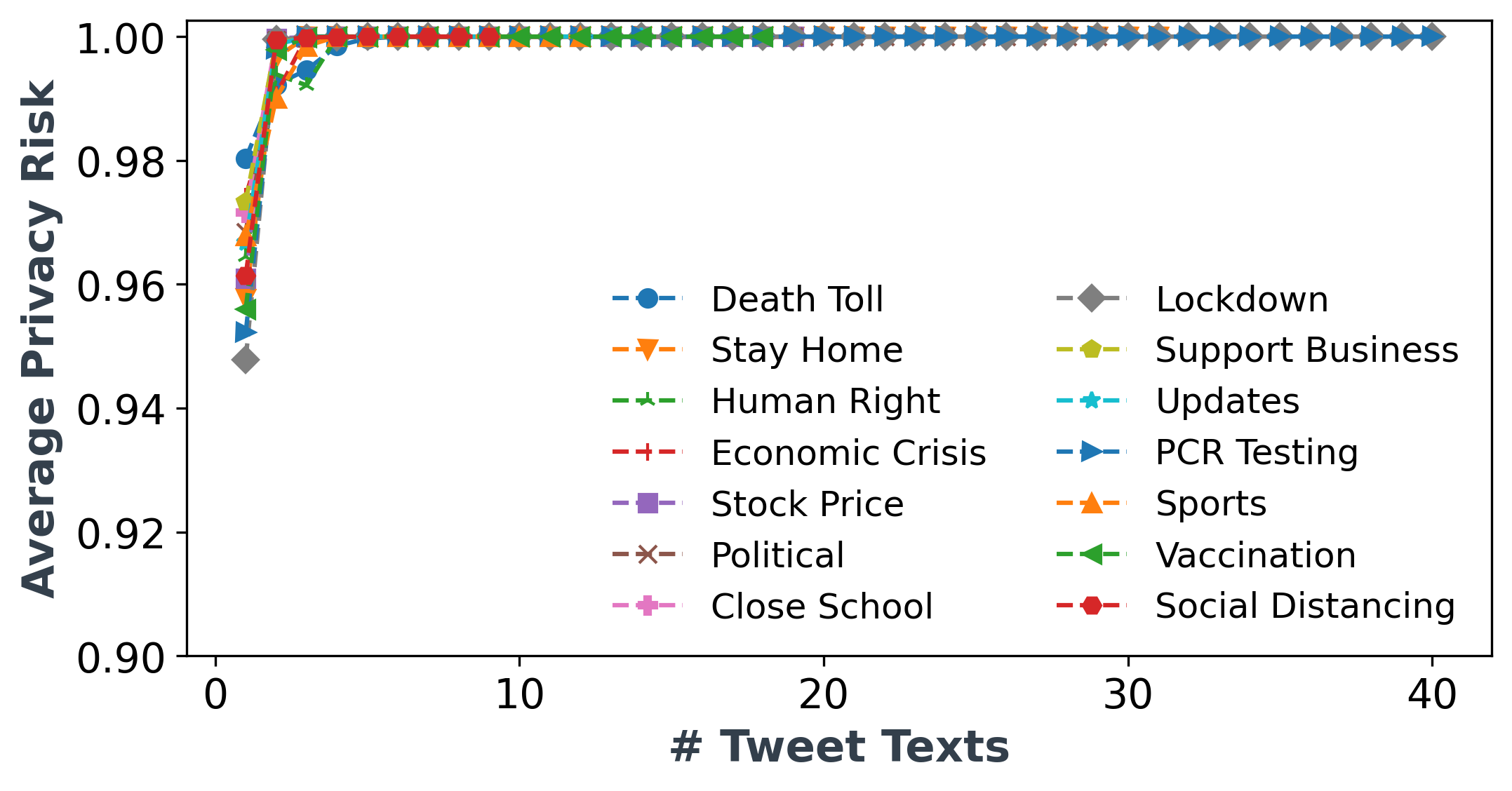} \label{fig:duringrisk}}

\subfloat[After]{\includegraphics[width=0.45\linewidth, keepaspectratio]{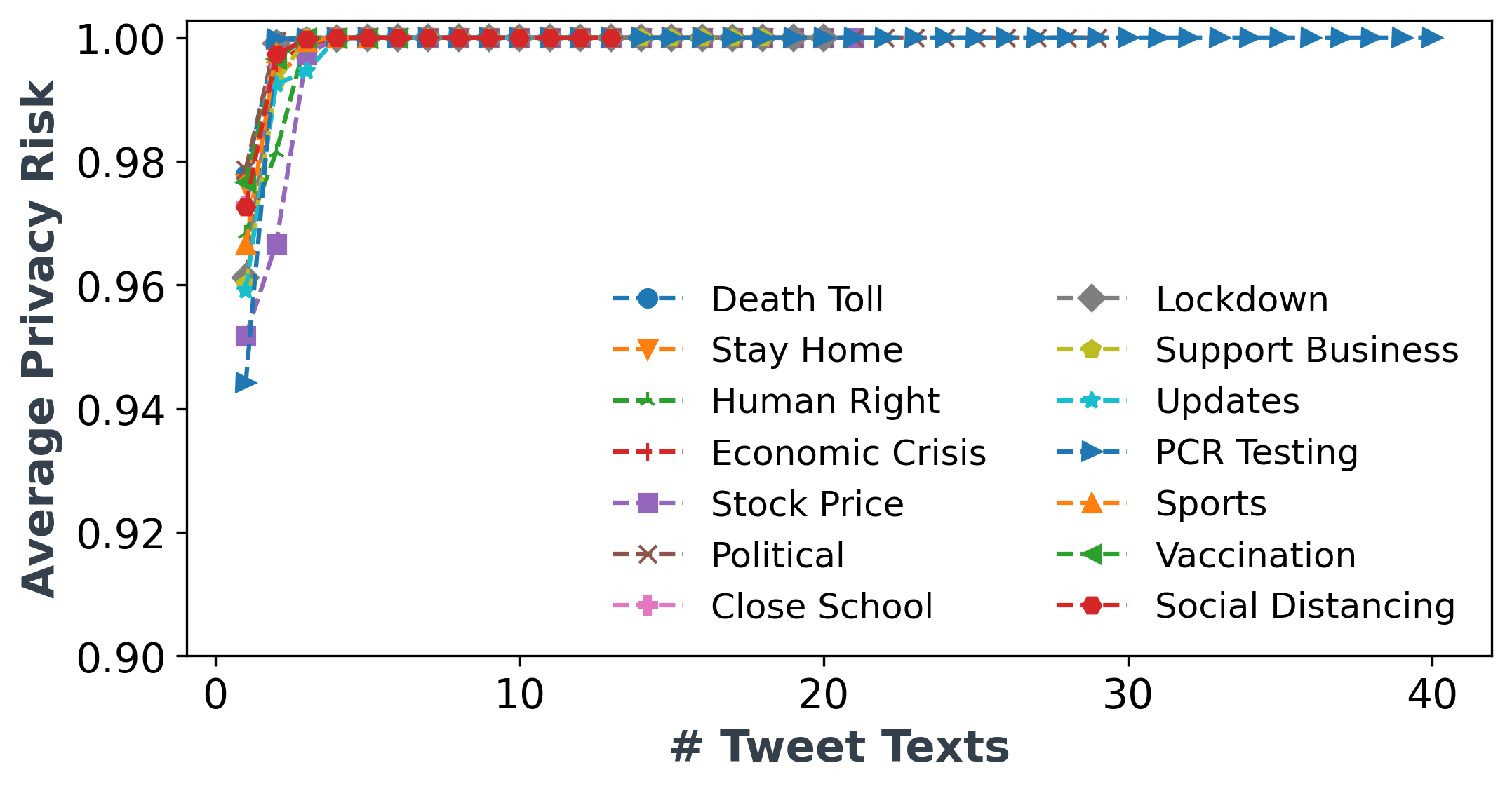}}
\caption{\small Average privacy risk with the increasing number of web entries (i.e., tweets) in three different period: {\bf Before}, {\bf During}, and {\bf After} lockdown.}
\label{fig:overallPrivacyRisk}
\end{figure*}

\subsection{Perception Analysis Toward COVID-19}

\begin{figure*}[h]
\centering
\subfloat[Australia]{
\includegraphics[width=0.4\textwidth, keepaspectratio]{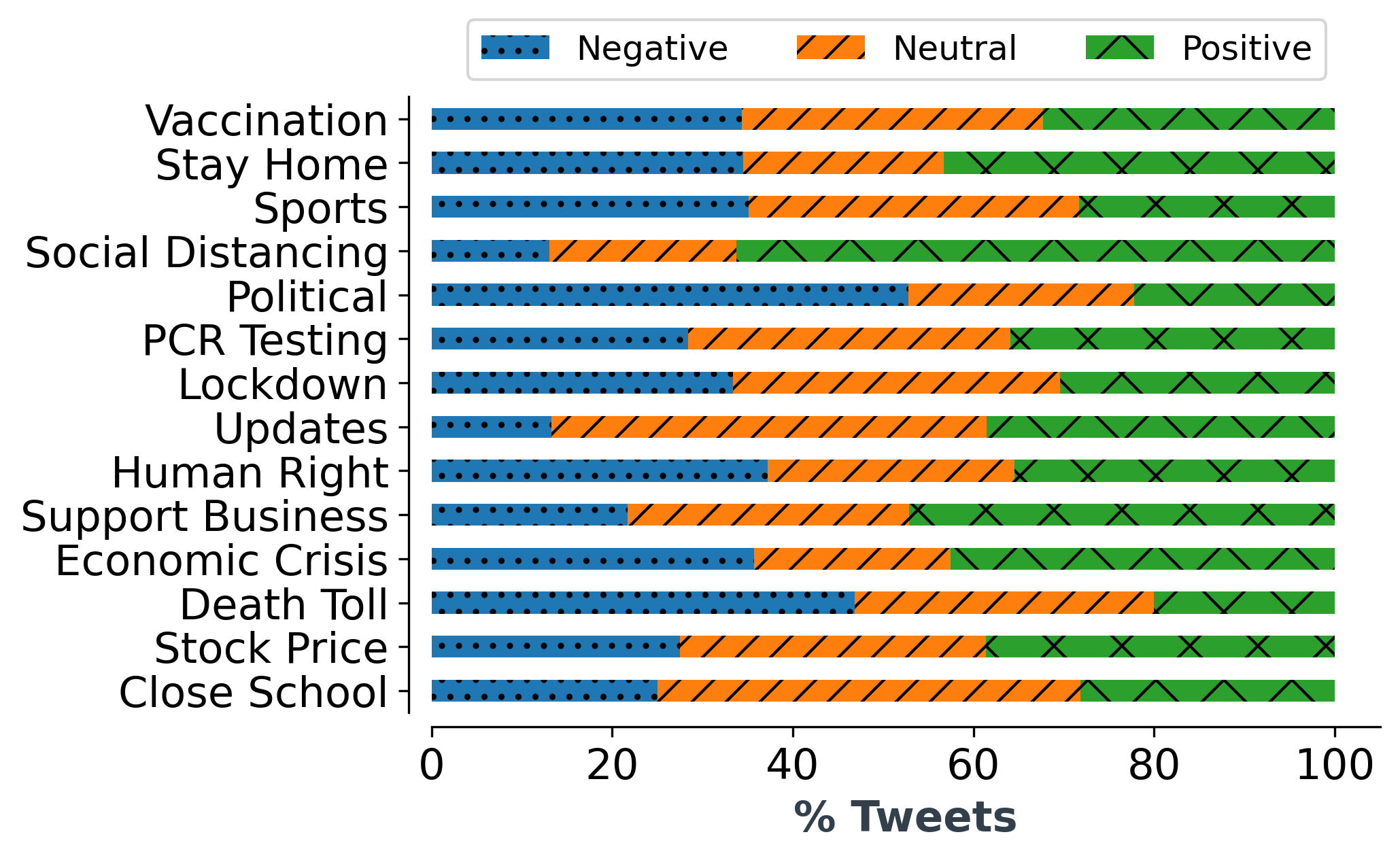}
}
\subfloat[India]{\includegraphics[width=0.4\textwidth, keepaspectratio]{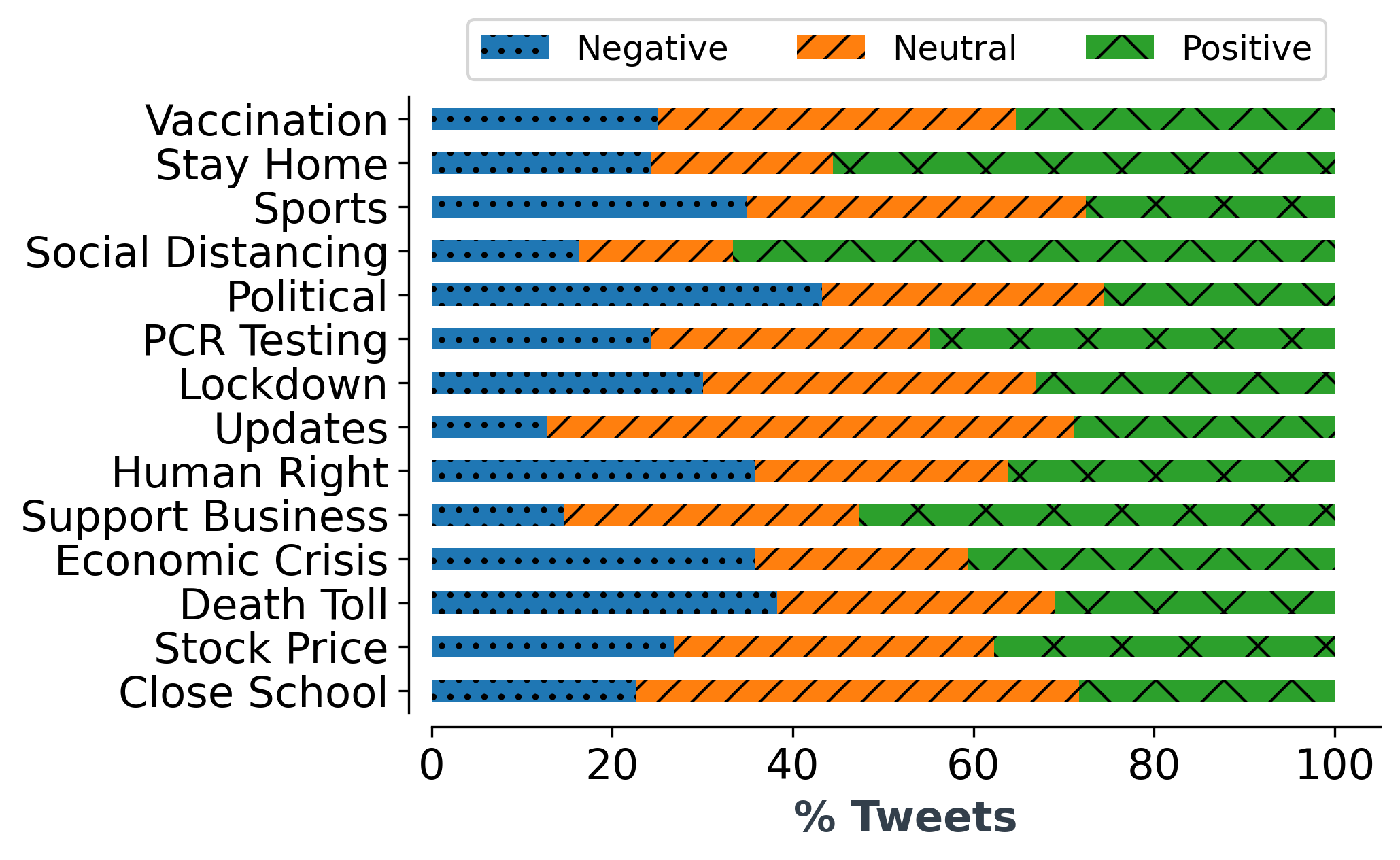}}\quad
\subfloat[UK]{\includegraphics[width=0.4\linewidth, keepaspectratio]{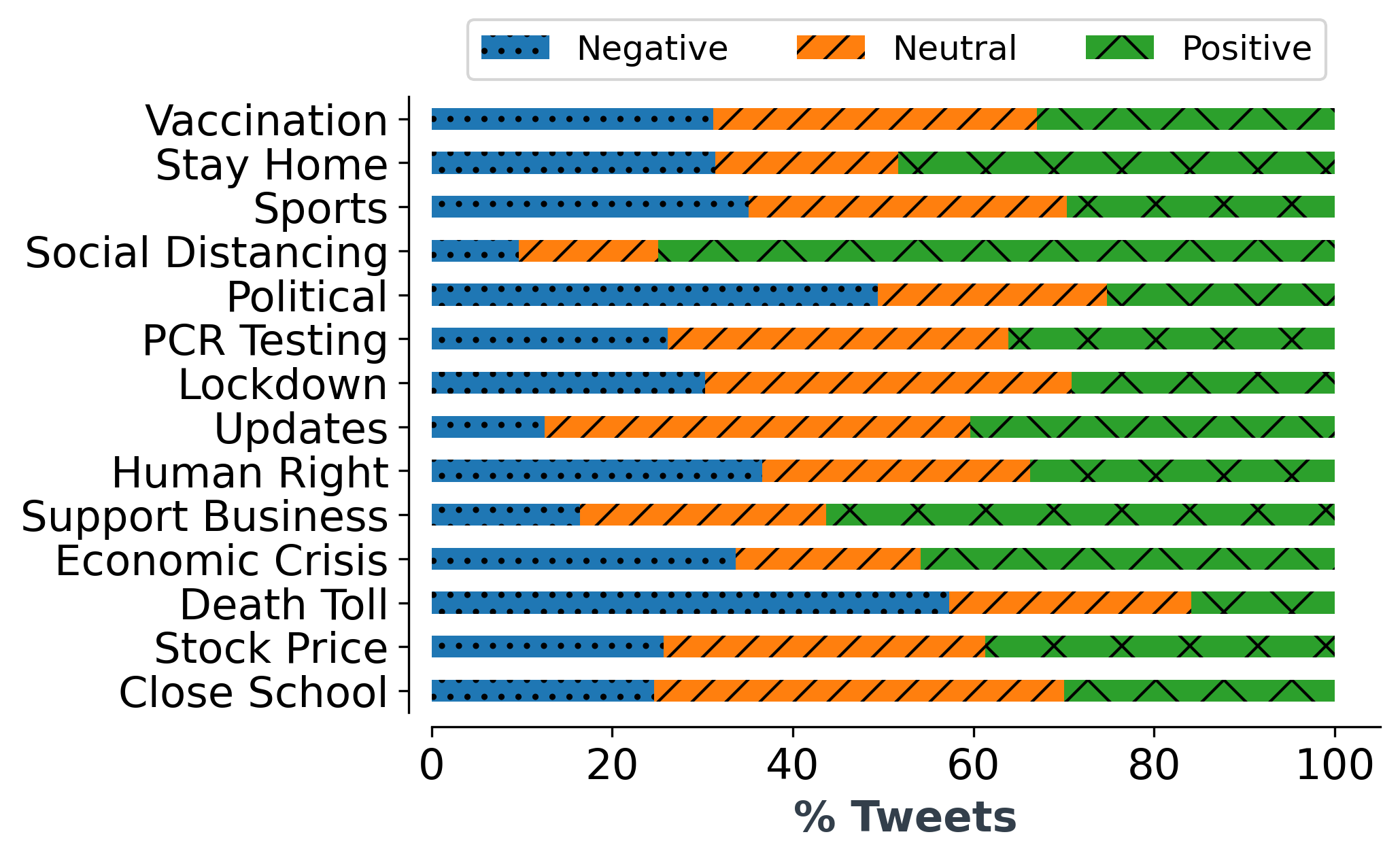}}
\subfloat[US]{\includegraphics[width=0.4\linewidth, keepaspectratio]{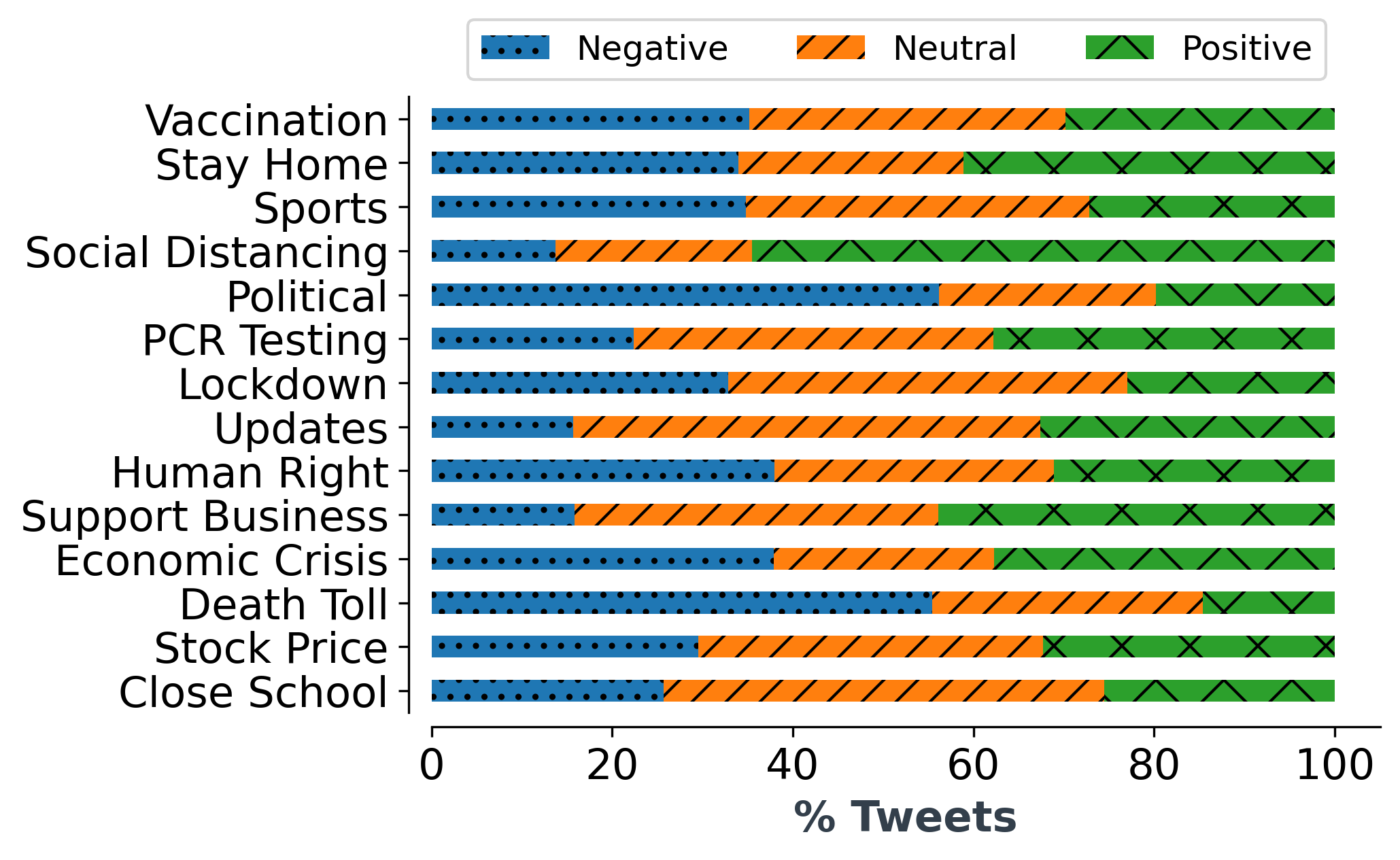}}
\caption{\small User Sentiment for each Topic in Australia, India, and UK, and US, for all the lockdown periods}
\label{fig:sentiment_countries}
\end{figure*}

In Section~\ref{sec:sentAnalysis}, we analyse the relation of people's sentiments with infection rates (IR) and COVID restrictions in a region. In general, we try to identify if there is an impact of social media tweets in managing the pandemic. In regards to this, Figure~\ref{fig:sentiment_countries} illustrates the trend in people's sentiments across four countries for all the lockdown periods. Clearly, \texttt{Death Toll} has received the highest negative sentiments from UK and US. In general, we observe similar trends for the topics across all the countries. For instance, topic \texttt{Social Distancing} has received approximately 70\% positive sentiments from all the countries.

\vspace{-0.3cm}
\subsection{Privacy Analysis}

In Section~\ref{sec:prvcAnalysis}, we quantify privacy risks against social media tweets and reveal interesting findings about user identification from just 3 sensitive tweets. In Figure~\ref{fig:overallPrivacyRisk}, we show an average privacy risk across various topics and \ik{an} increasing number of tweets. It is clear from the figure that COVID-19-related tweets are capable of re-identifying users with at least 94\% privacy risk. Similarly, in Table~\ref{tab2:case1}, we provide a few examples where users mention personal identifiable information (PII) in their tweets. 

\begin{table*}[!htbp]
\footnotesize
\caption{Examples of tweets from all lockdown periods.}
\vspace{-0.15cm}
\label{tab2:case1}
\begin{center}
\resizebox{1.0\textwidth}{!}{
\begin{tabular}{|c|p{12cm}|c|}
\hline
%\toprule
\centering \textbf{User Anonymised ID} & \centering \textbf{Tweets} & \textbf{Topic} \\
\hline
345767 & 'A friend of mine was issued a \$1,652 on-the-spot coronavirus fine this morning for entering Costco in Docklands as it was beyond the 5 km radius of her house. Sad! Be careful out there, folks.' & Stay Home (1 Tweet)    \\
\hline
168973 & '@JamesMelville Is your point that they should be worried about catching Covid-19 here? ', According to ONS data, 24.9\% of people who died of covid-19 died in a care home. Versus 26.3\% of people who died of covid-19 who had diabetes. It would seem a comparable blind spot.  (Of course some of those may have been diabetic \&amp; also in a care home but the point still stands)', '@GOVUK But I can go to my parents with my kids whilst Im sick with coronavirus right?', '@richardhyland @jonthepon Selfishness vs altruism. I was tutted at last weekend for trying to avoid three people walking side by side! (My 6 year old daughter yelled coronavirus! at them after they passed )' & Stay Home (4 Tweets)   \\ 
\hline
789654 & 'Im helping to fight \#COVID19. We only need 164 more people on the app to get a COVID estimate for Jefferson County. Please help by taking 1 min daily to report how you feel . You also get an estimate of COVID in your area. ' & DeathToll (1 Tweet)    \\
\hline
904365 & '\#Lewisham up 12   \#COVID19 confirmed cases', '\#Bromley up 40 confirmed \#COVID19 cases   as of 22 Oct', '\#Bromley up 46 confirmed \#COVID19 cases as of 23 Oct', 'Another positive \#COVID19 cases in my   son's year at school.', ''\#Lewisham up 26 confirmed \#COVID19 cases   as of 3 Nov. Sadly, an increase of 1 death registered to 23 Oct'& Death Toll (4 Tweet)   \\
\hline
167843 & '@PuneCityPolice i have stuck here in pune with mom and my family at native place , so i have apply pass at covid19.mhpolice and token no is PUN05522058080679 And i have apply gujarat approval letter and medical certificate with it. Pls give me pass.' & LockDown (1 Tweet) \\ 
\hline
369225 & 'We got a call from a UAE-based airlines in Frankfurt that our daughter Miha has been accepted on the flight and they are bringing her to Dubai' & LockDown (1 Tweet) \\
\hline
666231 & 'My blood naturally wants to clot. (Factor V Leiden is a genetic mutation. Yes, I am *lucky* enough to have blood clotting as my mutant powers.) In addition to being old and fat, now I Iearn another reason 'Rona REALLY wants to kill me.'& Vaccination (1 Tweet)  \\
\hline
541209 & 'OK! Now I'm really   angry. The current evidence from Israel is that the single dose is only about   30\% effective at best. I know you think that's plenty but you're playing with   our lives and not very successfully. So listen to the real science. 2 doses. ', 'And I am still feeling the effects of   *suspected* Covid19  - 6 weeks later.   Very very mild. I am 99\% better. But this virus lingers people  ( and I have an excellent immune system -   despite other health issues)', 'Genetic quirk could explain how pangolins   can tolerate coronaviruses' & Vaccination (3 Tweets) \\ 
\hline
905643  & `@NHSBSolCCG my wife is currently 33 weeks pregnant and we are now going to hospital on a weekly basis for checks and monitoring. I am worried about giving my unborn child coronavirus due to these frequent visits. Would I now be able to get the vaccine as they are at high risk?', `Obstetricians and Gynaecologists are recommending that women who are offered a Covid vaccine have if before they get pregnant. There is no need to delay pregnancy after the vaccine.' & Vaccination (2 Tweets) \\
\hline
\end{tabular}
}
\end{center}
\end{table*}

\end{document}